\documentclass[aps,twocolumn,showpacs,superscriptaddress,floatfix]{revtex4-1}
\usepackage{color}
\usepackage{comment}
\usepackage{graphics}
\usepackage{amsthm}
\usepackage{mathtools}
\usepackage{lipsum}
\usepackage{tikz}
\usepackage{listings}
\usetikzlibrary{shapes,arrows}

\def\gee        {\varepsilon}
\def\gl         {\lambda}
\def\go         {\omega}

\def\kk         {{\bf k}}
\def\jj         {{\bf j}}
\def\dd         {{\bf d}}
\def\cc         {{\bf e}}

\def\qq         {{\bf q}}
\def\xx         {{\bf x}}
\def\rr         {{\bf r}}

\def\GG         {{\bf G}}

\def\PP         {{\bf P}}

\def\efield{\boldsymbol{\mathcal{E}}}

\def\nt         {\tilde{n}}
\def\mt         {\tilde{m}}
\def\kkt        {\tilde{\kk}}

\renewcommand{\[}{\left[}

\newcommand{\be}{\begin{equation}}
\newcommand{\ee}{\end{equation}}
\newcommand{\ben}{\begin{equation*}}
\newcommand{\een}{\end{equation*}}
\newcommand{\bea}{\begin{eqnarray}}
\newcommand{\eea}{\end{eqnarray}}
\newcommand{\bean}{\begin{eqnarray*}}
\newcommand{\eean}{\end{eqnarray*}}

\newcommand{\yambo} {{\normalfont\ttfamily yambo}}
\newcommand{\Yambo} {{\normalfont\ttfamily Yambo}}
\newcommand{\ypp}  {{\normalfont\ttfamily ypp}}
\newcommand{\ay}  {{\normalfont\ttfamily a2y}}
\newcommand{\py}  {{\normalfont\ttfamily p2y}}
\newcommand{\ey}  {{\normalfont\ttfamily e2y}}

\newcommand{\wannier} {{\normalfont\ttfamily wannier90}}
\newcommand{\want} {{\normalfont\ttfamily WanT}}

\newcommand{\yambopj} {{\normalfont\ttfamily yambo\_PJ}}
\newcommand{\ypppj} {{\normalfont\ttfamily ypp\_PJ}}
\newcommand{\yamboph} {{\normalfont\ttfamily yambo\_ph}}
\newcommand{\yppph} {{\normalfont\ttfamily ypp\_ph}}
\newcommand{\yambort} {{\normalfont\ttfamily yambo\_rt}}
\newcommand{\ypprt} {{\normalfont\ttfamily ypp\_rt}}
\newcommand{\yambonl} {{\normalfont\ttfamily yambo\_nl}}
\newcommand{\yppnl} {{\normalfont\ttfamily ypp\_nl}}
\newcommand{\yambokerr} {{\normalfont\ttfamily yambo\_kerr}}

\newcommand{\Yambopy} {{\normalfont\ttfamily Yambopy}}
\newcommand{\yambopy} {{\normalfont\ttfamily yambopy}}
\newcommand{\aiida} {{\normalfont\ttfamily AiiDA}}

\newcommand{\abinit} {{\normalfont\ttfamily Abinit}}
\newcommand{\pwscf} {{\normalfont\ttfamily pwscf}}
\newcommand{\qe} {{\normalfont\ttfamily Quantum ESPRESSO}}

\newcommand{\HXCkind} {{\small\ttfamily HXC\_kind}}

\newcommand{\netcdf} {{\normalfont\ttfamily NetCDF}}
\newcommand{\etsfio} {{\normalfont\ttfamily ETSF-IO}}
\newcommand{\xc} {{\normalfont\ttfamily XC}}

\newcommand{\rhom}{\underline{\rho}}
\newcommand{\hm}{\underline{h}}
\newcommand{\Um}{\underline{U}}
\newcommand{\rmat}{\underline{r}}
\newcommand{\Sm}{\underline{\Sigma}}

\newcommand{\Kmm}{\underline{\underline{K}}}

\newcommand{\TOV}{{ Dipartimento di Fisica, Universit\`a di Roma ``Tor Vergata'', Via della Ricerca Scientifica 1,I--00133 Roma, Italy}}
\newcommand{\UK}{{School of Mathematics and Physics, Queen's University Belfast, Belfast BT7 1NN, Northern Ireland, UK}}
\newcommand{\NANO}{{Centro S3, Istituto Nanoscienze---Consiglio Nazionale delle Ricerche (CNR-NANO), via Campi 213/A, 41125 Modena, Italy}}
\newcommand{\ISM}{{Istituto di Struttura della Materia---Consiglio Nazionale delle Ricerche (CNR-ISM), via Fosso del Cavaliere 100, 00133 Rome, Italy}}
\newcommand{\ISMMLIB}{{Istituto di Struttura della Materia---Consiglio Nazionale delle Ricerche (CNR-ISM), Division of Ultrafast Processes in Materials (FLASHit), Via Salaria Km 29.5, CP 10, I-00016 Monterotondo Stazione, Italy}}
\newcommand{\CNRS}{{Aix Marseille Universit\'e, CNRS, CINaM UMR 7325, Campus de Luminy Case 913, 13288 Marseille, France}}
\newcommand{\AIX}{{Aix-Marseille Universit\'e, Laboratoire de Physique des Interactions Ioniques et Mol\'eculaires (PIIM), UMR CNRS 7345, F-13397 Marseille, France }}
\newcommand{\CIN}{{CINECA National Supercomputing Center, Casalecchio di Reno, I--40033 Bologna, Italy }}
\newcommand{\MOD}{{Universit\`a di Modena e Reggio Emilia, Via Campi 213/A, 41125 Modena, Italy}}
\newcommand{\ETSF}{{European Theoretical Spectroscopy Facility (ETSF)}}
\newcommand{\EPFL}{Theory and Simulations of Materials (THEOS) and National Centre for Computational Design and Discovery of Novel Materials (MARVEL),
\'Ecole Polytechnique F\'ed\'erale de Lausanne, 1015 Lausanne, Switzerland}
\newcommand{\DISC}{Dipartimento di Scienze Chimiche, University of Padova, I-35131, Padova, Italy}
\newcommand{\IMCN}{Institute of Condensed Matter and Nanoscience (IMCN), Universit\'{e} catholique de Louvain, B-1348, Louvain-la-Neuve, Belgium}
\newcommand{\LIEGE}{nanomat/Q-mat/CESAM, Universit\'e de Li\`ege, Institut de Physique, B-4000 Sart Tilman, Li\`ege, Belgium}

\newcommand{\ICMUV}{Institute of Materials Science (ICMUV), University of Valencia, Catedr\'{a}tico Beltr\'{a}n 2, E-46980, Valencia, Spain}
\newcommand{\PHYMS}{Physics and Materials Science Research Unit, University of Luxembourg, 162a avenue de la Fa\"iencerie, L-1511 Luxembourg, Luxembourg}

\newcommand{\mydrop}[1]{}

\tikzstyle{decision} = [diamond, draw, , 
    text width=1.5em, text badly centered, node distance=1cm, inner sep=0pt]
\tikzstyle{decision_recheck} = [diamond, draw, , 
    text width=3.5em, text badly centered, node distance=1cm, inner sep=0pt]
\tikzstyle{block} = [rectangle, draw, , 
    text width=5em, text centered, rounded corners, minimum height=2em]
\tikzstyle{line} = [draw, -latex']
\tikzstyle{cloud} = [draw, ellipse, node distance=2cm,
    minimum height=2em]
\usepackage{array}
\usepackage{makecell}
\usepackage{soul}

\begin{document}

\title[Many-body perturbation theory calculations using the yambo code]{Many-body perturbation theory calculations using the yambo code}

%\author{D. Sangalli,$^{1,2}$ C. Hogan,$^3$ A. Ferretti,$^{4,2}$ D. Varsano,$^{4,2}$ M. Gr\"uning,$^{5}$ M. Palummo,$^3$ C. Attaccalite,$^{6}$ E. Cannuccia,$^{3,7}$ M. Marsili,$^8$ F. Affinito,$^9$ P. Melo,$^{10}$ A. Molina-S\'{a}nchez,$^{11}$ I. Marri,$^{4,12}$ H. Miranda,$^{13}$ A. Marrazzo,$^{14}$ G. Prandini,$^{14}$ P. Bonf\`a,$^9$ F. Paleari,$^{15}$ M.O. Atambo,$^{4,12}$, A. Marini$^{1,2}$}
%
%\address{$^1$ \ISM}
%\address{$^2$ \ETSF}
%\address{$^3$ \TOV}
%\address{$^4$ \NANO}
%\address{$^5$ \UK}
%\address{$^6$ \CNRS}
%\address{$^7$ \AIX}
%\address{$^8$ \DEF}
%\address{$^9$ \CIN}
%\address{$^{10}$ \DEF}
%\address{$^{11}$ \DEF}
%\address{$^{12}$ \MOD}
%\address{$^{13}$ \DEF}
%\address{$^{14}$ \EPFL}
%\address{$^{15}$ \DEF}

\author{D. Sangalli}
\affiliation{\ISMMLIB}
\affiliation{\ETSF}
\author{A. Ferretti}
\affiliation{\NANO}
\affiliation{\ETSF}
\author{H. Miranda}
\affiliation{\IMCN}
\author{C. Attaccalite}
\affiliation{\CNRS}
\affiliation{\ETSF}
\author{I. Marri}
\affiliation{\NANO}
\author{E. Cannuccia}
\affiliation{\TOV}
\affiliation{\AIX}
\author{P. Melo}
\affiliation{\LIEGE}
\affiliation{\ETSF}
\author{M. Marsili}
\affiliation{\DISC}
\author{F. Paleari}
\affiliation{\PHYMS}
\author{A. Marrazzo}
\affiliation{\EPFL}
\author{G. Prandini}
\affiliation{\EPFL}
\author{P. Bonf\`a}
\affiliation{\CIN}
\author{M.O. Atambo}
\affiliation{\NANO}
\affiliation{\MOD}
\author{F. Affinito}
\affiliation{\CIN}
\author{M. Palummo}
\affiliation{\TOV}
\affiliation{\ETSF}
\author{A. Molina-S\'{a}nchez}
\affiliation{\ICMUV}
\author{C. Hogan}
\affiliation{\ISM}
\affiliation{\TOV}
\affiliation{\ETSF}
\author{M. Gr\"uning}
\affiliation{\UK}
\affiliation{\ETSF}
\author{D. Varsano}
\affiliation{\NANO}
\affiliation{\ETSF}
\author{A. Marini}
\affiliation{\ISMMLIB}
\affiliation{\ETSF}
\email[corresponding author: ]{andrea.marini@cnr.it}

\begin{abstract}
\yambo{} is an open source project aimed at studying excited state properties of condensed matter systems from first principles using many-body methods. As input, \yambo{} requires ground state electronic structure data as computed by density functional theory codes such as \qe{} and \abinit. \yambo{}'s capabilities include the calculation of linear response quantities (both independent-particle and including electron-hole interactions), quasi-particle corrections based on the GW formalism, optical absorption, and other spectroscopic quantities.
Here we describe recent developments ranging from the inclusion of important but oft-neglected physical effects such as electron-phonon interactions to the implementation of a real-time propagation scheme for simulating linear and non-linear optical properties. 
Improvements to numerical algorithms and the user interface are outlined. 
Particular emphasis is given to the new and efficient parallel structure that makes it possible to exploit modern high performance computing architectures. 
Finally, we demonstrate the possibility to automate workflows by interfacing with the \yambopy{} and AiiDA software tools. 
%These new features are supported by reference and validation results. 
%Besides the new physical features this work describes the numerous improvements of the numerical algorithms and user interaction. 
%Particular emphasis is given to the new and efficient parallel structure that makes it possible to exploit massively-parallel architectures. 
%Finally, we discuss the possibility to take advantage of automation tools such as \yambopy{} and AiiDA to perform automatic flows of calculations. 
\end{abstract}

%\maketitle
%\newpage

\maketitle
%
% comment the following to get rid of the TOC
%
\tableofcontents
%
%=======================
% FIND HERE THE LIST OF AUTHOR ABBREVIATIONS
%=======================
%
\mydrop{
\noindent{\bf Author abbreviations} 
\begin{itemize}
\item Andrea Marini\,{\bf (AM)}
\item Davide Sangalli\,{\bf (DS)}
\item Conor Hogan\,{\bf (CH)}
\item Andrea Ferretti\,{\bf (AF)}
\item Daniele Varsano\,{\bf (DV)}
\item Myrta Gr\"uning\,{\bf (MG)}
\item Claudio Attaccalite\,{\bf (CA)}
\item Elena Cannuccia\,{\bf (EC)}
\item Maurizia Palummo\,{\bf (MP)}
\item Margherita Marsili\,{\bf (MM)}
\item Fabio Affinito\,{\bf (FA)}
\item Pedro Melo\,{\bf (PM)}
\item Alejandro Molina-S\'{a}nchez\,{\bf(AMS)}
\item Ivan Marri\,{\bf(IM)}
\item Henrique Miranda\,{\bf(HM)}
\item Antimo Marrazzo\,{\bf(AMO)}
\item Gianluca Prandini\,{\bf(GP)}
\item Pietro Bonf\`a\,{\bf(PB)}
\item Fulvio Paleari\,{\bf(FP)}
\item Michael Ontita Atambo\,{\bf(MOA)}
\end{itemize}
}

%%%%%%%%%%%%%%%%%%%%%%%%%%%%
\section{The Yambo Project}
%%%%%%%%%%%%%%%%%%%%%%%%%%%%

Computational materials science based on first principles atomistic methods plays a key role in the discovery, characterization, and engineering of novel and complex materials.
While density functional theory (DFT) is the established workhorse for ground state properties of a wide range of systems ranging from atoms and molecules to solids and nanostructures containing thousands of atoms, there is an increasing demand for an \emph{accurate} description of \emph{excited state} properties in even the most challenging materials.
% as interest moves towards optoelectronics etc
Within the framework of solid state physics, the Green's function formulation of many-body perturbation theory (MBPT) --- specifically the GW approach to quasiparticles (QP) for charged excitations and the Bethe-Salpeter equation (BSE) for neutral excitations --- offers a quantitatively accurate solution~\cite{onida2002}.
The GW-BSE approach has been implemented in a number of free and commercially available codes~\cite{Deslippe2012,Umari2010,Marini2009,Gonze2005,Shishkin2006,Schlipf2018,govoni2015} and applied to a wide range of materials
(we cited works where the GW-BSE approach is coded in plane-waves, for a recent and more comprehensive review, see Ref.~\onlinecite{Leng2016}).
Nonetheless, the complexity and relatively poor scaling of the GW-BSE method, and often of its implementation, constitutes a barrier towards its application to realistic systems of large size or to physical phenomena that lie outside the scope of most state-of-the-art approaches.

Tackling these challenges in a software environment requires a fourfold strategy:
\begin{itemize}
    \item First, the description of underlying physical phenomena must be regularly advanced, both in terms of extensions of existing tools and by devising new methods. Oft-neglected terms such as electron-phonon and spin-orbit coupling play a crucial role in several physical phenomena. Examples are the finite temperature properties (dictated by the electron--phonon interaction) or the study of novel  materials like topological insulators, perovskites and layered transition metal dichalcogenides. In addition to extensions of existing tools \yambo{} implements brand new methods like  real--time tools to tackle the calculation of nonlinear optical properties.
    \item Second, algorithms must be refined and augmented in order to improve technical precision and numerical efficiency. This includes tricks for accelerating convergence as well as implementing alternatives to standard GW-BSE approximations such as plasmon-pole models of electronic screening and the Tamm-Dancoff approximation to exciton coupling.
    \item Third, codes must be designed to follow current trends in high-performance computing towards massively parallel, distributed memory architectures, while allowing for flexibility and control over tasks, memory, and disk usage in order to keep simulations efficient.
    \item Fourth, as the codes themselves become more complex and harder to maintain, modern software practices must be adopted. These include a wide range of aspects including improved documentation, use of modules and standard libraries, and automation of tasks for convergence, benchmarking and reproducibility.
\end{itemize}

\begin{figure}%[h!]
    \centering
    \includegraphics[clip,width=0.50\textwidth]{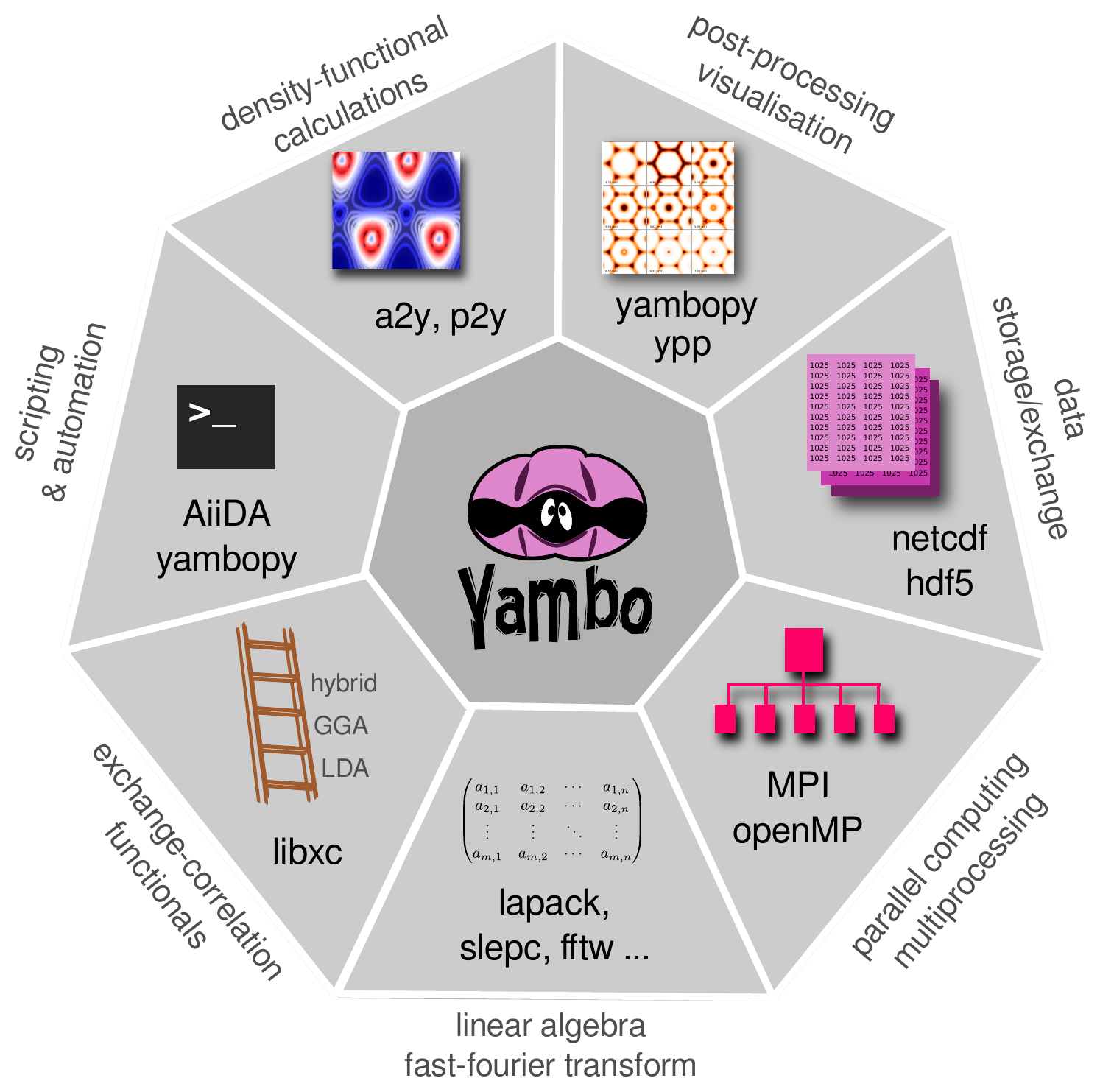}  
    \caption{
%    Options:
%    \begin{enumerate}
%        \item The \yambo\ project in a (heptagonal) nutshell
%        \item The \yambo\ project: physics, features, fast, and fun
%        \item BEHOLD THE SEVEN FINGERS OF YAMBO
%        \item 
        The \yambo{} project combines cutting edge computational materials science within a beyond-DFT framework with high performance algorithms, tools, and libraries
%    \end{enumerate}
    \label{fig:int_scheme}
    }
\end{figure}

In this paper we describe how the \yambo{} project has embraced this broad strategy.
\yambo{} is an open-source code based on many-body perturbation theory for computing electronic and optical excitations within a high performance environment 
%that combines cutting edge materials science with high performance algorithms, tools and libraries 
(Fig.~\ref{fig:int_scheme}).
Since its first public release in 2008, the project has evolved in a dramatic fashion and its development and user base has greatly expanded. Within the following ten years, the original paper was cited more than 500 times --- considerable for a pure MBPT code --- and the code has been used in many high impact studies spanning a wide range of novel materials and exciting technologies.
The highest cited applications cover graphene derivatives~\cite{Ruffieux2016,Samarakoon2011,Peelaers2011b,Vanin2009},
metal-halide perovskites~\cite{Filip2014,Wright2016},
van der Waals bonded layered compounds~\cite{Bernardi2013,Molina-Sanchez2013,Gao2012c,Palummo2015d},
Li-air and K-ion batteries~\cite{Hummelshoj2010,Luo2011},
and
TiO\textsubscript{2} photocatalytic surfaces~\cite{Chiodo2010,Kang2010a},
to select just a handful.
\yambo{} has moreover helped advance  fundamental understanding of  physical phenomena such as excitonic Bose-Einstein condensation~\cite{Cudazzo2010}, 
excitonic insulators~\cite{Varsano2017}, 
the influence of zero point motion~\cite{Cannuccia2011},  
charge transfer excitations~\cite{Hogan2013a}, etc. %\textbf{Select our best!}
A full list of publications can be found through our website~\cite{yambo_web}, \url{http://www.yambo-code.org}.

Part of \yambo's popularity and success may be ascribed to the code's user-friendliness: thanks to an intelligent command line interface, a full GW-BSE calculation on an unfamiliar material can in principle be carried out launching a single command. Extensive user documentation is provided on our website~\cite{yambo_web}. This includes descriptions of the fundamental theory, command line interface, and input variables, and provides a wide range of tutorials directed at explaining different functionalities of the code across a number of systems with different dimensionalities. 
Support is given by the developers through a forum.
In addition to the website~\cite{yambo_web,yambo_web_note}, the theory and use of \yambo{} has been disseminated through a number of international schools and workshops including a dedicated biennial CECAM event run by the developers and aimed at showcasing the latest developments.

The first major release, version 3.2.0, was described in detail in Marini et. al.~\cite{Marini2009} (henceforth referred to as CPC2009), and therefore the basic methodology, formalism, and code structure will not be repeated here.
Instead we describe the main additions made to the code up to and including version 4.4.0.
% unless otherwise specified.
Much development of the code has been driven by its status as a key \textit{ab initio} spectroscopy code of the European Theoretical Spectroscopy Facility (ETSF)~\cite{etsf} and as a flagship code of the {\em MaX European Centre of Excellence for Materials Design at the Exascale}~\cite{max} and of the {\em Nanoscience Foundries and Fine Analysis - Europe} user infrastructure~\cite{nffa}.

With regard to the broad strategy outlined above, \yambo{} now includes the possibility to compute the following state-of-the-art physical phenomena discussed later:
\begin{itemize}
\item Electron-phonon and exciton-phonon interaction: influence of temperature on electronic structure and optical spectra (section~\ref{sec:elph});
\item Real-time propagation of the density matrix (section~\ref{sec:RT_RHO}) and Bloch states for nonlinear optics (section~\ref{sec:RT_NL});
\item Spin-orbit coupling and Kerr effect within a fully noncollinear BSE framework (section~\ref{sec:BSE_features}).
\end{itemize}

Numerous methodological advances have been incorporated in the code in the last decade. We will discuss in more detail the following key features:
\begin{itemize}
\item Alternative approaches for computing dipole matrix elements and commutators (section \ref{subsec:dipoles});
\item Incorporation of empty state terminators in the linear response (section \ref{subsec:terminators_linear_resp}) and self-energy (section~\ref{subsec:terminators_GW});
\item Full frequency GW, including computation of lifetimes (section~\ref{subsec:real-axis});
\item Double grid approach and Krylov algorithm for improved BSE efficiency (section \ref{sec:BSE_numerical}).
%\item Post-processing tools for analysis of excitonic wavefunctions and Wannier interpolations of GW bandstructures (sections \ref{ss:excan} and \ref{subsec:wannier90})
\end{itemize}

Regarding parallelism, section~\ref{sec:parallel} outlines 
the code's strategies for exploiting massively-parallel architectures through the use of a highly user-tunable mixed MPI-OpenMP coding paradigm and the use, where possible, of external parallel libraries for linear algebra and I/O tasks. As different quantities (i.e. linear response, GW, BSE) computed by \yambo\ have very different behaviours in terms of performance, scalability, and memory distribution, it is important to outline the different approaches --- ultimately controlled by the user --- adopted by the code in each case. 
% installed on several HPCs

Last, \yambo\ has been almost completely rewritten since the first major release in order to follow modern software design practices such as modularity, reuse of routines and libraries, and so on, and the project as a whole has been expanded to include rigorous self-testing and automation frameworks. Here we highlight a few key features:
\begin{itemize}
\item Test-suite and benchmarking scripts (section~\ref{subsec:test-suite}); 
\item The \yambopy{} python scripts for automation and analysis (section~\ref{subsec:yambopy});
\item Plugin for workflow management via AiiDA (section~\ref{subsec:aiida});
\item Wide use of standard libraries (section~\ref{subsec:installation}).
\item Maintenance and distribution through \texttt{GitHub}.
\end{itemize}

In the following section we recall the structure of the \yambo{} software package and outline new features in its installation environment and interface with external codes and libraries. Sections~\ref{sec:linear_resp}--\ref{sec:real-time} outline new features implemented relating to improved algorithms and new capabilities. Section~\ref{sec:parallel} discusses the new parallelism paradigm and performance issues. %Sections~\ref{sec:ypp} and
Sec.~\ref{sec:scripting} introduces new scripting and automation tools. 
% and pre- and post-processing tools.
Following some general conclusions, various technical information is presented in the appendices along with a useful glossary of acronyms.

%%%%%%%%%%%%%%%%%%%%%%%%%%%%
%\section{Installation, libraries and interfaces \label{sec:general} }
\section{Technical Overview \label{sec:general} }
%%%%%%%%%%%%%%%%%%%%%%%%%%%%
%\mynote{
%{\bf Responsible}: DS\\
%{\bf Contributors}: AF, CH, MG, HM
%}
%

The \yambo{} package is released under the GNU GPL (v2) license and is hosted on {\tt GitHub} in a set of public and private repositories at \url{https://github.com/yambo-code}. Snapshots of major releases are also available for direct download through the \yambo{} website~\cite{yambo_web}.

\begin{figure*}%[h!]
    \centering
    \includegraphics[clip,width=0.90\textwidth]{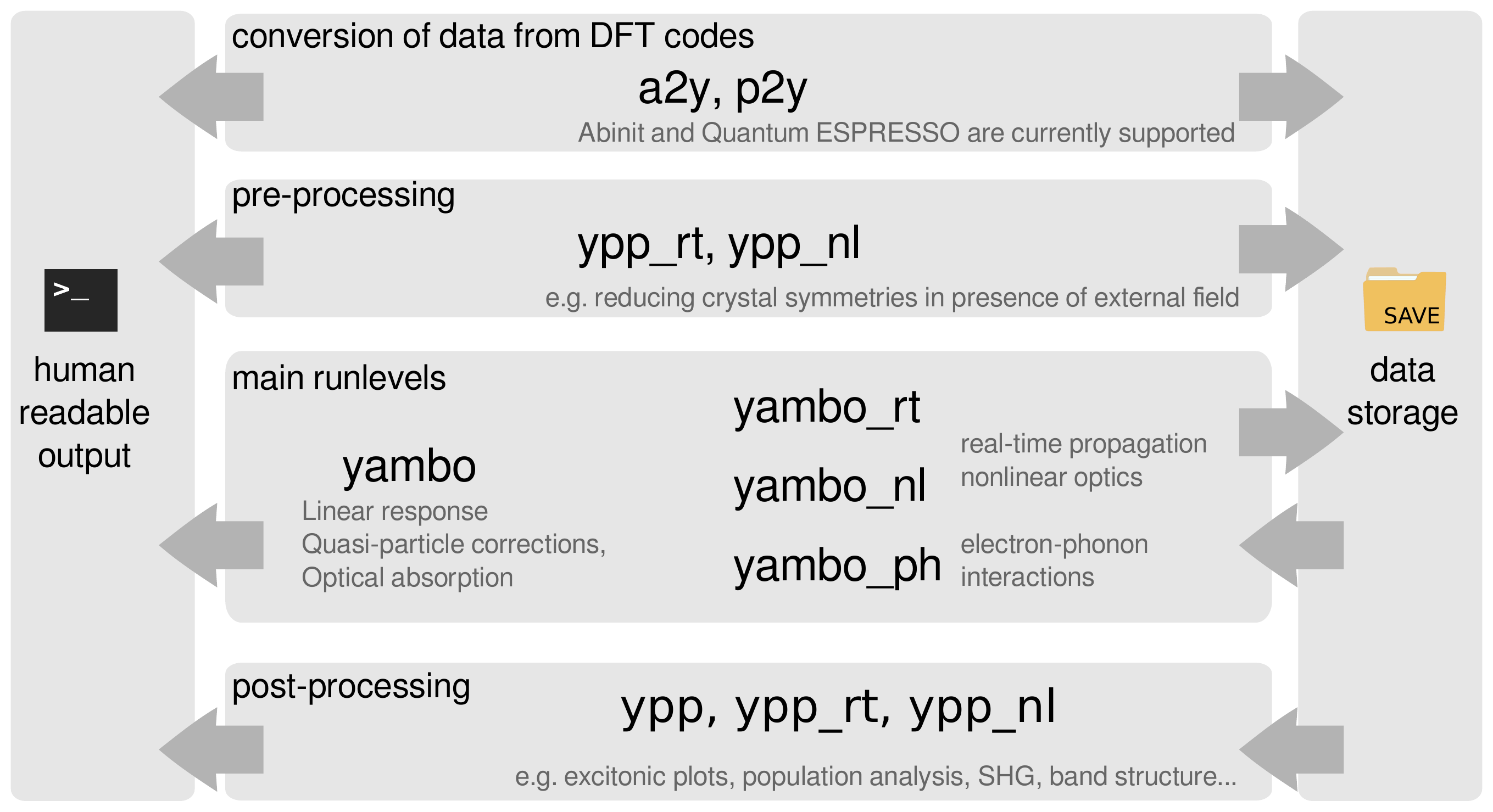}  
    \caption{Main software components and their general function.}
    \label{fig:yambo_exec}
\end{figure*}

The general structure of the \yambo{} software is laid out in Fig.~\ref{fig:yambo_exec}. The software consists of three kinds of executable that generally reflect the order in which the code is run. First, the output from standard DFT codes are converted into NetCDF `database' files (\texttt{ns.db1} and \texttt{ns.wf}) within a \texttt{SAVE} directory using the \ay\ and \py\ routines (see Section~\ref{sec:interfaces} below).
Second, the main calculations (`runlevels') of linear response, GW, and BSE are performed using the standard executable \yambo\ or the project-specific  executables. These include \yambort\ for real-time propagation (Section~\ref{sec:RT_RHO}), \yambonl\ for nonlinear optics (Section~\ref{sec:RT_NL}), and \yamboph\ for electron-phonon simulations (Section~\ref{sec:elph}). 
Running these codes results in the reading and writing of further databases (\texttt{SAVE/ndb.*}), as well as generation of text files for reading or plotting. 
Third, post-processing routines (\ypp\ and runlevel specific \yppnl\ or \ypprt\  executables) are used to manipulate and analyze the computed quantities stored in the databases. In special cases \ypp, \yppnl\ or \ypprt\ executables are needed as pre-processing tools to further manipulate the core databases (i.e. to remove some symmetries before real time simulations) or to create new databases (i.e. an \texttt{ndb} containing a mapping between core databases on two different k-grids), before actually running the main calculation.

\subsection{Installation \& projects}
\label{subsec:installation}
\yambo{} is compiled using the standard autotools procedure: \verb|./configure; make all| will generate the main executables listed in Fig.~\ref{fig:yambo_exec}.
Since the first release the configure script has been wholly upgraded to reflect the widespread availability of high performance software libraries and to aid portability across a wider range of system architectures.

By running \verb|./configure; make|, the list of possible executables is returned
\begin{verbatim}
 [all projects] all
 [project-related suite] project 
                  (core, rt-project, ...)

 [core] yambo
 [core] ypp
 [core] a2y
 [core] p2y

 [ph-project] yambo_ph
 [ph-project] ypp_ph

 [rt-project] yambo_rt
 [rt-project] ypp_rt

 [nl-project] yambo_nl
 [nl-project] ypp_nl

 [kerr-project] yambo_kerr

\end{verbatim}
While \yambo{} and \ypp{} are the main code components, a series of projects appear in the form of {\yambopj}/{\ypppj} with $PJ$ being the specific project identifier (\verb|ph,rt,nl,kerr|). These projects correspond to pre--processor flags that, during the compilation, activate lines of code and procedures that are project-specific. In \yambo{} several different codes coexist in the same source.

\subsection{Configuration}
\label{subsec:configuration}
In many cases \verb|configure| will manage to detect the compilation environment and external libraries automatically.
For more control, a flexible list of options is available (see  \verb|./configure --help|).
A wide range of optional features can be activated via \verb+--enable-FEATURE[=ARG]+ flags, e.g.
\begin{verbatim}
./configure --enable-open-mp
\end{verbatim}
including options controlling serial/parallel linear algebra, timing/memory profiling, type of fast Fourier transform (FFT) library, etc.
External libraries can be linked to by specifying either the installation directory including the ``libs'' and ``include'' folders,
\begin{verbatim}
  --with-libname-path=<path>      
\end{verbatim}
or the ``libs'' and ``include'' paths directly
\begin{verbatim}
  --with-libname-libdir=<path>    
  --with-libname-includedir=<path>
\end{verbatim}
or finally the libraries and the include command
\begin{verbatim}
  --with-libname-libs=<libs>      
  --with-libname-incs=<include command>
\end{verbatim}
This is an important improvement for allowing installation on machines with non-standard system directories.

Choice of compilers and preprocessors can be overridden via the environmental variables \verb|FC, CPP| etc. Finally, the generated \verb|config/setup| file can be tweaked by hand prior to compilation.

% Appendix for installation?
% if not the two subsections should be merged
%\subsection{Libraries}

\yambo\ can make use of several external libraries for improving performance and portability (see Table~\ref{tab:configure}). 
In addition to standard MPI (openmpi, Intel MPI, etc.)  and OpenMP for parallel computation,
these include standard scientific computation libraries such as BLAS and LAPACK (including the Intel MKL and IBM ESSL), scalable versions of these (BLACS, ScaLAPACK; {{\small\ttfamily--enable-par-linalg}}), as well as advanced parallel numerical libraries (SLEPc, PETSc; {{\small\ttfamily--enable-slepc-linalg}}). 
Use of the latter in \yambo{} is discussed in detail in Section~\ref{subsec:PAR_LinearAlgebra}.  
Heavy use is made of fast Fourier transforms (FFTs). 
\yambo{} supports many FFT implementations: Goedecker ({{\small\ttfamily--enable-internal-fftsg}}), FFTW (internal default) and 3D or standard FFT implementation of \qe{} ({{\small\ttfamily--enable-3D-fft}} or {{\small\ttfamily--enable-internal-fftqe}}) can be compiled while MKL and ESSL can be externally linked.
% Please check this! DS, I changed a bit. More checks by AF needed
Regarding internal I/O, linking to NetCDF or HDF5 format libraries is a requirement. The exchange-correlation functional library \verb+libxc+ is also required. 
Interfacing with the \yambopy\, and AiiDA platforms is explained thoroughly in Section~\ref{sec:scripting}.
Libraries related to porting data from DFT codes are discussed in the following section.

\begin{table}[!h]
    \centering
    \begin{tabular}{l|l}
       Library & Flag \\ \hline
       \multicolumn{2}{c}{} \\
       \multicolumn{2}{c}{Fourier transform} \\
        FFTW (2.0)         & Default \\
        Goedecker & {{\small\ttfamily--enable-internal-fftsg}} \\
        QE standard & {{\small\ttfamily--enable-internal-fftqe}} \\
        QE 3D & {{\small\ttfamily--enable-3d-fft}} \\
        MKL, ESSL, FFTW(3.x) & {{\small\ttfamily--with-fft-libs=<libs>}} \\ \hline
       \multicolumn{2}{c}{} \\
       \multicolumn{2}{c}{Linear Algebra} \\
       BLAS, LAPACK & {{\small\ttfamily--with-blas-libs=<libs>}} \\
       MKL, ESSL & {{\small\ttfamily--with-lapack-libs=<libs>}} \\  \hline
       \multicolumn{2}{c}{} \\
       \multicolumn{2}{c}{Parallel Linear Algebra} \\
        BLACS \&, & {{\small\ttfamily--enable-par-linalg}} + \\
        ScaLAPACK & {{\small\ttfamily--with-blacs-libs=<libs>}} \\ \hline
        \multicolumn{2}{c}{} \\
        \multicolumn{2}{c}{Sparse Linear Algebra} \\
        SLEPC \& & {{\small\ttfamily--enable-slepc-linalg}} \\ 
        PETSC & {{\small\ttfamily--with-slepc-libs=<libs>}} \\ 
          & {{\small\ttfamily--with-petsc-libs=<libs>}} \\        \hline
        %\multicolumn{2}{c}{} \\
        %\multicolumn{2}{c}{More} \\
        ... & ... 
    \end{tabular}
    \caption{Illustrative list of some of the configuration command line options. More options are available and can be listed by using \texttt{./configure --help}.}
    \label{tab:configure}
\end{table}

\subsubsection{External Libraries}
\label{subsec:ext_libs}
An important feature of the new configuration procedure in \yambo{} is that all required libraries can be automatically downloaded, configured and compiled at the compilation time.

Indeed, if \verb+configure+ does not find a required library (dependency), it will automatically download and compile it. A useful option is the
%{{\small\ttfamily--with-extlibs-path=<full\_path>}
\begin{verbatim}
--with-extlibs-path=<full_path>
\end{verbatim}
where one can provide a path of choice where \yambo{} will install all the automatically downloaded libraries, once and for all. The content of the folder is never erased. In subsequent compilation the library will be automatically re-used just specifying the same option.

\subsection{Interfaces with DFT codes \label{sec:interfaces}}

\yambo{} is interfaced with two widely used plane-wave first-principles codes: \pwscf~from the {\tt Quantum ESPRESSO} (QE) distribution~\cite{Giannozzi2009,Giannozzi2017} and \abinit~\cite{Gonze2002,Gonze2005,Gonze2009}. The two interfaces have been introduced in Ref.~\cite{Marini2009} (sections 5.1 and 5.3).  Both work with norm conserving pseudo-potentials and import Kohn-Sham (KS) eigen-energies $\epsilon_{n\kk}$ and eigen-functions $\psi_{n\kk}$ as well as information needed to compute the non-local part of the pseudo-potential $V^{nl}(\xx,\xx')$. Since the publication of Ref.~\cite{Marini2009} both interfaces have been largely improved and extended. Two are the most relevant changes. All interfaces are now able to deal with both collinear and non-collinear spin systems.
%, following the full support for spin made available with \yambo-3.4.
All interfaces take advantage of the {\xc} library~\cite{Marques2012,Lehtola2018}, thus a very broad class of functionals is supported. A more detailed summary of the changes follows.
%
%\begin{figure}%[h!]
%    \centering
%    \includegraphics[clip,width=0.40\textwidth]{figures/Figure_Interface_scheme.pdf}  
%    \caption{MG: let me know if it is fine and I will write a caption and cross ref in the text}
%    \label{fig:int_scheme}
%\end{figure}

\subsubsection{Interface with \qe}
%\mynote{{\bf Contributors}: AF, MG, CH \\}
\py\ (\pwscf-2-\yambo) is the \yambo{} interface with \qe{}. Its development line followed two routes, one related to the developments of QE I/O and one aimed at adding new features to \py{}.

A wider class of pseudo-potentials (psps) is now supported, including UPF  version 2, and multi-projector psps -- i.e. with more then one projector per angular momentum channel. In the same direction the \xc\ library~\cite{Marques2012,Lehtola2018} allows for the support of most of the LDA and GGA functionals as a starting point for the MBPT (quasiparticle or response function) calculations. In addition, hybrid functionals, with fractions of exchange and screened exchange interaction, are also supported within \py. To keep compatibility with all versions of QE within a user--friendly approach, \py\ has now an automatic detection of the I/O format used in the ground--state calculation and is able to read different xml data-file formats ({\tt qexml} and {\tt qexsd}, in the QE language), also supporting the more recent HDF5 binary files.

Spin is now fully supported both in collinear and non-collinear frameworks. For example, the use of magnetic symmetries allows to take advantage of composite symmetries, i.e. which contain time-reversal, in systems which are not invariant under pure time-reversal. Work is in progress to extend the support of ultra-soft pseudopotentials (USPP). %\mynote{DS: This last sentence can be changed depending on the status of the interface at submission time}.

Other important changes were carried out to optimize the interface, first of all with an improved parallelization (implemented over the writing of wavefunction fragments). Moreover the Kleinman-Bylander (KB) form factors are now converted in a \yambo-like database, while the calculation of the commutator $[\mathbf{r},V^{nl}]$, which was previously done at the \py\ level, 
is now more efficiently done by \yambo\ while computing the dipoles.

%=================================
\subsubsection{Interface with \abinit}
%\mynote{{\bf Contributors}: DS, MG, HM}
%==================================
%
\ay\ and \ey\ are the \abinit-2-\yambo\ interfaces.
The original \ay\ implementation reads data in Fortran binary format. \ey\ was developed later and is based on the \etsfio~\cite{Caliste2008} and \netcdf~\cite{Rew1990,Brown1993} libraries. Both interfaces are based on the \abinit\ KSS file and were developed following the evolution of \abinit. However, since the support to the KSS file was dropped by the \abinit{} team, the development and maintenance of interfaces based on it became difficult. As an example, the support for multi-projector pseudo-potentials was first released via a patch for the \abinit\ code, which allows the printing of the relevant data into the \abinit\ KSS file.~\footnote{The patch works from \abinit-6.12 to \abinit-7.4 and it is meant to be used with \ay.} As a consequence, the development of the KSS-based interfaces was also dropped by the \yambo{} team. The old \ay\ implementation works up to \abinit\ version 7, while \ey\ is supported up to the very recent \abinit\ 8 releases.

Starting with \yambo\ 4.4, we will release a new version of the \ay\ interface, which is based on the direct reading of the \abinit\ wave-function files (WFK files)  written in \netcdf\ format.
A preliminary version of \ey\ based onto the WFK file was also released with \yambo\ 4.0. However, since the support to the \etsfio\ library is not developed anymore, the WFK based \ey\ interface was never finalized.
The new strategy (i) avoids the need for the KSS file, (ii) is numerically more efficient and (iii) reduces the I/O, since wave functions are stored on the smaller $\kk$-centred spheres in reciprocal space (as opposed to the KSS file which relied on a larger gamma-centred sphere). Finally, since WFK files are fully supported by the \abinit\ team, the new interface will be compatible with all recent \abinit\ developments (also including multi-projectors pseudo-potentials) and naturally portable to work with future \abinit\ releases.
 
%\subsection{Main executables}

%\subsubsection{\yambo}

\begin{widetext}
\begin{table}[t]
\begin{center}
\begin{tabular}{ll|ll}  
 \hline 
\multicolumn{2}{c|}{Common to \yambo, \ypp, and \ay\ / \py\ / \ey}&\multicolumn{2}{c}{Common to \yambo\ and \ypp} \\
 \hline
                   &                                     & &  \\
{\tt -h}& Short Help                                     & {\tt -J <opt>} & Job string identifier \\
{\tt -H}& Long Help                                      & {\tt -V <opt>} & Input file verbosity \\
{\tt -M}& Switch-off MPI support (serial run)            & {\tt -F <opt>} & Input file \\
{\tt -N}& Switch-off OpenMP support (single thread run)  & {\tt -I <opt>} & Core I/O directory \\
        &                                                & {\tt -O <opt>} & Additional I/O directory \\
        &                                                & {\tt -C <opt>} & Communications I/O directory \\
                   &                                     & &  \\
\hline
\multicolumn{2}{c|}{\yambo}&\multicolumn{2}{c}{\ypp} \\
\hline
&&& \\
{\tt -D      }     &DataBases properties                 & {\tt -q <opt> }    &(g)enerate-modify/(m)erge quasi-particle DBs\\
{\tt -W <opt>}     &Wall Time limitation (1d2h30m format)& {\tt -k <opt> }    &BZ Grid generator  \\
{\tt -Q      }     &Don't launch the text editor         & {\tt -i}           &Wannier 90 interface  \\
{\tt -E <opt>}     &Environment Parallel Variables file  & {\tt -b}           &Read BXSF output generated by Wannier90  \\
                   &                                     & {\tt -s <opt>}     &Electrons,[(w)ave,(d)ensity,(m)ag,do(s),(b)ands]  \\
                   &                                     & {\tt -e <opt>}     &Excitons, [(s)ort,(sp)in,(a)mplitude,(w)ave] \\
{\tt -i      }     &Initialization                       & {\tt -f}           &Free hole position [excitonic plot]\\
{\tt -r      }     &Coulomb potential                    & {\tt -m}           &BZ map fine grid to coarse \\
{\tt -a      }     &ACFDT Total Energy                   & {\tt -w <opt> }    &{\makecell[l]{WFs:(p)erturbative SOC map\\ \ \ \ \ \ \ \ or (c)onvertion to new format}} \\
{\tt -s      }     &ScaLapacK test                       & {\tt -y}           &{\makecell[l]{Remove symmetries not consistent \\ \ \ \ \ \ \ \ with an external potential}} \\
                   &                                     & & \\ \cline{3-4}
{\tt -o <opt>}     &Optics [opt=(c)hi/(b)se]             & \multicolumn{2}{c}{Common to \ay\ / \py\ / \ey} \\ \cline{3-4}
{\tt -y <opt>}     &\ \ BSE solver [opt=h/d/s/(p/f)i]    &  &  \\
                   &\ \ \ \ \ \ (h)aydock/(d)iagonalization/(i)nversion & {\tt -U} & Do not fragment the DataBases \\
{\tt -k <opt>}     &\ \ Kernel [opt=hartree/alda/lrc/hf/sex] & {\tt -O <opt>} & Output directory \\
                   &                                     & {\tt -F <opt>} & PWscf xml index/Abinit file name \\
{\tt -d      }     &Dynamical Inverse Dielectric Matrix  & & \\
{\tt -b      }     &Static Inverse Dielectric Matrix     & {\tt -b <int>}  & Number of bands for each fragment  \\
                   &                                     & {\tt -a <real>} & Lattice constant rescaling factor \\
{\tt -x      }     &Hartree-Fock Self-energy and Vxc     & {\tt -t      }  & Force no TR symmetry \\
{\tt -g <opt>}     &Dyson Equation solver                & {\tt -n      }  & Force no symmetries \\
                   &\ \ [opt=(n)ewton/(s)ecant/(g)reen]  & {\tt -w      }  & Force no wavefunctions \\
{\tt -p <opt>}     &GW approximations,[opt=(p)PA/(c)OHSEX] & & \\
{\tt -l      }     &G$_0$W$_0$ Quasiparticle lifetimes   & {\tt -d      }  & States duplication \ \ \ \ \ \ \ \ \ \ \ [a2y only] \\
                   &                                     & {\tt -v      }  & Verbose wfc I/O reporting   [p2y only] \\
{\tt -q <opt>}    &Compute dipoles [available from v4.4] & & \\ \cline{3-4}
\multicolumn{2}{c|}{ }&\multicolumn{2}{c}{\yppph} \\ \cline{3-4}
&&& \\
&&{\tt -p <opt>}     &Phonon [(d)os,(e)lias,(a)mplitude] \\
&&{\tt -g}           &gkkp databases \\
&&& \\

\hline
\multicolumn{2}{c|}{\yambort}&\multicolumn{2}{c}{\ypprt} \\
\hline
&&& \\
{\tt -v <opt>}    &Self-Consistent Potential  & {\tt -t}  & TD-polarization [(X)response] \\
 &\ opt=(h)artree,(f)ock,(coh),(sex),(cohsex),(d)ef,(ip) & & \\
{\tt -q <opt>}    &Real-time dynamics [replaced by {\tt -n <opt>} in v4.4] & & \\
%& \ \ [opt=(p)ump or probe, (pp)ump \& probe] & & \\
{\tt -e}          &Evaluate Collisions & & \\
&&& \\

\hline
\multicolumn{2}{c|}{\yambonl}&\multicolumn{2}{c}{\yppnl} \\
\hline
&&& \\
{\tt -u}    &Non-linear spectroscopy& {\tt -u}  & Non-linear response analysis  \\
&&& 

\end{tabular}
\caption{Command line options for the various \yambo{} tools.\label{tab:synopsis}}
\end{center}
\end{table}
\clearpage
\end{widetext}
\subsection{Data post- (and pre-) processing}

\ypp\, is the \yambo\, postprocessing and preprocessing tool. It has several capabilities which can be used to prepare \yambo{} simulations (preprocessing) or subsequently analyse (postprocessing) the outcome.

As one of the preprocessing options, \ypp\, can generate random grids of $\kk$-points to be used as input for a DFT code to compute the corresponding KS energies. The same \ypp\, can then generate an auxiliary database with a map linking the KS energies on the random grid to the uniform grid used to compute, for example, spectral properties. The approach is useful to speed up convergence as discussed in Sec.~\ref{sec:BSE_numerical_dgrid}. Another preprocessing option is the removal of a specific set of symmetries and thus the expansion of the wave-functions from the IBZ associated to the full set of symmetries to the resulting IBZ. This is needed to perform real--time simulations as described in Sec.~\ref{sec:real-time}. Finally preprocessing can also be used to map DFT calculations with and without spin-orbit coupling (SOC) to compute absorption spectra with SOC corrections included in a perturbative way as described in the supplemental material of Ref.~\cite{Qiu2013}.

Most of the postprocessing features involve data analysis.
\ypp\, can prepare readable ascii files to plot several single-particle properties such as wave--functions, charge density, density of states (DOS), magnetization, current and band structures. In particular, it can be used to obtain the QP-DOS and to interpolate QP-energies to plot the resulting band-structure along high-symmetry paths. A mixed feature (i.e. which can be used both for preprocessing and for postprocessing) is the ability of \ypp\, to manipulate QP-databases (ndb.QP). Indeed, this is useful both for QP plots or for using ndb.QP files as input in the BSE calculations. Finally, it can be used as a tool to analyse the  excitonic wave-function.
As examples of postprocessing, we discuss in detail ($i$) how to plot the QP band structure starting from calculations on a regular grid in Sec.~\ref{subsec:qp_pp} and ($ii$) how to plot the excitonic wave-function in Sec.~\ref{ss:excan}.

\subsection{Usage}
\yambo{} relies on a powerful and user friendly command line interface for generating and modifying input files as well as for launching the executables. 
The basic functionality is unchanged from that described in CPC2009; however, some flags have been changed since the initial release.
Several new options have been added to aid usage or debugging on parallel clusters or cross-compiled architectures.  
For instance, \verb+yambo -M+ and \verb+yambo -N+ switch off the MPI and OpenMP functionalities, respectively, \verb+yambo -Q+ stops the text editor from launching, and \verb+yambo -W <opt>+ places an internal wall clock limit on the runtime. 
Launching \verb+yambo -H+ shows the fully updated list of command options: see Table~\ref{tab:synopsis}.

%%%%%%%%%%%%%%%%%%%%%%%%%%%%
%\section{Dipoles, Coulomb Cutoff, Linear Response \label{sec:dipoles} }

\section{Linear Response \label{sec:linear_resp} }

%%%%%%%%%%%%%%%%%%%%%%%%%%%%
%\mynote{
%{\bf Responsible}: AF\\
%{\bf Contributors}: AM, IM (terminators in LinRes), DS (dipoles), DV (cutoff)
%}
%

In the independent particle (IP) approximation, the density-density response function can be written as:
\begin{widetext}
\begin{eqnarray}
%\begin{multline}
 \chi_{ {\bf{G}}{\bf{G}'}}^0 (\qq,\omega)=\frac{f_s}{N_{\bf{k}}\Omega}
\sum_{nm\kk} \rho_{nm\kk}(\qq, \GG)  \rho^{\star}_{nm\kk}(\qq,\GG^{\prime}) \times
    \Big[ \frac{f_{m\kk}(1-f_{n\kk-\qq})}{\omega -(\epsilon_{m\kk} - \epsilon_{n\kk-\qq}) -i\eta}
         -\frac{f_{m\kk}(1-f_{n\kk-\qq})}{\omega -(\epsilon_{n\kk-\qq} -\epsilon_{m\kk}) +i\eta}  \Big],
\label{eq:X_IP}
%\end{multline}
\end{eqnarray}
\end{widetext}
where $n,m$ indexes represent band indexes (which also include the spin index in case of spin collinear calculations and which refer to spinors in case of non--collinear calculations), 
$f_{n\kk}$ and $\epsilon_{n\kk}$ are the occupations and the energies of the KS states,
$f_s=1$ for spin--polarized calculations, $f_s=2$ otherwise.
In practice, the sum in Eq.~\eqref{eq:X_IP} is split into two terms as described in Appendix~\ref{App:Resp-function-sum}.
The matrix elements
\begin{equation}
   \label{eq:rhotw}
   \rho_{nm\kk}(\qq,\GG)=\langle n\kk | e^{i(\qq+\GG)\cdot\hat{\mathbf{r}}} | m \kk-\qq \rangle,
\end{equation}
 have been already introduced in Ref.~\cite{Marini2009} and constitute one of the core quantities computed by the \yambo\, code. Their evaluation is done via the Fourier transform of the wave-function product in real space, $\psi^*_{n\kk}(\rr)\psi_{m\kk-\qq}(\rr)$, 
 and has been strongly optimized being one of the most common operations performed by \yambo{}
 %in the last \yambo\, releases 
 (see discussion in sec.~\ref{subsec:PAR_LinearResp}).

%======================================
\subsection{Dipole matrix elements}
%======================================
\label{subsec:dipoles}
Despite the computational cost, the numerical algorithm to compute the terms in Eq.~\eqref{eq:rhotw} is straightforward. Since absorption is defined as the macroscopic average of the density-density response function, $\chi(\qq\to 0) $, the knowledge of $\rho_{nm\kk}(\qq\to 0,0)$ is also needed.
%For $\qq=\GG=0$, $\varrho_{nm}(\kk,0,0)=\delta_{nm}$
To this end, the dipole matrix elements ${\rr_{nm\kk}=\langle n\kk |{\mathbf{r}} | m \kk \rangle}$ are commonly computed~\cite{delsole-girlanda93prb,onida2002} within periodic boundary conditions (PBC) using the relation ${[\rr,H]=\mathbf{p}+[\rr,V_{nl}]}$. Explicitly, this gives
\begin{equation}
\langle n\kk |\rr | m \kk \rangle=\frac{\langle n\kk | \,{\mathbf{p}}+[\rr,V_{nl}] \,| m \kk \rangle} {\epsilon_{n\kk}-\epsilon_{m\kk}}\, .
\label{eq:dip}
\end{equation}
The direct  evaluation of Eq.~\eqref{eq:dip} (\emph{G-space v} approach) is quite demanding due to the $[\rr,V_{nl}]$ term, and  is evaluated 
%by starting 
from the KB form factors loaded by the interfaces, see also Sec.~\ref{sec:interfaces}. This implementation has been strongly optimized and extended to account for projectors with angular momentum $l>2$.

We have also made available alternative strategies for computing the dipoles. The \emph{shifted grids} approach is based on the idea of numerically evaluating $\rho_{nm \kk}(\qq_\epsilon,0)$ for a very small $q_\epsilon=|\qq_\epsilon|$. Thus the wave--function at $\kk$ and the wave--functions at $\kk-\qq_\epsilon$ are needed. Since the $\qq\rightarrow 0$ limit may be direction dependent, this is done in practice by means of wave--functions computed on four different grids in $\kk$-space, i.e. a starting $\kk$-grid plus three grids with $\kk+q_\epsilon\,\cc_i$ slightly shifted along the three Cartesian directions $\cc_x, \cc_y, \cc_z$. Such approach is computationally more efficient, although it requires to generate a larger set of wave--functions. However, there exists a random phase associated to the wave--functions on each of the four $\kk$--grids, since they are obtained by independent diagonalizations of the KS Hamiltonian. Because of this, \emph{shifted grids} dipoles have inconsistent phases among different directions and it is not possible to use them when the dipole matrix elements are needed (instead of their square modulus only) as for example in the evaluation of the Kerr effect (see Sec.~\ref{sec:BSE_soc}) or for non-linear optics (see Sec.~\ref{sec:RT_NL}).

The \emph{G-space v} approach assumes that the only non-local terms in $H$ are the kinetic energy and the pseudopotentials. There are however cases, for example when the Hamiltonian contains non-local hybrid functionals, Hubbard U terms, or nonlocal self-energies, in which the evaluation of the commutator may become very cumbersome. To solve this issue one could in principle use the \emph{shifted grids} approach. However, also this approach may become impractical because of the calculation of wave-functions on the shifted grids.

For those cases we have implemented in \yambo{} two alternative strategies, one for extended and one for isolated systems.
For extended systems the \emph{Covariant} approach exploits the definition of the position operator in $\kk$ space: $\hat{\rr} = i \partial_\kk$. The dipole matrix elements are then evaluated as finite differences between the $\kk$-points of a single regular grid. A five-point midpoint formula is used, with a truncation error $O(\Delta \kk^4)$. The \emph{shifted grids} and the \emph{Covariant} approach are very similar, however in the latter the arbitrary phase of the wave-functions at different $\kk$-points is correctly accounted for. To this aim, $i \partial_\kk$ is implemented as a covariant derivative which cancels the relative phase factor (see Appendix~\ref{App:cov-dips} for details). For these reasons the \emph{Covariant} approach overcomes the limitations of the \emph{shifted grids} approach. The main drawbacks of the \emph{Covariant} approach is that the numerical value of the dipoles needs to be converged against the size of the $\kk$ grid and the present implementation does not work for metals. However, in practice the convergence of dipole matrix elements is usually faster than that of the absorption spectrum.\\
For finite systems, finally, the dipole matrix elements can be directly evaluated in real space (\emph{R-space x} approach).

We underline that in the case of a local Hamiltonian all approaches are equivalent. 
The desired strategy can be selected via the input variable: \\[5pt]
{{\small\ttfamily DipApproach="G-space v"}} \\
{{\small\ttfamily \#[Xd] [G-space v/R-space x/Shifted grids/Covariant]}}
\\[5pt]
(\emph{G-space v} being the default value).

%======================================
\subsection{Coulomb interaction}
%======================================
The Coulomb interaction enters in many sections of the \yambo{} code, such as linear response, self-energy, and BSE kernel calculation. In reciprocal space, the bare Coulomb interaction for bulk systems is defined as $v({\bf q+G})=4\pi/|{\bf q+G}|^2$. For the calculation of quantities requiring integration over transferred momenta in the Brillouin zone (BZ), such as the self-energy, the integrals are evaluated by summations over regular $\qq$-grids. In order to remove divergencies in systems of reduced dimensionality, i.e. in the presence of a 2D or 1D sampling of $\kk$-points, or to speed up the convergences in 3D systems, \yambo{} offers the possibility to evaluate Coulomb integrals by using the {\it random integration method} (RIM), which consists of evaluating these integrals by Monte Carlo sampling (as already discussed in detail in Sec.~3.1 of Ref.~\cite{Marini2009}), dividing the full BZ in small regions around each $\kk$-point of the chosen uniform grid.

In order to avoid spurious interaction between replicas when dealing with low-dimensional materials such as clusters, slabs, or wires, \yambo\ can also use Coulomb cutoff truncation techniques. These consists of truncating the Coulomb interaction beyond a certain region (depending on the chosen geometry):
\begin{equation}
       \tilde{v}(\mathbf{r}) = 
    \begin{cases}
      1/r & \text{if } \it \mathbf{r} \in D \\
      0 & \text{if } \it \mathbf{r} \notin D.
    \end{cases}
\end{equation}
Different geometrical choices are available. Spherical and cylindrical shapes, suitable to treat 0D and 1D systems, respectively, have been already described in details in Ref.~\cite{rozzi2006exact}. In addition, a box-like cutoff obtained by performing a numerical Fourier Transform of the real space expression is available for 0D systems. By defining only one or two sides of the box, it is possible to treat 2D or 1D systems within the same numerical approach. It is important to stress that, as the construction of such potential requires integration over the BZ, the RIM method discussed above must be activated. 

Finally, a Wigner-Seitz truncation scheme, similar to the one discussed in Ref.~\cite{Beigi2006} is also available. In this scheme the Coulomb interaction is truncated at the edge of the Wigner-Seitz super-cell compatible with the $\kk$-point sampling. This truncated Coulomb potential turns out to be suitable for finite systems as well as for 1D and 2D systems, provided that the supercell size, determined by the adopted $\kk$-point sampling, is large enough to get converged results~\cite{cutoff_inprep}. 

%======================================
\subsection{Sum-over-states terminators in IP Linear Response}
%======================================
\label{subsec:terminators_linear_resp}

The independent particle polarizability $ \chi_{ {\bf{G}}{\bf{G'}}}^0 ({\bf{q}}, \omega)$, Eq.~\eqref{eq:X_IP}, and the correlation part of the GW self-energy $\Sigma_c(\omega)$, Eq.~\eqref{eq:sigma_c} in section~\ref{subsec:real-axis},  are evaluated through sum-over-states (SOS) expressions obtained by applying an energy cutoff to the infinite sum over virtual states. These expressions are, however, slowly convergent and, especially for large systems, require the inclusion of a large number of empty states ($N_b$). This condition makes GW calculations computationally  demanding, both in terms of time-to-solution and memory requirements.
%Independent particle-polarizability $ X_{ {\bf{G}},{\bf{G'}}}^0 ({\bf{q}}, \omega)$ and  correlation part of the GW %self-energy $\Sigma_c^{n {\bf{k}} }(\omega)$, Eq. \ref{eq:sigma_c}, involve infinite sums over Kohn-Sham (KS) virtual %states. In order to make possible the calculation of both $ X_{ {\bf{G}},{\bf{G'}}}^0 ({\bf{q}}, \omega)$  and $\Sigma_c^{n %{\bf{k}} }(\omega)$, the sum over empty states is commonly  cutoff at a certain energy. 
In order to overcome this limitation, a number of approaches have been proposed to reduce \cite{bruneval2008,berger2010,deslippe2016,gao2016} or remove~\cite{rocca2008,Umari2010,govoni2015} sum over states; among them, we have implemented in \yambo\ the  extrapolar correction scheme  proposed by Bruneval and Gonze (BG) \cite{bruneval2008}.

This scheme, here referred as
X--terminator,  permits to accelerate GW convergence by reducing of a sensible amount the number of  virtual orbitals necessary to calculate both polarizability and self-energy.
In this procedure extra terms, whose calculation imply a small computational overhead, are introduced to correct both polarizability and self-energy  by approximating the effect of the states not explicitly taken into account. The method consists in replacing the energies of empty states  that are above a certain threshold, and that are not explicitly treated, by a single adjustable parameter defined as extrapolar energy.
%. This parameter, that represents a common energy, is defined extrapolar energy. The method, therefore, provides for replacement of poles arising from energies of empty states above of an arbitrary energy threshold $E_{N_b^{\prime}}$ with a single average pole. 
%Noticeable, in the denominator of the additional term  we have eliminated the dependence on the energy of the empty states that are above $E_{N_b^{\prime}}$, so that we can apply the closure relation to the numerator. As a consequence, the additional term depends on a finite number of elements and  results computationally quite inexpensive. 
When the method of terminators is applied, the independent-particle polarizability can be written as~\cite{bruneval2008}:
\begin{equation}
    \chi_{{\bf{G}}{\bf{G}'}}^0 (\qq,\omega)= \chi^{0,\text{trunc}}_{ {\bf{G}}{\bf{G}'}} (\qq,\omega) + \Delta\chi_{{\bf{G}}{\bf{G}'}} (\qq,\omega, \bar{\epsilon}_{\chi_0})
    \label{eq:X_term_def}
\end{equation}
where the first term on the r.h.s. is truncated at the $N_b^{\prime}$ state (in general $N_b^{\prime} \ll N_b$)  and the second term depends on the extrapolar energy for the polarizability $\bar{\epsilon}_{\chi_0}$. The explicit expression for $\Delta\chi_{{\bf{G}}{\bf{G}'}}(\qq,\omega, \bar{\epsilon}_{\chi_0})$ is provided in App.~\ref{App:terminators}.

In the present implementation of \yambo, 
%GW calculations can be performed without including  terminator corrections, by applying terminators only on polarizability (X-terminator), only on the self-energy (G-terminator) or both on the polarizability and the self-energy. 
the input parameter governing the use of the terminator corrections on the response function (X-terminator) is \\[5pt]
{\small\ttfamily{XTermKind= "none"}}  
{\small\ttfamily{  \# [X] X terminator ("none","BG")}}
\\[5pt]
(default: {\tt "none"}).
%{\small\ttfamily{GTermKind}} rule in  the use of X- and G- terminators in \yambo.
When the variable is set to {\small\ttfamily{none}} (default option), the X-terminator is not applied. On the contrary when {\small\ttfamily{XTermKind}}  assumes the value {\small\ttfamily{BG}}, the extrapolar corrective term is calculated. 
The extrapolar energy $\bar{\epsilon}_{\chi_0}$, see Eq.~\eqref{X-term}, is defined by the input variable (default: 1.5 Ha)
\\[5pt]
{\small\ttfamily{XTermEn= 1.5 Ha}}
{\small\ttfamily{  \# [X] X terminator energy}}
\\[5pt]
The value means $\bar{\epsilon}_{\chi_0}=\epsilon_{N_{b'} \kk}+1.5\ \text{Ha}$, 
with $\epsilon_{N_{b'} \kk}$ the highest energy state included in the calculation. 
%By default,  {\small\ttfamily{XTermEn}} and {\small\ttfamily{GTermEn}} are placed to 1.5 Ha.

\begin{figure}
    \centering
    \includegraphics[width=0.48\textwidth]{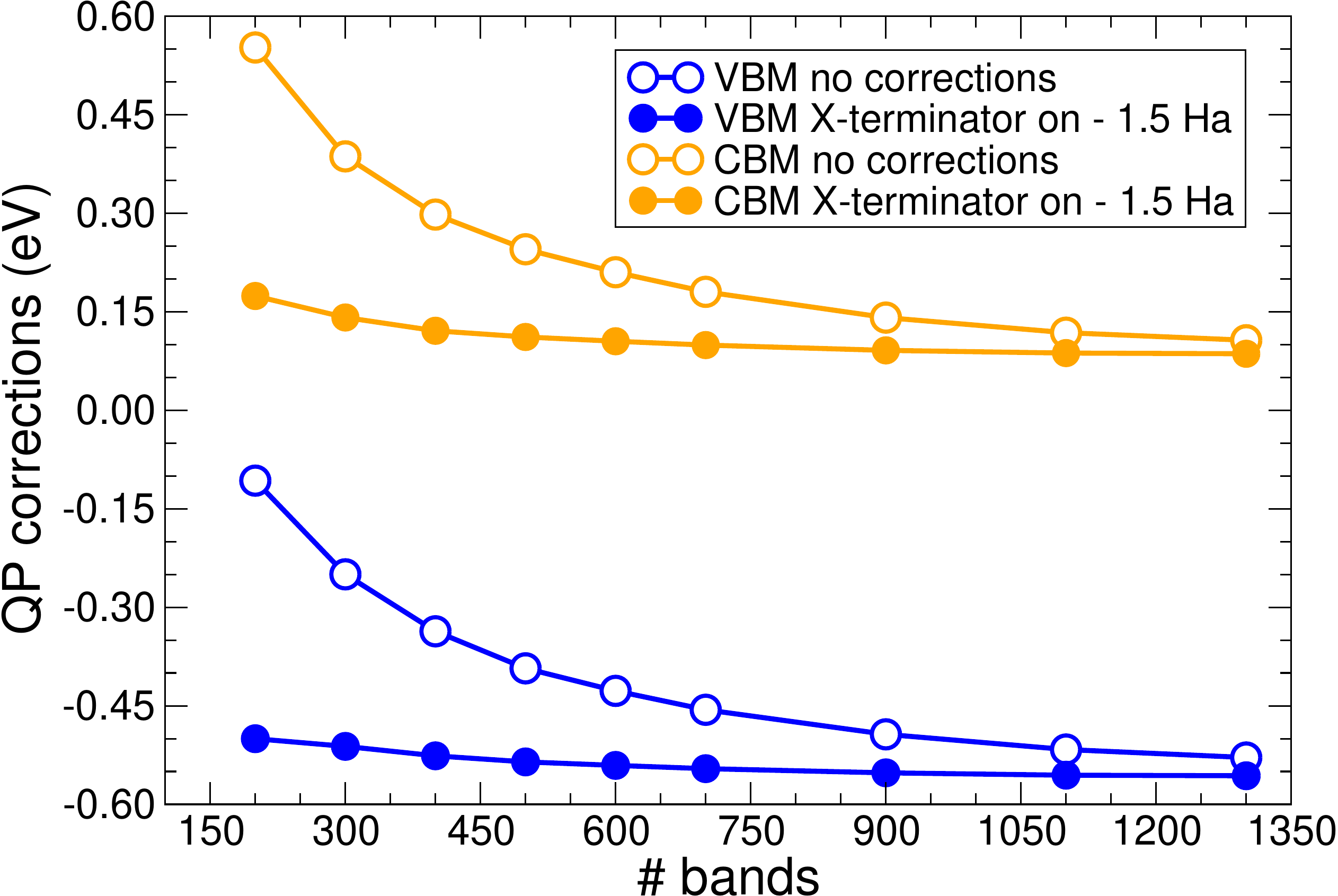}
    \caption{Effect of the X-terminator on the convergence (vs number of bands included in the response function) of the VBM and CBM GW-corrections for a bulk Si described in a supercell. 
    %Calculations have been performed using a $18 \times 18 \times 1$ {\bf{k}}-mesh  to sample the BZ, 7 Ry to represent the response function in reciprocal space, and  1350 bands for the sum-over-states to evaluate self-energy.  
    }
    \label{fig:Xterm}
\end{figure}
For  demonstration  purposes, in Fig.~\ref{fig:Xterm} we report 
the calculated QP corrections for the valence band maximum (VBM) and the conduction band minimum  (CBM)  of a bulk Si described in a supercell (36 Si atoms, 72 occupied states).  
Results are obtained by increasing the  number  of  bands  explicitly included in the calculation of the response function $\chi$  and by imposing a very high number of bands in the self-energy, that is therefore converged.
Empty circles connected with solid lines
  denote  the  results obtained  for  the
VBM and CBM states without applying any correction.
Improvements induced by the use of the X-terminator are
depicted by solid circles connected with solid lines
 that have been obtained imposing {\small\ttfamily{XTermEn}}=1.5 Ha.
We can  observe that the X-terminator leads to a relevant reduction in the number of bands necessary to converge
the polarizability and thus the GW corrections.
%For demonstration purposes, in Fig.~\ref{fig:Xterm} we report the convergence with respect to the number of bands included in the calculation of $\chi$ of GW corrections for the valence band maximum (VBM) and the conduction band minimum (CBM)  of  a silicon supercell. 
%(bulk-like system) obtained by duplicating 9 times the unit cell along the (100) direction (36 Si atoms, 72 occupied states).
%Black and red lines  denote the quasiparticle (QP) corrections for the VBM and CBM states  without applying any correction.  
%Results are obtained by increasing the number of bands used to calculate the polarizability and adopting 
%1350 bands in the self-energy SOS. 
%Improvements induced by the use of the X-terminator are depicted by the green and blue lines that have been obtained imposing {\small\ttfamily{XTermEn}}=1.5 Ha. Setting for this specific case a target accuracy of 20 meV on the calculation of the QP corrections for the VBM and CBM states, we observe that the  X-terminator leads to a reduction of about a factor 2.5 in the number of bands necessary to converge the polarizability.

%%%%%%%%%%%%%%%%%%%%%%%%%%%%
\section{Quasi--particle corrections\label{sec:GW} }
%%%%%%%%%%%%%%%%%%%%%%%%%%%%
%\mynote{{\bf Responsible}: DV\\
%{\bf Contributors}: MP, CA, IM, AMS, MM}
%

Accurate quasi-particle energies can be obtained by calculating self-energy corrections to KS energies  \cite{aryasetiawan1998}. 
In general, the non-local, non-Hermitian and frequency dependent electronic self-energy operator can be expressed as the sum of a bare, energy independent exchange term and a screened, dynamic correlation term:
\begin{equation}
    \Sigma(\rr,\rr^\prime,\omega)=\Sigma_x(\rr,\rr^\prime)+\Sigma_c(\rr,\rr^\prime,\omega) .
\end{equation} 
In this section we describe features implemented in \yambo\ aimed at improving the accuracy of GW calculations by going beyond the  commonly used plasmon-pole approximation~\cite{Larson2013} for the dielectric matrix and in speeding up calculations by reducing the number of empty states needed to get converged results. 
GW energies on top of KS eigenvalues are commonly calculated by considering one-shot corrections using the G$_0$W$_0$ approximation. Nevertheless in \yambo\ is also possible to perform partial self consistent calculations (evGW), where the energies of the Green's function, and the polarizability are iterated until self-consistency is reached, while wave functions are kept frozen.
This approach generally reduces the starting point dependence and it has been shown to provide reliable results for molecular systems~\cite{coccia2017theoretical,faber2013many}, wide band-gap materials \cite{2DThy2017} and perovskites \cite{Filip14, Palummo_JPCL18}.  
In the following we will just refer in general to the GW approach and discuss how the GW self--energy is computed.

%===============
\subsection{Full frequency GW \label{subsec:real-axis} }
%\subsection{Full frequency GW: real axis integration, lifetimes \label{subsec:real-axis} }
%===============

%\subsection{Correlation part of the Self Energy}

Within the GW approximation, the matrix elements of the correlation self-energy over the KS basis are expressed as:
\begin{widetext}
\begin{equation}
\langle n\kk |\Sigma_c (\omega)| n^\prime \kk \rangle = 
\sum_{m\qq} \int \frac{d\omega'}{2\pi i} I^{nn'\kk}_{m\qq} (\omega^\prime)
\left[
\frac{f_{m \kk-\qq}\theta(\omega^\prime)}{\omega-\omega'-\epsilon_{m \kk-\qq} -i\eta}+
\frac{(1-f_{m \kk-\qq})\theta(-\omega^\prime)}{\omega-\omega'-\epsilon_{m \kk-\qq} +i\eta}
\right].
\label{eq:sigma_c}    
\end{equation}
$I$ is linked to the self--energy spectral function. From a computational point of view the definition of $I$ is really critical as, in Eq.~\eqref{eq:sigma_c}, it is connected to the self--energy via a complex Hilbert transformation. In \yambo{} $I$ is defined as
\begin{align}
I^{nn'\kk}_{m\qq} (\omega^\prime)=
-\frac{1}{N_k \Omega}\sum_{\GG \GG'} W^{\delta}_{\GG \GG'}(\qq,\omega') 
 \times \rho_{nm\kk}( \qq, \GG) \rho^*_{n'm\kk}(\qq, \GG'). 
\label{eq:I_c}
\end{align}
In Eq.~\eqref{eq:I_c}, $W^{\delta}$ is the {\it delta}--like part of the screened interaction. This is defined by
\begin{align}
W^{\delta}_{\GG \GG'}(\qq,\omega)=\left[\frac{1}{2}\Im\left(W_{\GG \GG'}(\qq,\omega)+W_{\GG' \GG}(\qq,\omega)\right)-\frac{i}{2}\Re\left(W_{\GG \GG'}(\qq,\omega)-W_{\GG' \GG}(\qq,\omega)\right)\right].
\label{eq:W_delta}
\end{align}
\end{widetext}
In Eq.~\eqref{eq:W_delta} $ W_{\GG \GG'}(\qq,\omega)$ is the screened Coulomb potential defined as 
\begin{align}
W_{\GG \GG'}(\qq,\omega)=\epsilon^{-1}_{\GG \GG'}(\qq,\omega)\frac{4\pi}{|\qq+\GG||\qq+\GG'|}.
\label{eq:W_def}
\end{align}
Note that, in the case of systems with both spatial and time reversal symmetry,
$W_{\GG \GG'}(\qq,\omega)=W_{\GG' \GG}(\qq,\omega)$ and $W_{\GG \GG'}(\qq,\omega)$ reduces to the imaginary part of $W$.

In order to take into account the frequency dependence of the self-energy, two different strategies are implemented in \yambo.
As already described in Ref.~\cite{Marini2009}, it is possible to adopt the plasmon-pole approximation (PPA) in order to model the dynamic screening matrix. This approximation essentially assumes that all the spectral weight of the dielectric function is concentrated at a plasmon excitation. 
Among different models present in the literature \yambo\ implements the Godby-Needs construction~\cite{godby1989PPA} where the parameters of the model are chosen in such a way that $\epsilon^{-1}_{{\bf G G^\prime}}({\bf q},\omega)$ is reproduced at two different frequencies: the static limit $\omega=0$ and another imaginary frequency $\omega=i\omega_p$ given in the input file by 
%PPAPntXp= 27.21138     eV    # [Xp] PPA imaginary energy
{\small\ttfamily{PPAPntXp}}  
(default: {\tt 1 Ha}).
Quasi-particle energy levels calculated within this approximation have been shown to agree to a large extent with numerical integration methods for materials with different characteristics including semiconductors and metal-oxides~\cite{Larson2013,stankovski2011PPA}. 
Moreover, it has the great advantage to avoid the computation of the inverse of the dielectric matrix for many frequency points and to make the frequency integral of Eq.~\eqref{eq:sigma_c} expressible in an analytic form.
Nevertheless the assumption made for the PPA breaks down in certain situations as when dealing with metals~\cite{cazzaniga2012g,marini2001,liu2016} or interfaces~\cite{giantomassi2011} and the frequency integral need to be solved numerically. In \yambo\ the integral is solved on the real-axis which implies the knowledge of the 
full frequency dependence of $ W_{\GG \GG'}(\qq,\omega)$. 
In practice, first the inverse dielectric function $\epsilon^{-1}_{\GG \GG'}(\qq,\omega)$ is evaluated for a number of frequencies set by the variable {\small\ttfamily{ETStpsXd}}, and uniformly distributed in the energy range given by the maximum electron-hole pairs included in the response function defined in Eq.~\eqref{eq:X_IP}. Next, the summation over $\GG$ and $\GG^{\prime}$ is performed computing $I^{nn'\kk}_{m\qq} (\omega^\prime)$ defined in Eq.~\eqref{eq:I_c},
and finally the correlation part of the self-energy is computed via a Hilbert transform 
defined in Eq.~\eqref{eq:sigma_c}.

In this scheme, the evaluation of Eq.~\eqref{eq:W_def} is the most time consuming step due the computation of the inverse dielectric matrix for a large number of frequencies (order of 100) in order to have converged results. Nevertheless as the calculations for each frequency are independent from each other, parallelization over frequencies provides a linear speedup.

Quasi-particle energies calculated by using the real-axis method have been demonstrated to provide the same level of accuracy of other beyond plasmon-pole techniques such as the contour deformation scheme~\cite{rangel2019reproducibility}.

%==============================================================================
\subsection{Electron-mediated lifetimes\label{subsec:lifetimes} }
The ability of \yambo{} to calculate the real--axis $GW$ self--energy allows direct access to the quasi--particle electron--mediated lifetimes. Indeed if we define $\Gamma^{e-e}_{n\kk}(\omega)\equiv \Im\left(\langle n\kk |\Sigma_c (\omega)| n \kk \rangle\right)$, from Eq.~\eqref{eq:sigma_c} it is easy to see that
%\begin{widetext}
%\begin{equation}
\begin{multline}
\Gamma^{e-e}_{n\kk}(\omega) = \frac{1}{2}\sum_{m\qq} 
I^{nn\kk}_{m\qq} (\omega-\epsilon_{m \kk-\qq}) \\
\Big[
\theta(\omega-\epsilon_{m \kk-\qq})f_{m \kk-\qq} \\
-\theta(\epsilon_{m \kk-\qq}-\omega)\left(1-f_{m \kk-\qq}\right)
\Big].
\label{eq:ee_life}    
\end{multline}
%\end{equation}
%\end{widetext}
\yambo{} can evaluate the quasi--particle lifetimes $\tau_{n\kk}(\omega)$, proportional to the inverse of $\Gamma^{e-e}_{n\kk}(\omega)$, either in the on--the--mass--shell approximation\,(OMS) or in the full GW approximation. The difference between the two is the inclusion of the renormalization factors, $Z_{n\kk}$. For details on the theory see, for example, Ref.\onlinecite{marini2001} and references therein.

\begin{figure}[t]
    \centering
    \includegraphics[clip,width=0.45\textwidth]{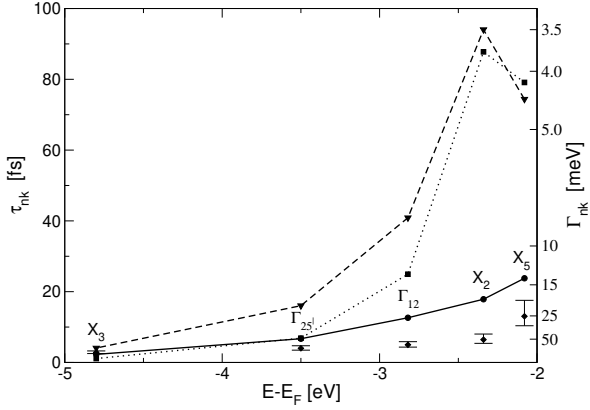}
    \caption{e--e linewidths\,($\Gamma_{n\kk}$) and lifetimes\,($\tau_{n\kk}$) of selected {\it d}-bands of copper. Different level of approximations are shown together with the experimental data\,(diamonds with error bars). The calculated lifetimes are: Full line; $G_0W_0$. Dotted line: OMS $G_0W_0$. Dashed line:  OMS $G_1W_0$. (reprinted with permission from Ref.~\onlinecite{marini2001}. Copyright (2001) by the American Physical Society).}
    \label{fig:cu_lifetimes}
\end{figure}

In the OMS we have that $\left.\Gamma^{e-e}_{n\kk}\right|_{OMS}=\Gamma^{e-e}_{n\kk}(\epsilon_{n \kk})$. The e--e lifetimes of bulk copper are shown in Fig.~\ref{fig:cu_lifetimes} using several flavours of GW approximations\cite{marini2001}.

An important numerical property of the electron--mediated lifetimes calculation is that they depend only on the $\kk$--grid. Indeed, as evident from Eq.~\eqref{eq:ee_life} the band summations are limited by the two theta functions that confine the scattering events in reduced regions of the BZ. This is the mechanism that, in simple metals, leads to the well known quadratic scaling of  $\left.\Gamma^{e-e}_{n\kk}\right|_\mathrm{OMS}$ near the Fermi level, as a function of distance of $\epsilon_{n \kk}$ from the Fermi level itself.

More physical insight in the electronic lifetimes will be given in Sec.~\ref{subsec:TdepLIFE} where the phonon--mediated case will be described.
%==============================================================================

%===============
\subsection{Reducing the number of empty states summation: terminators \label{subsec:terminators_GW} }
%===============
In Sec.~\ref{subsec:terminators_linear_resp} we have discussed the X-terminator procedure. A similar scheme can be  adopted to study the correlation part of the GW self-energy, as from Eq.~(16) of Ref.~\cite{bruneval2008}.
Also in this case the approximation implies the introduction of an extra term 
that takes into account contributions arising from states not explicitly included in the calculation. 
The input parameter governing the use of the terminator corrections on the self-energy (G-terminator) is \\[5pt]
{\small\ttfamily{GTermKind= "none"}}  
{\small\ttfamily{     \# GW terminator ("none","BG")}}
\\[5pt]
(default: {\tt "none"}).
When the variable is set to {\small\ttfamily{none}}, the G-terminator is not applied. On the contrary when it  assumes the value {\small\ttfamily{BG}}, the extrapolar corrective term is calculated. 
The extrapolar energy for the self-energy  is defined by the tunable input variable \\[5pt]
{\small\ttfamily{GTermEn= 1.5 Ha}}
{\small\ttfamily{     \# [X] X terminator energy}}
\\[5pt]
(default: 1.5 Ha).
\begin{figure}
    \centering
    \includegraphics[width=0.48\textwidth]{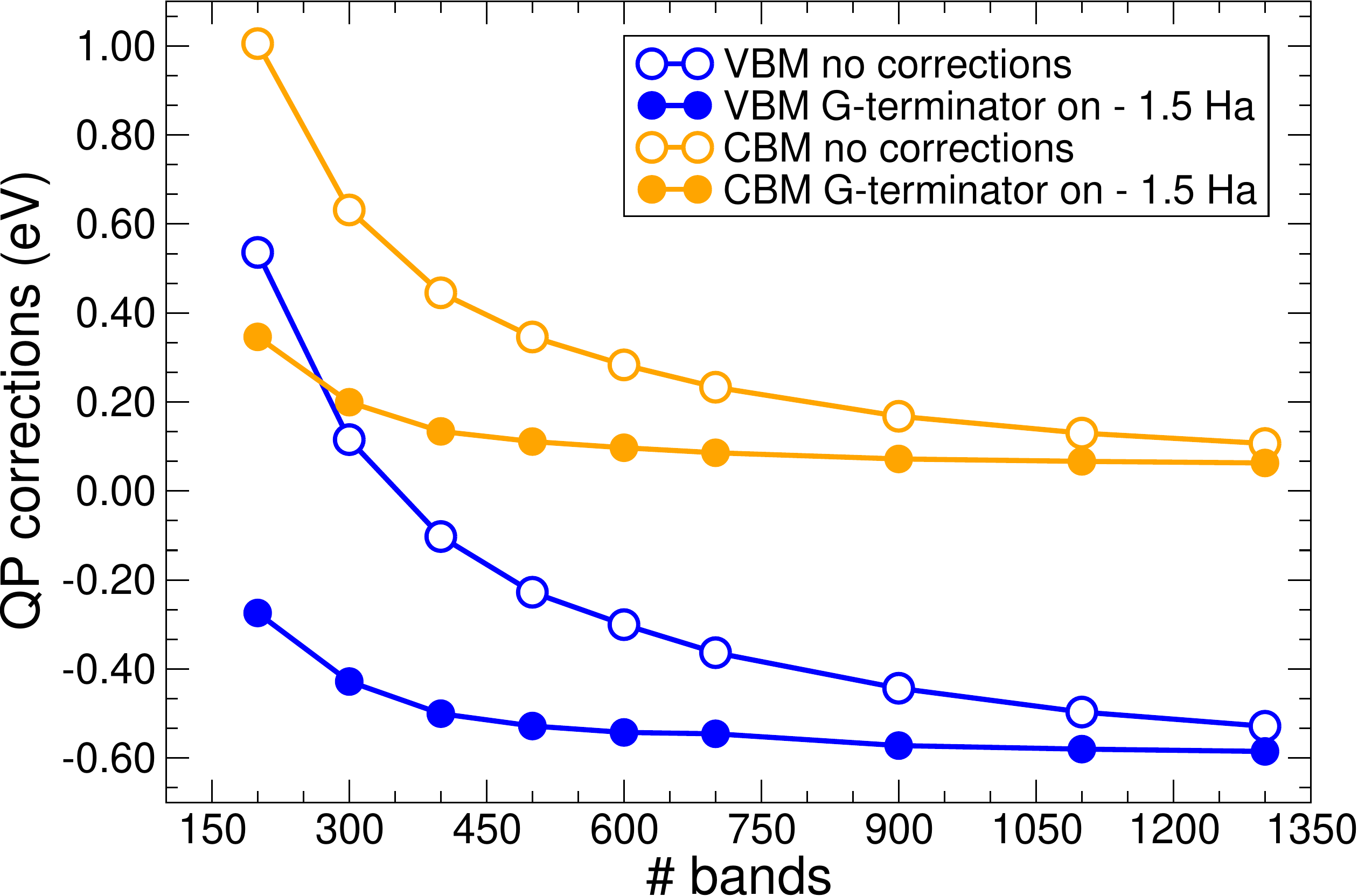}
    \caption{(Color online). Effect of the G-terminator on the convergence vs number of bands included in the self-energy on the VBM and CBM GW-corrections for a bulk Si described in a supercell. 
    }
    \label{fig:Gterm}
\end{figure}

Also in this case, the value is referenced to the highest band included in the calculation. 
In Fig.~\ref{fig:Gterm} we reconsider the system discussed in the example of Section~ \ref{subsec:terminators_linear_resp}, Fig.~\ref{fig:Xterm}. In this case, however, we study  the  convergence of the  self-energy  by exploiting  the G-terminator procedure.
 Empty circles connected with solid lines  show the usual GW convergence for the VBM and CBM states (no corrections applied). 
Calculations have been performed by imposing a high number of  bands in the polarisability (that is therefore converged) and by increasing the number of bands  included  in  the  self-energy.  We set {\small\ttfamily{GTermEn}}=1.5 Ha, that represents the best choice  for  this  system.  
Improvements provided by the use of the G-terminator procedure are represented by solid circles connected with solid lines;
it is evident that the application of this scheme  accelerates the  convergence by leading to a significant reduction  in the number of states necessary to converge the GW self-energy and therefore the calculated QP correction.

\begin{figure}%[t]
    \centering
    \includegraphics[width=0.48\textwidth]{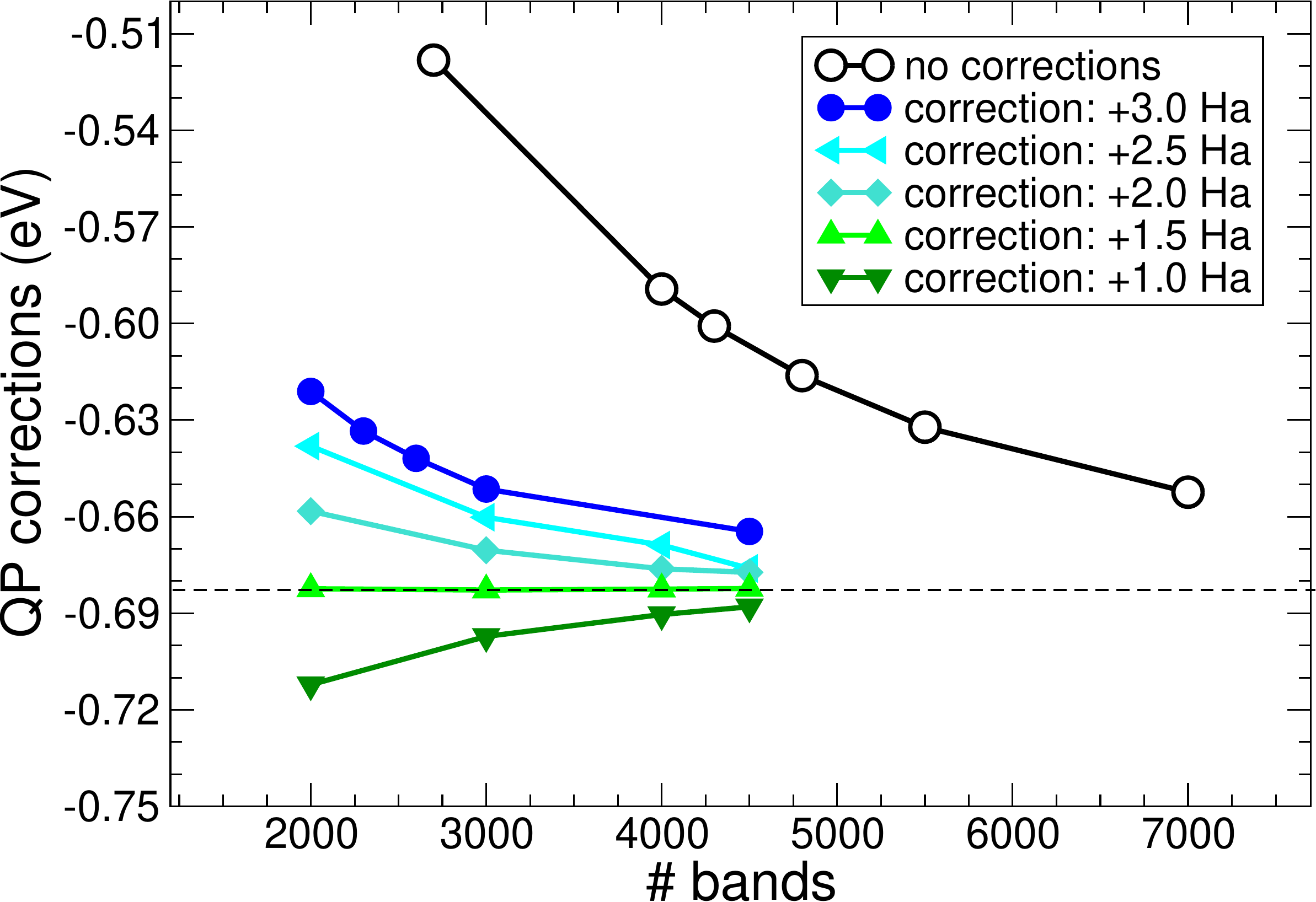}
    \caption{(Color online). Convergence plots of GW-corrected data  for the VBM of a  $\text{TiO}_2$ NW  (27 atoms, 108 occupied states) as a function of the number of bands included in the calculation. Response  and self-energy terminators are simultaneously applied. Calculations have been performed using the same number of bands for the polarizability and the self-energy.   The black line show the usual GW convergence with no corrections.  Coloured lines  are obtained applying the method of Ref.~\cite{bruneval2008} with different values of the extrapolar energy, ranging from  1.0 to 3.0 Ha above the last explicitly calculated KS state.}
    \label{fig:Term_TiO2}
\end{figure}
In order to elucidate the role played by the extrapolar parameter,  
we report in Fig.~\ref{fig:Term_TiO2} a convergence study of the VBM GW correction for a ${\text{TiO}}_{2}$ nanowire (NW). The  black line is obtained  without applying  any  correction. Coloured lines are instead obtained by applying both X- and G-terminators, 
moving the extrapolar energy from  1.0 to 3.0 Ha.  Results are reported  as a function of the number
of states explicitly included in the calculation of both polarisability and self-energy.
As pointed out in Ref.~\cite{bruneval2008}, the extrapolar energy for the self-energy can be
safely taken equal to the extrapolar energy introduced in
Eq.~\eqref{X-term} for the polarizability; for this reason we impose
{\small\ttfamily{XTermEn}} = {\small\ttfamily{GTermEn}}.
Consistently with  the  study  of  Fig.~\ref{fig:Term_TiO2},  
the  convergence  of  the  VBM
without terminators is very slow and requires the inclusion of a large number of bands to be achieved; this condition makes the calculation cumbersome also  on modern HPC-machines.
When the terminator technique is adopted to correct both polarisability and self-energy,  the convergence becomes much faster; especially for some   values of the extrapolar energy (about 1.5  Ha), we observe a significant reduction in the number of bands necessary to converge the calculation, with  a strong  reduction of  both the time-to-solution and the allocated memory. Noticeably, the correction is almost independent on the selected extrapolar energy (terminators are convergence accelerators and the extrapolar correction vanished in the limit of infinite bands included); this parameter therefore influence  the number of bands necessary to converge the calculation (and thus the computational cost of the simulation) but not the final result.

\subsection{Interpolation of the QP band structure\label{subsec:qp_pp}}

In DFT the eigenvalues of the Hamiltonian at every $\mathbf{k}$-point can be obtained by the knowledge of the ground-state charge density, allowing one to perform non-self-consistent calculations on an arbitrary set of $\mathbf{k}$-points. Instead at the HF or GW level, to obtain QP corrections for a given $\mathbf{k}$-point it is necessary to know the KS wave--functions and eigen-energies on all $(\mathbf{k+q})$-points, having chosen a regular grid of $\mathbf{q}$-points as convergence parameter.  In practice \yambo{} computes QP corrections on a regular grid. As a consequence the evaluation of band structures along high-symmetry lines can be computationally very demanding.

A simple strategy which is implemented in \ypp\, is to interpolate the QP corrections from such regular grid to the desired high symmetry lines. The approach implemented is based on a smooth Fourier interpolation~\cite{pickett1988}, which is particularly efficient for 3D grids. The interpolation scheme can also take, as additional input, the KS energies computed along the high symmetry lines to better deal with bands crossing and regions with non analytic behaviour, such as cusp-like features.

A more involved strategy is instead based on the Wannier interpolation scheme as implemented in the \wannier~\cite{Mostofi} and \want~\cite{Ferretti2012} codes, where electronic properties computed on a coarse reciprocal-space mesh can be used to interpolate onto much finer meshes at low cost~\cite{MarzariRMP}.
In the context of GW calculations, the Wannier interpolation scheme can be used to interpolate the QP energies and other band structure properties~\cite{Ferretti2012} (e.g.~effective masses) from QP corrections computed only on selected $\mathbf{k}$-points.
Wannier interpolation of GW band structures requires two sets of inputs: on one side quantities computed at the DFT level such as KS eigenvalues, overlaps between different KS states, and orbital and spin projections of KS states, that are imported from \qe, and on the other side the QP corrections computed by \yambo.
In fact, \wannier~works with uniform coarse meshes on the whole BZ, while \yambo~uses symmetries to compute quantities on the IBZ. In addition, converging the GW self-energy typically requires denser meshes with respect to what is needed for the charge density or Wannier interpolation. 
To address this issue, \ypp{} allows one to unfold the QP corrections from the IBZ to the whole BZ, as required by \wannier~ for interpolation purposes~\footnote{A tutorial on the Wannier interpolation of the GW band structure of silicon is available in the \wannier~package on {\tt GitHub}.}.
Finally the \wannier~code yields a GW-corrected Wannier Hamiltonian and interpolates the GW band structure. A similar procedure is implemented in \want.

\begin{figure}%[t]
    \centering
    \includegraphics[width=0.48\textwidth]{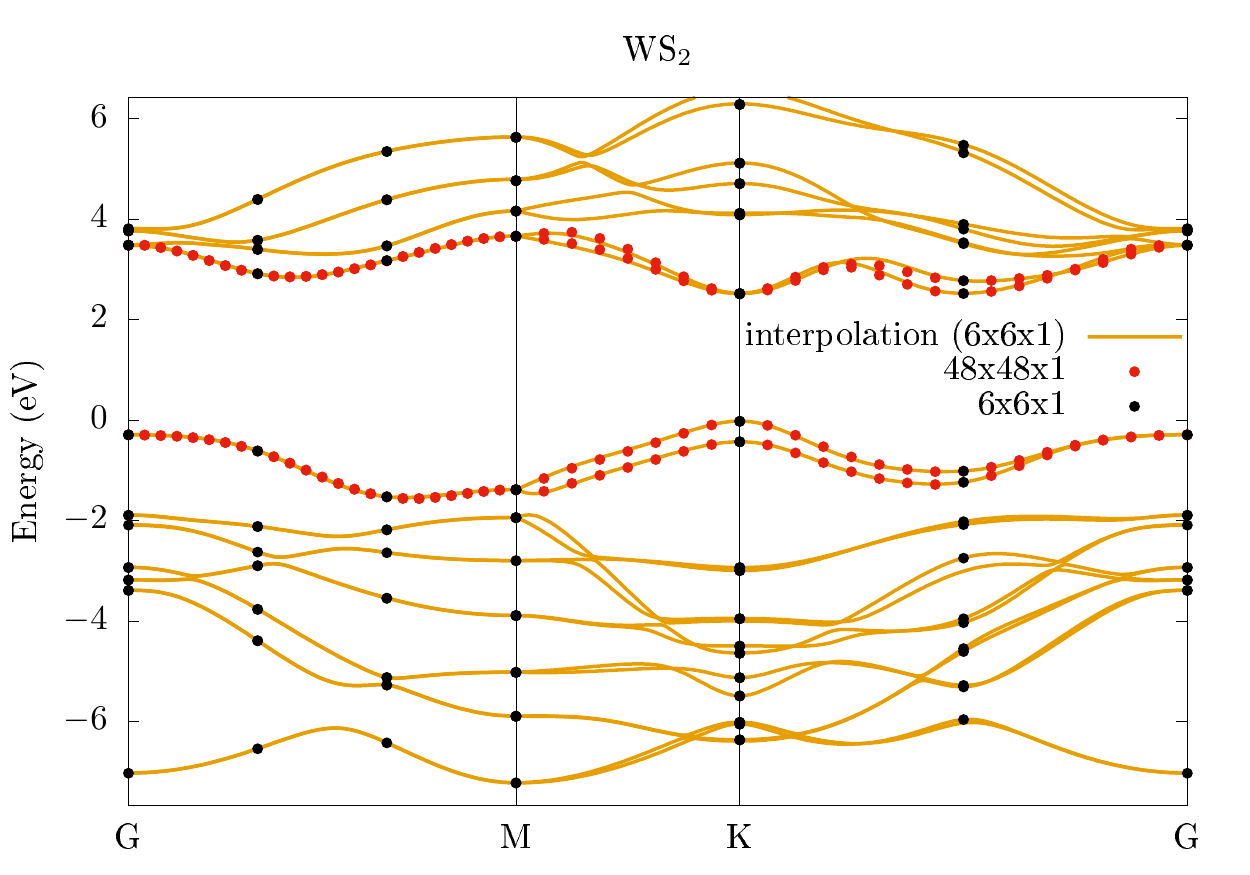}
    \caption{(Color online). GW band structure of monolayer WS$_2$ including spin-orbit coupling and using 48$\times$48$\times$1 $\mathbf{k}$-points grid for the self-energy. The orange lines represent Wannier-interpolated bands obtained from 7 QP energies corresponding to a 6$\times$6$\times$1 grid (black dots), while the red dots shows the QP energies of the full 48$\times$48$\times$1 grid.}
    \label{fig:w90_ws2}
\end{figure}

For example, in monolayer WS$_2$ a grid of 48x48x1 (or denser) is required to converge the GW self-energy. In this case, the band structure can be obtained either by explicitly computing the QP corrections on all $\mathbf{k}$-points of the 48x48x1 grid, or it can be Wannier-interpolated from the QP corrections computed onto coarse subgrids, such as a 6x6x1 corresponding to 7 symmetry-nonequivalent $\mathbf{k}$-points only in the IBZ (see Fig.~\ref{fig:w90_ws2}). The second approach requires substantial less CPU time.

%%%%%%%%%%%%%%%%%%%%%%%%%%%%
\section{Optical absorption\label{sec:BSE} }
%%%%%%%%%%%%%%%%%%%%%%%%%%%%

The solution of the Bethe-Salpeter equation on top of DFT-GW is the state-of-the-art first principles approach to calculate neutral excitations in solid-state systems~\cite{onida2002}, with successful applications to, 
molecules~\cite{coccia2014ab,coccia2017theoretical}, surfaces~\cite{varsano2017role,Hogan2011}, two-dimensional materials~\cite{Molina2018,Hogan2018b}, and nanostructures~\cite{varsano2008,prezzi2008optical},
including biomolecules in complex environments~\cite{varsano2016,varsano2014ground}. The BSE is a Dyson equation for the four point response function $L$. It can be rewritten as an eigenproblem for a two-particle effective Hamiltonian $H^\text{2p}$ in the basis of electron and hole pairs $|eh\rangle$. $H^\text{2p}$ is the sum of an independent-particle Hamiltonian $H^\text{IP}$---i.e. the e--h energy differences corresponding to the independent-particle four-point response function $L_0$ --- and the exchange $V$ and direct contributions $W$ accounting for the e--h interaction.  
The original implementation of the BSE in \yambo\ (See Sec.~2.2 and 3.2 of CPC2009) has been extended in the past decade to ($i$) improve its numerical efficiency (Sec.~\ref{sec:BSE_numerical})---allowing one to treat systems with a large number of electron-hole pairs (i.e. above $10^{5}$)   ---and ($ii$) to capture physical effects (Sec.~\ref{sec:BSE_features}) that were originally neglected---e.g. allowing for the description of the Kerr effect in magnetic materials (Sec.~\ref{sec:BSE_soc}).
Finally, a range of tools have been developed to analyse the exciton localization both in real and reciprocal space (Sec.~\ref{ss:excan}).

\subsection{Numerical efficiency}\label{sec:BSE_numerical}

The computational cost of the BSE grows as a power of the number of electron-hole pairs. As this number can be as large as $10^5-10^6$, it is crucial to devise numerically efficient algorithms for the calculation of the $V$ and $W$ matrix elements and the solution of the BSE. The massive parallelization and memory distribution which contributes in making these calculations possible for very large systems are discussed in Sec.~\ref{sec:parallel}. Here we discuss the use of the double grid for the sampling of the BZ ~\cite{Kammerlander2012}, where the BSE is solved (Sec.~\ref{sec:BSE_numerical_dgrid})---which aims at reducing the number of degrees of freedom involved---and the use of Lanczos-based algorithms together with the interface to the SLEPC library (Scalable Library for Eigenvalue Problem Computations)~\cite{hernandez_slepc} (Sec.~\ref{sec:BSE_numerical_slepc})---which aims at avoiding the full diagonalization of $H^\text{2p}$.

\subsubsection{Double-grid and the inversion solver}~\label{sec:BSE_numerical_dgrid}

The BSE implementation in  \yambo\,
is based on an expansion of the relevant quantities in the basis of electron-hole states. This expansion often requires a very dense
$\mathbf{k}$-point sampling of the Brillouin zone
(BZ). Typically, the number of electron-hole states used in the
expansion can be relatively small if one is only interested in the absorption spectra, but the number of $\mathbf{k}$-points can easily reach
several thousands. Different approaches have been proposed in the literature to solve this problem. A common approach is  the  use  of  arbitrarily  shifted $\mathbf{k}$-point grids,  that often  yield  sufficient  sampling  of  the  BZ  while keeping the number of $\mathbf{k}$-points manageable. Such a shifted grid,  indeed,  does  not  use  the  symmetries  of  the  BZ and guarantees a maximum number of nonequivalent $\mathbf{k}$-points thereby accelerating spectrum convergence. However, it may
induce  artificial  splitting  of  normally  degenerate  states, thus producing artifacts in the spectrum. In \yambo\, we introduced a strategy to solve the BSE equation that alleviates the need for dense $\mathbf{k}$-point grids and does not break the BZ symmetries. Such approach takes into account the fast--changing independent--particle contribution~\cite{Marini2001_PhDthesis,Kammerlander2012}. Indeed the independent-particle term of the BSE, $L_0$, is evaluated on a very dense $\kk$-grid and then the BSE is solved on a coarse $\kk$-grid.
This means in practice that $L_0$ remains defined on the coarse grid, but each matrix element of $L_0$ contains the sum of the nearby poles on the dense grid.  The dense grid can be generated by means of DFT and read using \texttt{ypp -m}, that creates a mapping between the coarse and the dense grid. Then BSE is solved by inversion setting \texttt{BSSmod=`i'}. A similar approach can be also used when computing the response function in ${\bf G}$ space, by replacing each transition in the $F_{nmk}({\bf q},\omega)$ term in eq.~\eqref{eq:X_IP_split} with a sum over the transitions in the dense grid.

\subsubsection{Spectra and exciton wavefunctions via Krylov subspace methods}\label{sec:BSE_numerical_slepc}

\begin{figure}
    \centering
    \includegraphics[clip,width=0.45\textwidth]{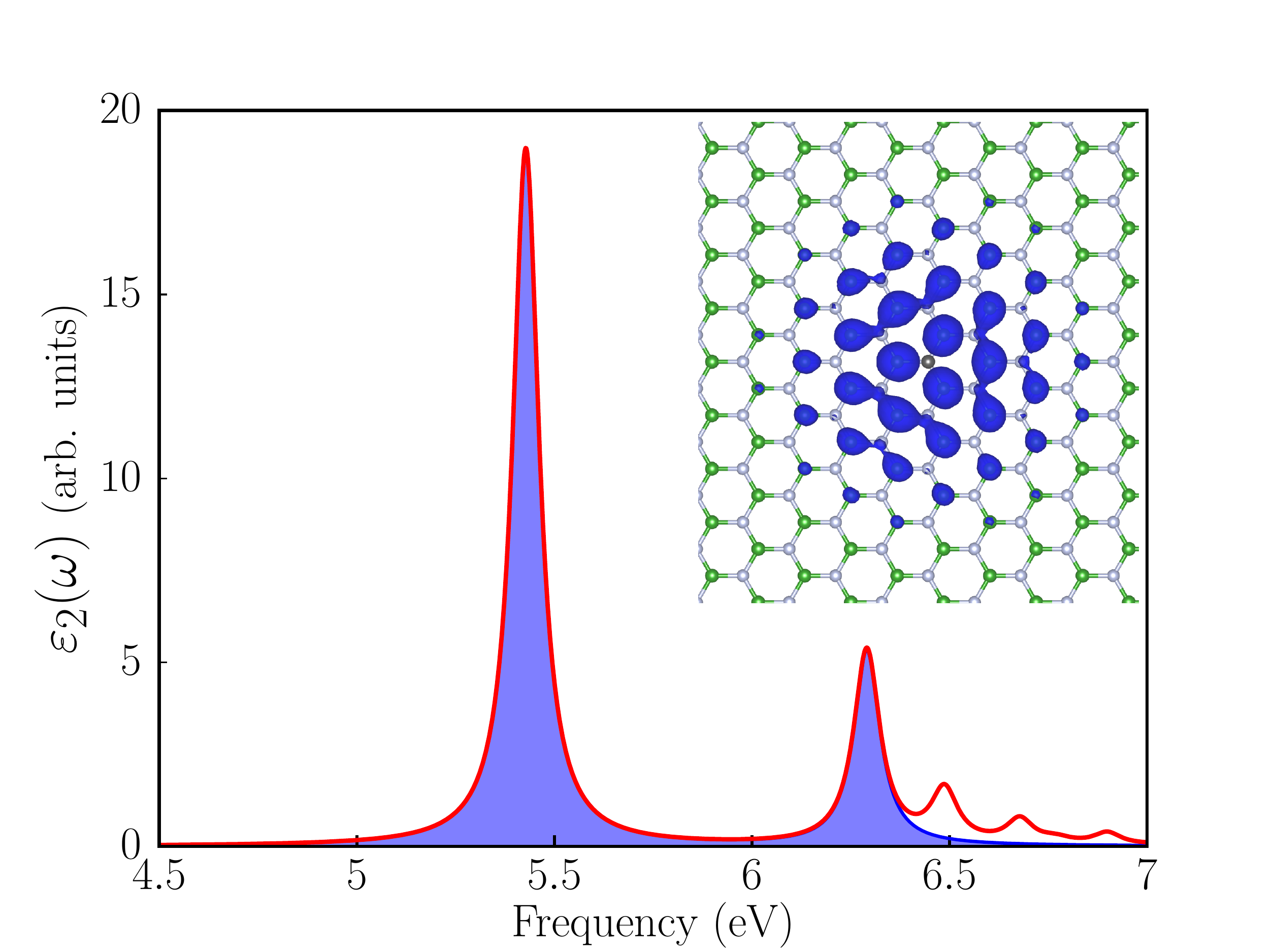}
    \caption{(color online) Optical absorption spectrum of monolayer hBN in a $\sqrt{3}\times\sqrt{3}\times 1$ supercell. The red line refers to an iterative solution using the Haydock solver. The blue shaded region corresponds to a SLEPC calculation where only the first two excitons were included. The inset shows the intensity of the exciton wave function corresponding to the main peak, based on the latter calculation (the hole position is fixed above a nitrogen atom and the resulting electron distribution is displayed).}
    \label{fig:slepc_wf}
\end{figure}

Solving the BSE implies the solution of an eigenvalues problem for the two-particle Hamiltonian that in the e--h basis can result in a matrix as large as $10^6\times 10^6$. The standard dense matrix diagonalization algorithm is available in \yambo\ through the interface with the LAPACK and the ScaLAPACK libraries~\cite{scalapack} (Sec.~\ref{sec:parallel}). Alternatively, when only the spectrum is required, \yambo\ provides the Haydock-Lanczos solver~\cite{haydock1980book}. The latter, originally developed for the Hermitian case only (see Section~3.2 of CPC2009)
---by neglecting the coupling between e--h at positive and negative energies within the Tamm-Dancoff approximation---has been extended to treat the full non-Hermitian two-particle Hamiltonian~\cite{Gruning2009,Gruning2011}. Cases in which considering the  full non-Hermitian two-particle Hamiltonian turns out to be important have been discussed in  Refs.~[\onlinecite{Ma2009,Gruning2009,Palummo2009}].

More recently, \yambo\ has been interfaced with the SLEPC library~\cite{hernandez_slepc}
which uses objects and methods from the PETSC library~\cite{petsc-efficient}
to implement Krylov subspace algorithms to iteratively solve eigenvalue problems.
These are used in \yambo\ to obtain selected eigenpairs of the excitonic Hamiltonian.
This allows the user to select a fixed number of excitonic states to be explicitly calculated thus avoiding the full dense diagonalization and saving a great amount of computational time and memory.
Two options are available for the SLEPC solver. The first, which is the default, uses the PETSC matrix-vector multiplication scheme; it is faster but duplicates the BSE matrix in memory when using MPI. The second, which is activated by the logical \texttt{BSSSlepcShell} in the input file, uses the internal \yambo\, subroutines (the same also used for the Haydock solver); it is slower but distributes the BSE matrix among the MPI tasks.
To select the part of the spectra of interest, the library allows one to use different extraction methods controlled by the variable \texttt{BSSSlepcExtraction}. The standard method, \texttt{ritz}, obtains the lowest lying eigenpairs, while the \texttt{harmonic} method obtains the eigenpairs closest to a defined energy.
The SLEPC solver makes it possible to obtain and plot exciton wave functions (\texttt{ypp -e w}) in large systems where the full diagonalization might be computationally too demanding. For example, the spectrum and the wave function of the lowest-lying exciton in monolayer hBN are shown in Fig. \ref{fig:slepc_wf}. The BSE eigenmodes were extracted only for the two lowest-lying excitonic states, and a $\sqrt{3}\times\sqrt{3}\times 1$ supercell was used in the calculation (the SLEPC spectrum is shown in blue). The full Haydock solution is displayed with a red line for comparison. 

\subsection{Physical effects}\label{sec:BSE_features}

\begin{figure}[t]
    \centering
    \includegraphics[clip,width=0.45\textwidth]{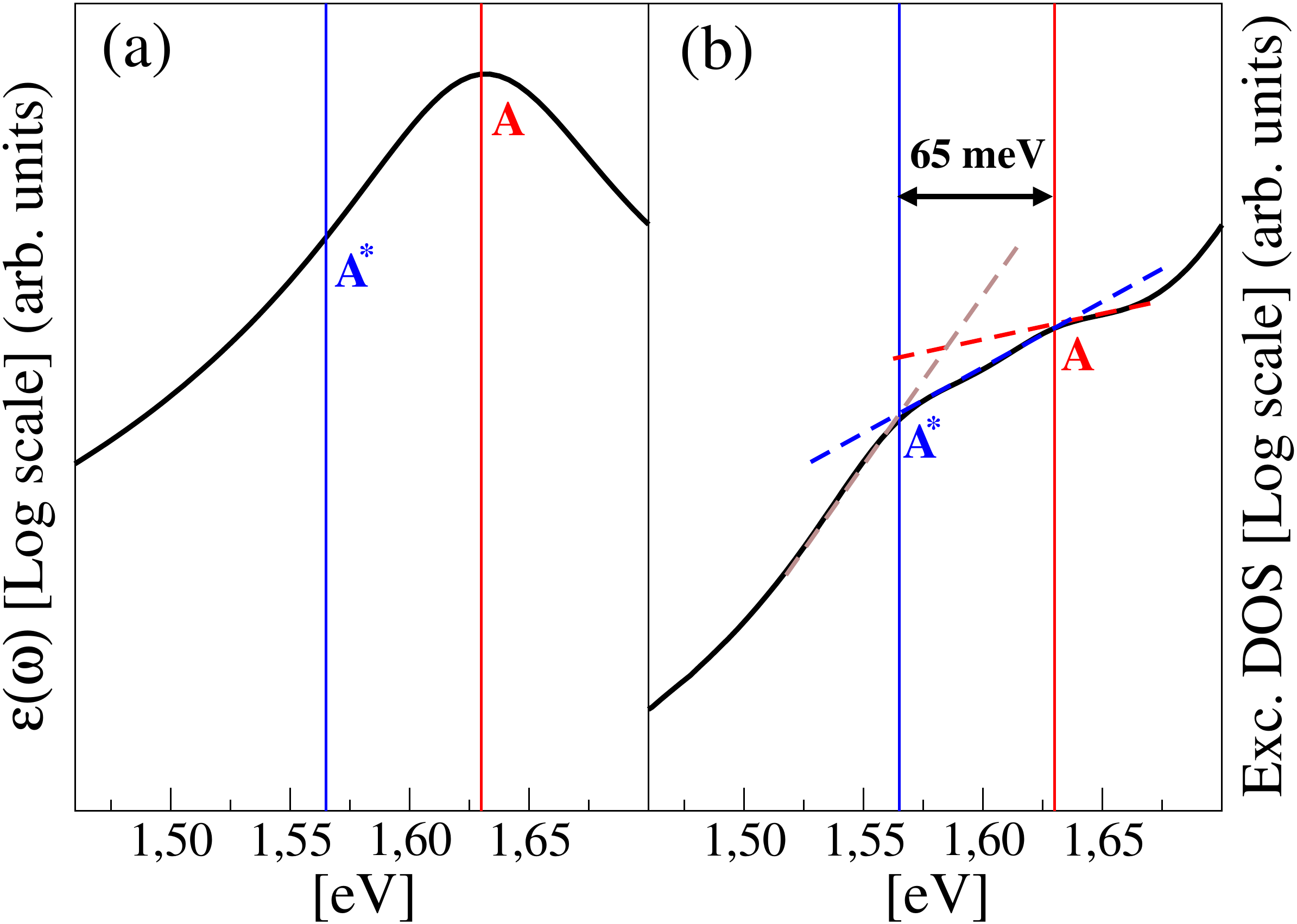}
    \caption{(color online) Optical absorption spectrum of a WSe\textsubscript{2} (panel a) from Ref.~\onlinecite{MolinaSanchez2015},
    compared with Excitonic DOS (panel b). Calculations are
    performed including SOC. In the Excitonic DOS both the bright $A$ and the dark $A^*$ exciton are visible as a change in the slope (the dashed lines are a guide for the eyes, while the vertical continuous lines mark the energy positions of the excitons). Only the bright $A$ exciton is instead visible in the absorption.}
    \label{fig:bse_Wse2_darkA}
\end{figure}

\subsubsection{Spin-orbit coupling and Kerr}\label{sec:BSE_soc}
%
% DS (+AMS/MP)
%
With the implementation and release of the full support for non--collinear systems it is now possible to account for the effects of spin-orbit coupling (SOC) on the optical properties at the BSE level. A detailed description of the implementation and a comparison with other simplified approaches (like the perturbative SOC) can be found in Ref.~\cite{PaperSOC2018}. Since the BSE is written in transition space, the definition of the excitonic matrix is not different from the collinear cases of both unpolarized and spin-collinear systems. For a given number of bands, the main difference is that in the unpolarized case the matrix can be blocked in two matrices of size $N\times N$, describing singlet and triplet excitations. Already in the spin-collinear case this is not possible and the matrix has twice the size $2N\times 2N$. In the non-collinear case, the z-component of the spin operator, $S_z$, is not a good quantum number and the size of the matrix becomes $4N \times 4N$. Since SOC is usually a small perturbation, this means in practice that in the non-collinear case there are peaks which are shifted in energy as compared to the collinear cases ($\Delta S_z=0$ transitions) plus the possible appearance of very low intensity peaks corresponding to spin flip transitions.

The ability of the BSE matrix to capture the interplay between absorption and spin, makes the approach suitable to describe magneto-optical effects. Indeed, starting from the BSE matrix, the off-diagonal matrix elements of the macroscopic dielectric tensor $\varepsilon_{ij}(\omega)$ can be derived, thus describing the magneto-optical Kerr effect~\cite{Sangalli2012}. Notice that in the definition of $\varepsilon_{ij}(\omega)$ the product of dipoles $x^*_{nm\kk}y_{nm\kk}$ enters, thus requiring approaches where the relative phases between different dipoles are correctly accounted for. To this end the \yambokerr\, executable must be used, activating the {{\small\ttfamily EvalKerr}} flag in the input file. The correct off--diagonal matrix elements of the dielectric tensor can be obtained in the velocity gauge (see sec.~\ref{sec:BSE_more}), and only for systems with a gap and Chern number equal zero in the length gauge~\cite{Sangalli2012}.

\subsubsection{Fractional occupations, gauges and more}\label{sec:BSE_more} 

Other extensions have been made available. The implementation has been modified so that the excitonic matrix is now Hermitian (or pseudo--Hermitian if coupling is included) also in the presence of fractional occupations in the ground state. This is done in practice by introducing a slightly modified four-point response function $\tilde{L}$  which is divided by the square root of the occupations as discussed in Eqs.~(14--16) of Ref.~\cite{Sangalli2016}.
The resulting excitonic Hamiltonian has the form
\begin{equation}
\tilde{H}_{ll'}=\Delta\epsilon_l \delta_{ll'} - \sqrt{\Delta f_l} \left( v_{ll'}-W_{ll'} \right) \sqrt{\Delta f_{l'}} 
\label{eq:BSE_symm}
\end{equation}
with $l=\{nk\kk\}$ a super--index in the transition space, with the square root of the occupation factors appearing on the left and on the right of the BSE kernel $v-W$. This has been used to compute absorption of systems out of equilibrium, but it is also important to describe metallic systems like graphene or carbon nanotubes where excitonic effects can be non--negligible due to the reduced dimensionality of the system.

Further, it is now possible to compute the dielectric tensor starting from the different response functions, as described in Ref.~\cite{Sangalli2017}. Indeed, starting from the excitonic propagator $L$, it is possible to construct the density--density response function $\chi_{\rho,\rho}$,
or the dipole--dipole response function  $\chi_{\dd,\dd}$ at $\qq=0$ (length gauge), and the current--current response function $\chi_{\jj,\jj}$ (velocity gauge).
This can be controlled by setting {\small\ttfamily{Gauge="length"}} or {\small\ttfamily{Gauge="velocity"}} in the input file (the length gauge is the default). In case the velocity gauge is chosen the conductivity sum rule is imposed unless the flag {\small\ttfamily{NoCondSumRule}} is activated in the input file. At zero momentum, changing response function is equivalent to change gauge. At finite $\qq$ instead the use of $\chi_{\jj,\jj}$ allows for the calculation of both the longitudinal and the transverse components of the dielectric function.
%Indeed, the $L$ propagator has both the correct transverse and longitudinal poles (while $\overline{L}$ has the correct transverse poles but not the correct longitudinal ones).
The finite--q BSE has been implemented and it is currently under testing before its final release.

Another extension is connected to the output of a BSE run, which also generates a file with the joint density--of--states, at the IP level, and the excitonic density of states, at the BSE level. These can be used for example to visualize dark or very small intensity peaks as shown in Fig.~\ref{fig:bse_Wse2_darkA}.

\subsection{Analysis of excitonic wavefunctions}\label{ss:excan}

Once a BSE calculation is performed using an algorithm which explicitly computes the excitonic eigenvectors $A^{\lambda}_{cv\kk}$, several properties of the excitons can be analyzed as shown in Sec.~\ref{sec:BSE_more} (see Fig.~\ref{fig:slepc_wf}). First of all, the excitonic eigenvalues $E_{\lambda}$ can be sorted and plotted. The so-called amplitudes and weights can also be calculated to inspect which are the main contributions in terms of single quasi-particles to a given excitonic state. The weights are defined as the squared modulus of the excitonic wavefunction  $|A^{\lambda}_{eh}|^2$ (by default only electron--hole pairs that contribute to the exciton more than 5$\%$ are considered; the threshold can be tuned by modifying the input file {\tt `MinWeight'}). The amplitudes are defined as $ \sum_{cv\kk} | A^{\lambda}_{cv\kk} |^2 \delta (\epsilon_{c\kk} -\epsilon_{v\kk} -\hbar \omega)$. 

Moreover the excitonic wavefunction written in real-space ${\Psi_\lambda(\rr_e,\rr_h)=\sum_{cv\kk} A^{\lambda}_{cv\kk} \psi^*_{v\kk}(\rr_h)\psi_{c\kk}(\rr_e)}$ can be computed. ${\Psi_\lambda(\rr_e,\rr_h)}$ is a two-body quantity or joint-correlation function.
Fixing the position of the hole $\rr_{h}=\bar{\rr}_h$, ${|\Psi_\lambda(\bar{\rr}_h,\rr)|^2}$ provides the conditional probability of finding the electron somewhere in space. This quantity is clearly nonperiodic and its spatial decay can change from material to material, marking the distinction between Frenkel and Wannier excitons. As an alternative it is also possible to plot ${|\Psi_\lambda(\rr,\rr)|^2}$ which is instead Bloch-like.

\begin{figure}[t]
    \centering
    \includegraphics[clip,width=0.48\textwidth]{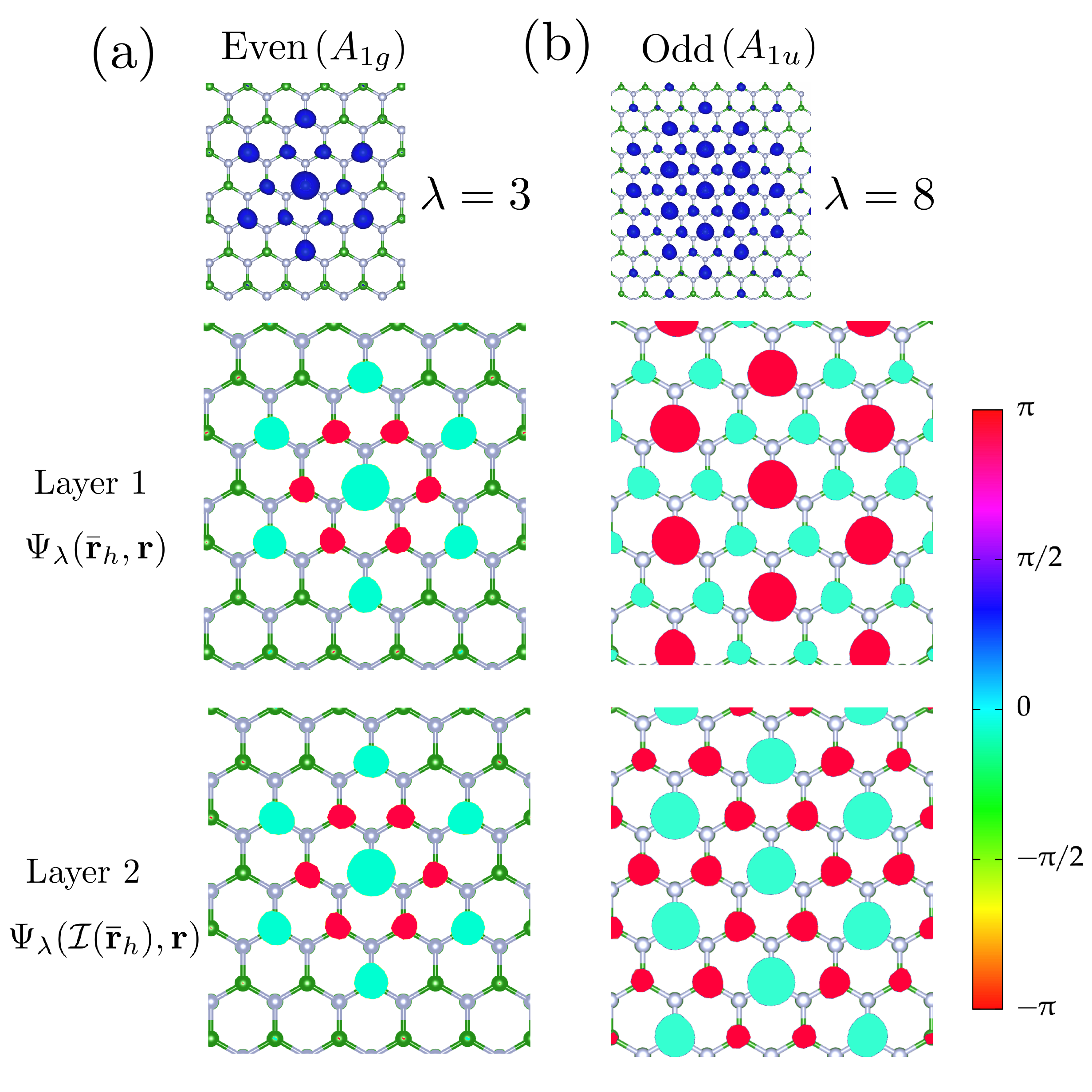}
    \caption{Exciton wave functions $\Psi_\lambda$ of states $\lambda=3$ (a) and $\lambda=8$ (b) of bilayer hBN (only the layer where the electron density is non-negligible is shown). The intensity of $\Psi_\lambda$ is shown in the top frames. Its phase is displayed in the lower frames for two inversion-symmetrical positions of the hole ($\bar{\rr}_h$ and $\mathcal{I}(\bar{\rr}_h)$). The hole is always fixed above a nitrogen atom of the layer not shown. Adapted from Ref. [\onlinecite{Paleari2018}].}
    \label{fig:2lhBN}
\end{figure}
In Fig.~\ref{fig:2lhBN} we focus on two interlayer excitonic states of bilayer hexagonal boron nitride ($\lambda=3$ and $\lambda=8$).
We first plot ${|\Psi_\lambda(\bar{\rr}_h,\rr)|^2}$ (top frames), then we proceed to extract more information by analyzing the phase of $\Psi_\lambda(\bar{\rr}_h,\rr)$ (lower frames).
By comparing the phase for two positions of the hole related by inversion symmetry ($\bar{\rr}_h$ and $\mathcal{I}(\bar{\rr}_h)$), we see that the first exciton (Fig.~\ref{fig:2lhBN}a) is even with respect to inversion symmetry, while the second one is odd (Fig.~\ref{fig:2lhBN}a).
Symmetry analysis of the wave function permits us to conclude that $\Psi_3(\rr_h,\rr_e)$ and $\Psi_8(\rr_h,\rr_e)$ transform as the $A_{1g}$ and $A_{2u}$ representations of point group $D_{3h}$ of the lattice, respectively.

%%%%%%%%%%%%%%%%%%%%%%%%%%%%
\section{Electron-Phonon Interaction \label{sec:elph} }
%%%%%%%%%%%%%%%%%%%%%%%%%%%%

The electron-phonon (EP) interaction is related to many materials properties~\cite{RevModPhys.89.015003} such as the critical temperature of superconductors, the electronic band gap and electronic carrier mobility of semiconductors~\cite{Ponce2018}, the temperature dependence of the optical spectra, the Kohn anomalies in metals~\cite{Forster2013}, and the relaxation rates of carriers \cite{MolinaSanchez2017, Wang2018}. \yamboph\, calculates fully \textit{ab initio} the EP coupling effects on the electronic states, on the excitonic states energies, and on the optical spectra.
The approach used is the many-body formulation which is the dynamical extension of the static theory of EP coupling originally proposed by Heine, Allen, and Cardona (HAC)\cite{allen_heine1976,HAC}. In this framework, the quasi-particle (QP) energies are the complex poles of the Green's function written in terms of the EP self-energy, $\Sigma^\text{el--ph}_{n\bf k}(\omega,T)$, composed of two terms, the Fan, $\Sigma^\mathrm{FAN}_{n\bf k}(\omega,T)$, and Debye-Waller $\Sigma^\mathrm{DW}_{n\bf k}(T)$  contributions~\cite{Fan1951,Antončík1955} (for the complete derivation see, for example, Ref.~\onlinecite{cannuccia_epjb2012,Marini2015}):
\begin{multline}
    \Sigma^\mathrm{FAN}_{n\kk}(i\omega,T) = \sum_{n'\qq \lambda} 
    \frac {|g^{\qq \lambda}_{nn'\kk}|^2}{N_q} \\
\left[ \frac{N_{\qq\lambda}(T)+1-f_{n'\kk-\qq}}{i\go-\gee_{n' \kk-\qq} -\omega_{\qq\lambda}} +\right.\\\left.
\frac{N_{\qq\lambda}(T)+f_{n' \kk-\qq}}{i\go-\gee_{n' \kk-\qq}+\omega_{\qq\lambda}}\right].
\label{eq:FAN}
\end{multline}
Similarly
\begin{align}
\Sigma^\mathrm{DW}_{n\kk}(T)=-\frac{1}{2N_q}\sum_{\qq \lambda} \frac{\Lambda^{\qq\lambda}_{n n' \kk}}{\gee_{n'\kk}-\gee_{n\kk}}  (2 N_{\qq\gl}(T) +1).
\label{eq:DW}
\end{align}
In Eqs.~\eqref{eq:FAN}-\eqref{eq:DW} $N_{\qq\gl}(T)$ is the Bose function distribution of the phonon mode $(\qq,\lambda)$ at temperature $T$.

The ingredients of $\Sigma^\mathrm{el-ph}_{n\bf k}(\omega,T)$, apart from the electronic states, are the phonon frequencies $\omega_{\qq\lambda}$ and the EP matrix elements: $g^{\bf q, \lambda}_{nn'\bf k}$ (first order derivative of the self--consistent and screened ionic potential) and $\Lambda_{nn'\bf k}^{\bf q \lambda}$ (a complicated expression written in terms of the first order derivative\cite{cannuccia_epjb2012,Marini2015}).

These quantities are currently calculated with \qe\, within the framework of DFPT \cite{Baroni_DFPT}. They are read and opportunely stored by the post-processing tool \yppph\, and then reloaded by \yamboph. The procedure is analogous to the one followed by \abinit\ \cite{ponce2014}. 

The HAC approach corresponds to the limit $\lim_{\omega_{\qq\lambda}\rightarrow 0}\Sigma^\mathrm{FAN}_{n\bf k}(\epsilon_{n\kk},T)$. In the HAC the Fan correction reduces to a static self-energy\cite{cannuccia_epjb2012}.

In the next subsections we will give more details about how \yamboph\ has been used to calculate the temperature dependence of the band structure (Sec.\ref{subsec:TdepBS}) and  of the optical spectrum (Sec. \ref{subsec:TdepBethe}). Finally, in Sec.\ref{subsec:dbl_epc} we will describe the way the $\qq\rightarrow 0$ divergence of EP matrix elements has been addressed. 

%***************************************************************************************
\subsection{Temperature-dependent electronic structure}\label{subsec:TdepBS}

The HAC approach is based on the static Rayleigh-Schr\"odinger perturbation theory, allowing one to calculate the temperature-dependent correction of the bare electronic energies, owing to the phonon field perturbation. 
In the QP approximation, the bare energy is instead renormalized because of the virtual scatterings described by the real part of the self-energy and it also acquires a finite lifetime due to the imaginary part of the self-energy. The eigenvalues $ E_{n\bf k}(T)$ are then complex and depend on the temperature. The more the QP approximation is valid the more the renormalization factors $Z_{n\bf k}$ are close to 1, analogously to the GW method. 

If the QP approximation holds, the spectral function $A_{n\bf k}(\omega,T)=\Im \left[G_{n\bf k}(\omega,T)\right]$ is a single-peak Lorentzian function centered at $\Re[E_{n\bf k}]$ with width $\Gamma_{n\bf k}=\Im{[E_{n\kk}]}$. In case of strong EP interaction it has been proven that the spectral function spans a wide energy range \cite{cannuccia_prl2011,gali2016} and the QP approximation is no longer valid. 

\begin{figure}[t]
    \centering
    \includegraphics[clip,width=0.4\textwidth]{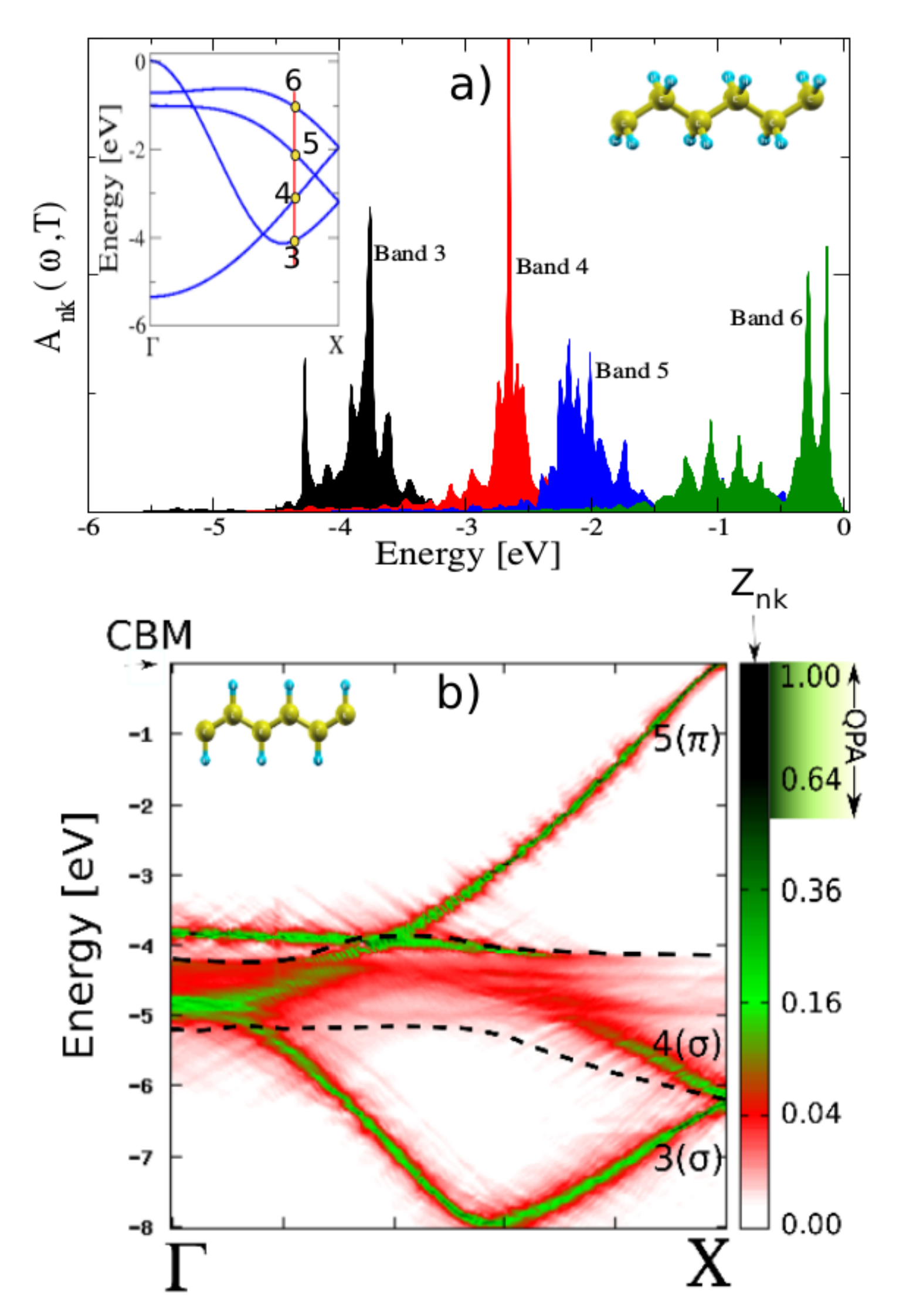}
    \caption{(a) Spectral functions (SFs) of few selected electronic states of trans-polyethylene. In the inset, the last four occupied bands are shown. The red line marks the $\kk$-point at which the corresponding SFs are presented. The selected states are marked with dots. (b) Two-dimensional plot of the SFs of trans-polyacetylene. The range of values of A($\kk$,$\omega$) are given in terms of dimensionless quantity $Z_{n\kk}$. (reprinted with permission of the European Physical Journal (EPJ) from Ref.[\onlinecite{cannuccia_epjb2012}] and with the permission of American Physical Society from Ref. [\onlinecite{cannuccia_prl2011}]). }
    %Copyright (2010) by the American Physical Society).}
    \label{fig:elph-sf}
\end{figure}

Figure~\ref{fig:elph-sf} shows the spectral functions (SFs) of trans-polyethylene and trans-polyacetylene, calculated at $0K$. In Fig.~\ref{fig:elph-sf}(a) multiple structures appear in the SFs. SFs are then spread over a large energy range. In Fig.~\ref{fig:elph-sf}(b) a two-dimensional plot of the SFs reveals a completely different picture with respect to the original electronic band structures. Since SFs are featured by a multiplicity of structures, each carries a fraction of the electronic charge $Z_{n\kk}$ depriving the dominant peak of its weight. A crucial aspect is that some SFs overlap, like in the case of trans-polyacetylene, and in the end it is impossible to associate a well defined energy to the electron and to state which band it belongs to. 

%***************************************************************************************
\subsection{Phonon--mediated electronic lifetimes}\label{subsec:TdepLIFE}
By following the same strategy used in the electronic case, the phonon--mediated contribution to the electronic lifetimes can be easily calculated from Eq.\ref{eq:FAN}. Indeed if we define 
$\Gamma^{e-p}_{n\kk}(\omega,T)\equiv \Im \Sigma^\mathrm{FAN}_{n\kk}(\omega,T)$ it is easy to see that
\begin{multline}
\Gamma^{e-p}_{n\kk}(\omega,T) = \frac{\pi}{N_{\qq}} \sum_{n'\qq\lambda}
|g^{\qq \lambda}_{n n'\kk}|^2 \\
\Big[
\delta(\omega-\epsilon_{n' \kk-\qq}-\omega_{\qq\lambda})
(N_{\qq\lambda}(T)+1-f_{n' \kk-\qq}) \\
+\delta(\omega-\epsilon_{n' \kk-\qq}+\omega_{\qq\lambda})
(N_{\qq\lambda}(T)+f_{n' \kk-\qq})
\Big].
\label{eq:ep_life}    
\end{multline}
In perfect analogy with the electronic case, within the OMS, we have that $\left.\Gamma^{e-p}_{n\kk}(T)\right|_\mathrm{OMS}=\Gamma^{e-p}_{n\kk}(\epsilon_{n \kk},T)$. Like in the electronic case the most important numerical property of the lifetimes calculation is that they depend only on the $\qq$--grid. 

\begin{figure}[t]
    \centering
    \includegraphics[clip,width=0.45\textwidth]{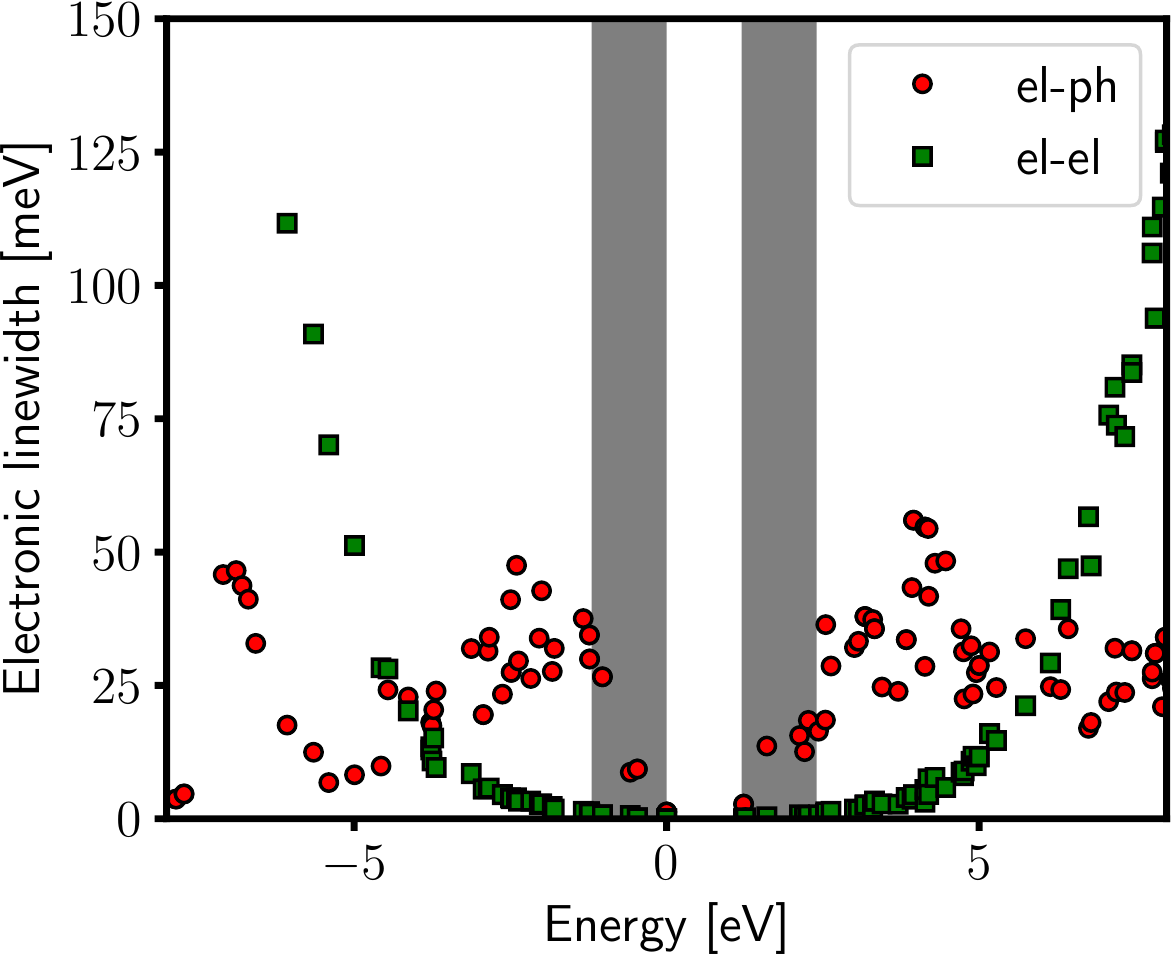}
    \caption{Quasiparticle linewidths of bulk silicon calculated by using the GW approximation for  the e--e scattering (green boxes) and the Fan approximation for the e--p scattering (red circles). The two gray areas denote the energy regions $\epsilon_\mathrm{VBM}-E_g$ and $\epsilon_\mathrm{CBM}+E_g$. Adapted from Ref. [\onlinecite{Marini2013}].}
    \label{fig:ee_vs_ep_lifetimes}
\end{figure}

It is very instructive to compare  $\left.\Gamma^{e-p}_{n\kk}(T)\right|_\mathrm{OMS}$ and the $\left.\Gamma^{e-e}_{n\kk}(T)\right|_\mathrm{OMS}$ for a paradigmatic material like bulk silicon.  This is done in Fig.\ref{fig:ee_vs_ep_lifetimes} in the zero temperature limit. The very different nature of the two lifetimes appear clearly. By simple energy conservation arguments, the electronic linewidth are zero by definition in the two energy regions $\epsilon_\mathrm{VBM}-E_g$ and $\epsilon_\mathrm{CBM}+E_g$, with $E_g$ the electronic gap (in silicon $E_g\approx$1.1\,eV) and $\epsilon_\mathrm{CBM}$ $(\epsilon_\mathrm{VBM})$ the conduction band minimum (valence band maximum).
In these energy regions the e--p contribution is stronger and the corresponding linewidths are larger then the  e--e ones. The quadratic energy dependence of the e--e linewidths inverts this trend at higher energies.

\begin{figure}[t]
    \centering
    \includegraphics[clip,width=0.49\textwidth]{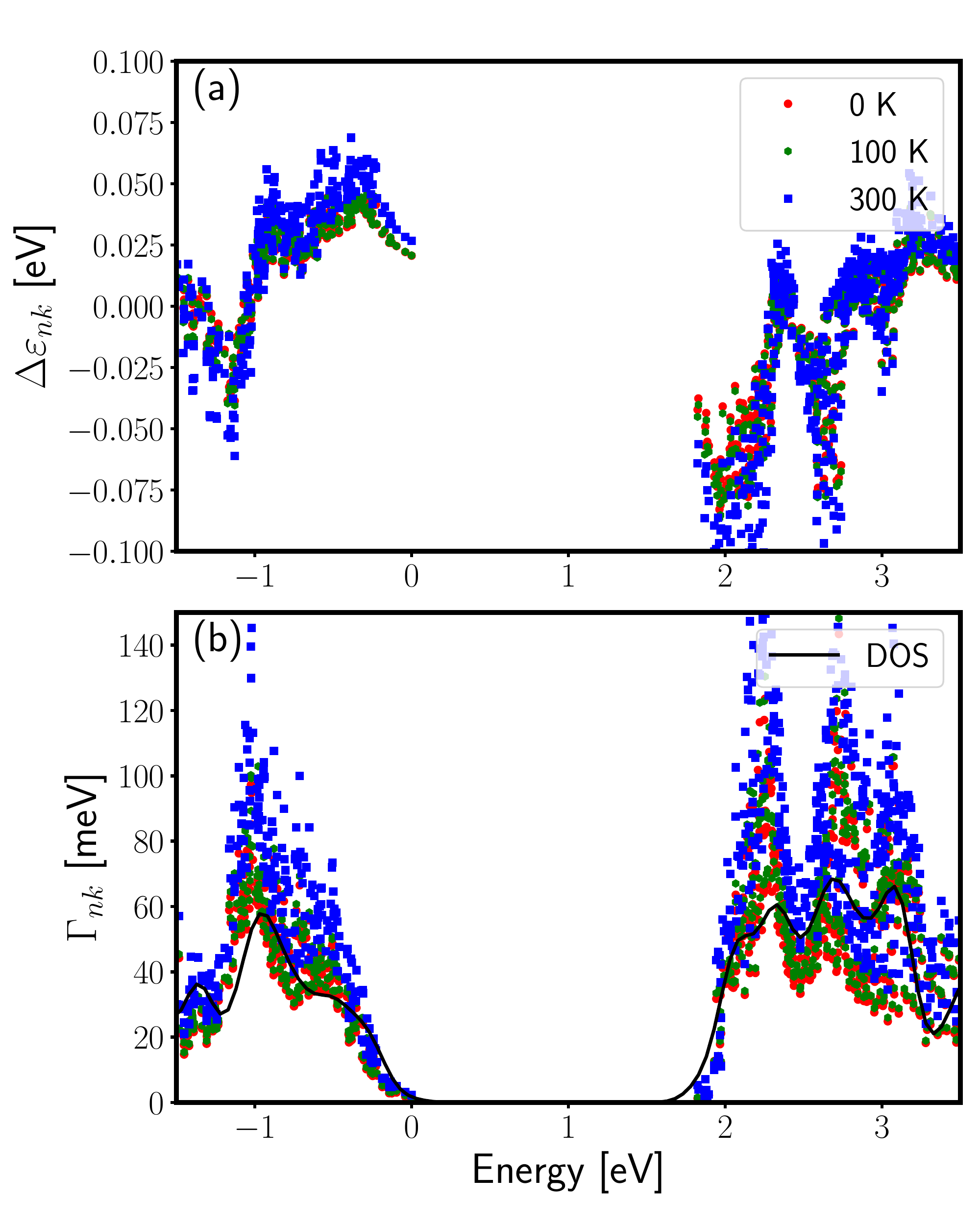}
    \caption{Single-layer MoS$_2$: (a) Electron-phonon correction of the eigenvalues $\varepsilon_{n\bf k}$ for several temperatures. (b) Electron-phonon widths of several temperatures and electronic density of states (black line). }
    \label{fig:elph-qp-w}
\end{figure}
While the e-e contributions grow quadratic in the energy dependence, the e-p ones follow the electronic density of states profile. This property is confirmed by Fig.~\ref{fig:elph-qp-w}$(b)$ (see also Ref.\cite{Bernardi2014}) and remains accurate when the temperature increases. Fig.~\ref{fig:elph-qp-w}$(a)$ shows instead the EP correction in single-layer MoS$_2$ of the valence and conduction band states for several temperatures, together with the widths and the density of states (DOS). In general, the EP correction tends to close the bandgap. This is visible in Fig. \ref{fig:elph-qp-w} (panels a and b), the conduction state energy decreases with temperature, while the valence one increases. Only in a few cases we find an opening of the bandgap when temperature increases \cite{Villegas2016}.

%***************************************************************************************
\subsection{Finite Temperature Bethe-Salpeter Equation}\label{subsec:TdepBethe}

\begin{figure}
    \centering
    \includegraphics[clip,width=0.45\textwidth]{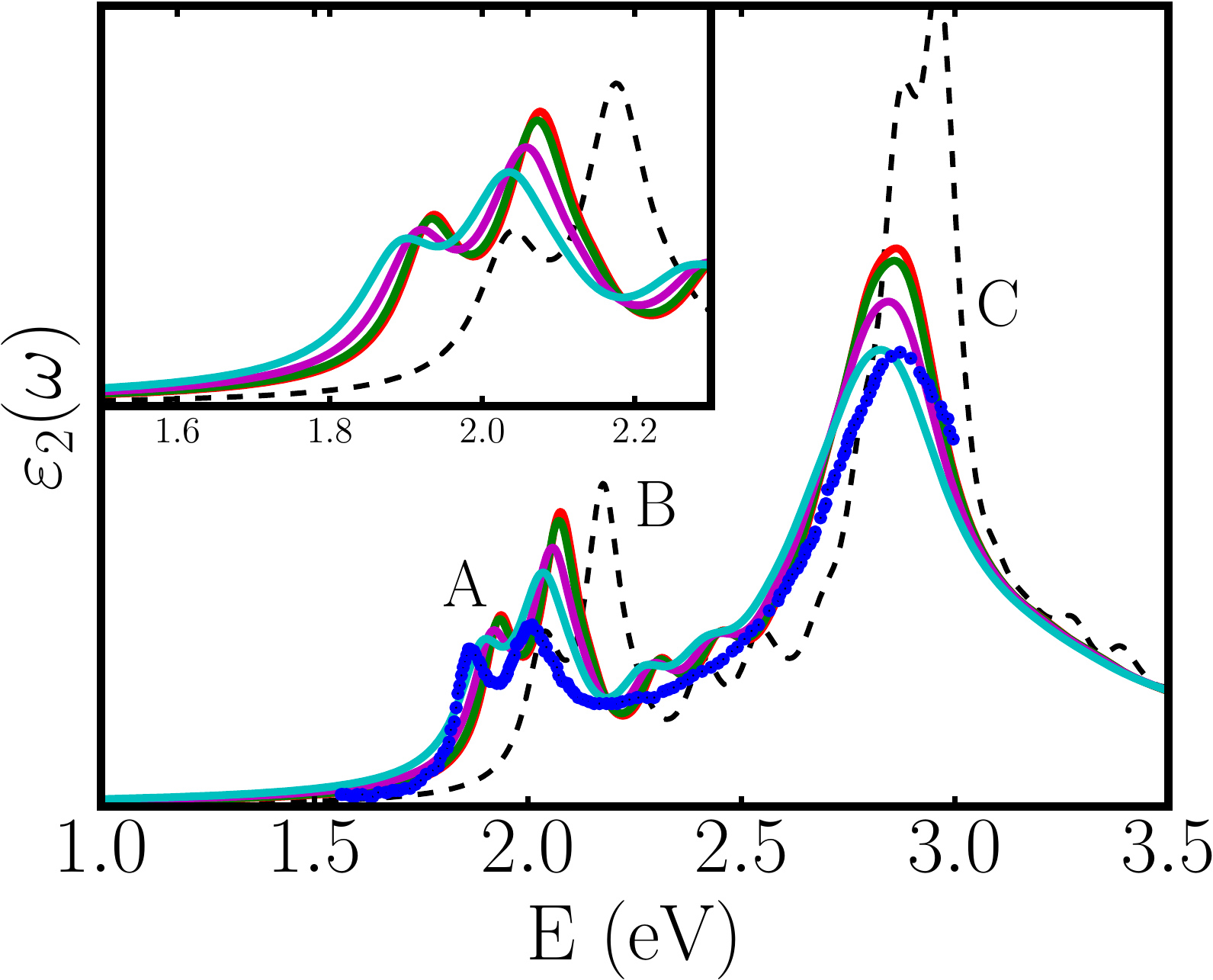}
    \caption{Optical spectra of single-layer  MoS$_2$  for temperatures of 0 (red), 100 (green), 200 (magenta) and 300 K (cyan). Spectrum without electron-phonon effects is also shown (dashed line). Blue dots represent experimental data from Ref.~\onlinecite{Li2014}
    (reprinted with permission from Ref.~\onlinecite{MolinaSanchez2016}. Copyright (2016) by the American Physical Society)}
    \label{fig:bse-temp}
\end{figure}

Once the temperature-dependent corrections to electron and hole states have been calculated, they constitute the key ingredients of the finite temperature excitonic eigenvalue equation. Since the electron and hole eigenvalues are complex numbers the resulting excitonic eigenvalues have a real part (the exciton binding energy) and an imaginary part (the exciton lifetime). The dielectric function then depends explicitly on $T$, $\epsilon_2(\omega,T)= -(8\pi/V) \sum_\lambda |S_\lambda(T)|^2\Im{[\omega-E_\lambda(T)]^{-1} } $, where
$S_{\lambda}(T)$ are the excitonic optical strengths and $E_\lambda(T)$ are the complex excitonic energies. As shown in Fig.~\ref{fig:bse-temp} for a single layer MoS$_2$, the main effect of the temperature on the optical spectra is the renormalization of the energy transitions along with a broadening of the spectrum lineshape related to the finite lifetime of the underlying excitonic states which increases with $T$~\cite{kawai2014,MolinaSanchez2016}. This picture is also valid when $T\rightarrow 0$ because of the zero-point vibrations. A remarkable effect of the exciton-phonon coupling has been observed in hexagonal BN. It has been proven that the optical brightness turns out to be strongly temperature-dependent such as to induce bright to dark (and viceversa) transitions~\cite{marini2007}. 

%========================================================================
\subsection{Double grid in the electron-phonon coupling: a way to deal with the $\bf{q}\rightarrow 0$ divergence}
\label{subsec:dbl_epc}
%========================================================================
A technically relevant issue is the slowing down of energy correction 
convergence at some high-symmetry points. Some EP matrix elements might be zero by symmetry and are not representative of the discretization of an integral. \yambo{} deals with this issue by computing the energy shift corrections on a random $\bf q$-wavevector grid of transferred momenta. The numerical evaluation of the EP self-energy on a dense $\bf q$-grid is a formidable task (see Eq. (5) of Ref.~\cite{cannuccia_epjb2012}). The reason is that such dense grids of transferred momenta are inevitably connected with the use of equally dense grids of $\bf k$-points. The solution implemented in \yambo\, is a double grid approach: matrix elements are calculated for a fixed $\bf k$-point grid while energies are integrated using a larger grid of random-points in the whole BZ.

To speed-up the convergence with the number of random points, the BZ is divided in small spherical regions $R_{\qq}$ centered around each $\qq$ point of the regular grid and the integral is calculated using a numerical Monte-Carlo integration technique. Furthermore, the divergence at $\qq \rightarrow 0$ of the $|g^{\qq \lambda}_{n'n\kk}|^2$ matrix elements is explicitly taken into account for the 3D case for which the $\qq$ integration compensates the $\qq^{-2}$ divergence. In the case of 2D materials the divergence of the EP matrix elements would not be lifted by the surface element $2\pi q$. In principle, an analytic functional form for the EP matrix elements can be envisaged as reported by Ref.~\cite{Kaasbjerg2012}.  

%%%%%%%%%%%%%%%%%%%%%%%%%%%%
\section{Real time propagation \label{sec:real-time} }

\newcommand{\ks}{Kohn-Sham}

A new feature in \yambo{}
%, recently introduced in the \yambo\, code, 
is the numerical integration of a time--dependent (TD) equation of motion (EOM), able to describe the evolution of the electronic system under the action of an external laser pulse.
Similarly to the equilibrium case, the most diffuse ab--initio approaches to real--time propagation are based on TD--DFT and there exist a number of GPL codes available to this end~\cite{octopus,salmon}. On the contrary the implementation of real time propagation within MBPT is an almost unique feature of the \yambo{} code.
Two different schemes are available. In one case, the density matrix of the system, $\rho(\rr,\rr',t)$, is propagated in time, as described in sec.~\ref{sec:RT_RHO}. In the second case, the valence bands $u_{n\kk}(\rr,t)$ are propagated by means of a time dependent Schrodinger equation, as described in sec.~\ref{sec:RT_NL}.

Standard TD--DFT codes often (but not always) implement real time propagation in real space or reciprocal space basis--set. Instead for the two schemes above mentioned, the EOMs in \yambo{} are represented in the space of the equilibrium KS wave--functions. Since a direct implementation of MBPT in real space (or in reciprocal space) is very cumbersome, the KS space offers a convenient alternative. The comparison between real--space vs KS--space has been extensively discussed in the literature TD--DFT where both approaches are feasible. Despite strict converge against the number KS states can be hard, very good results are obtained already with very few basis functions.
The philosophy is similar to the one used to compute equilibrium QP corrections and BSE spectra, where both the self--energy and the excitonic matrix are written in KS space.

\subsection{Time--dependent Screened Exchange}\label{sec:RT_RHO}
%================================================================

The EOM for the density matrix projected in the space of the single particle wave--functions, $\rhom$, is derived from non--equilibrium (NEQ) many--body perturbation theory and reads
\begin{equation}
\label{eq:td_rho}
i\hbar  \partial_t\, \rhom_\kk(t) =
\left[ \hm^{rt}_\kk[\rhom]+\Um^{ext}_\kk(t), \rhom_\kk(t)  \right] -i \Gamma_\kk \rhom_\kk \;.
\end{equation}
Here we underline quantities which are vectors in the transition space (and we will underline twice matrices in transition space). $h^{rt}$ contains the equilibrium eigenvalues $\epsilon_{n\kk}$ plus the variation of the self-energy $\Delta\Sm^{Hxc}[\rhom]$; $\epsilon_{n\kk}$ can be the KS energies or the QP corrected energies. QP corrections can be loaded either from a previous calculation or by adding a scissor operator from input. For $\Delta\Sm^{Hxc}[\rhom]$ different levels of approximation can be chosen, setting the \HXCkind\, input variable.
$\Um^{ext}=-e\efield\cdot {\bf \rmat}$ represents the external potential written in the length gauge; shape, polarization, intensity (and eventually frequency) of the field  $\efield$ can be selected in input. ${\bf \rmat}$ is the position operator. The coupling to the external field is exact up to first order. From the knowledge of the density matrix, the first order polarization $\PP(t)=-e\sum_{i\neq j{\bf k}} {\bf r}_{ij{\bf k}} \rho_{ij{\bf k}}$ is computed at each time step. The spectrum of the system can then be obtained by the Fourier transform of the polarization, which can be done as a post-processing step. Absorption is thus obtained via the dipole--dipole response function (equivalent to the length gauge in linear response). A delta-like external field is convenient to obtain the spectrum for all frequencies. 

\begin{figure}[t]
\centering
\includegraphics[width=0.48\textwidth]{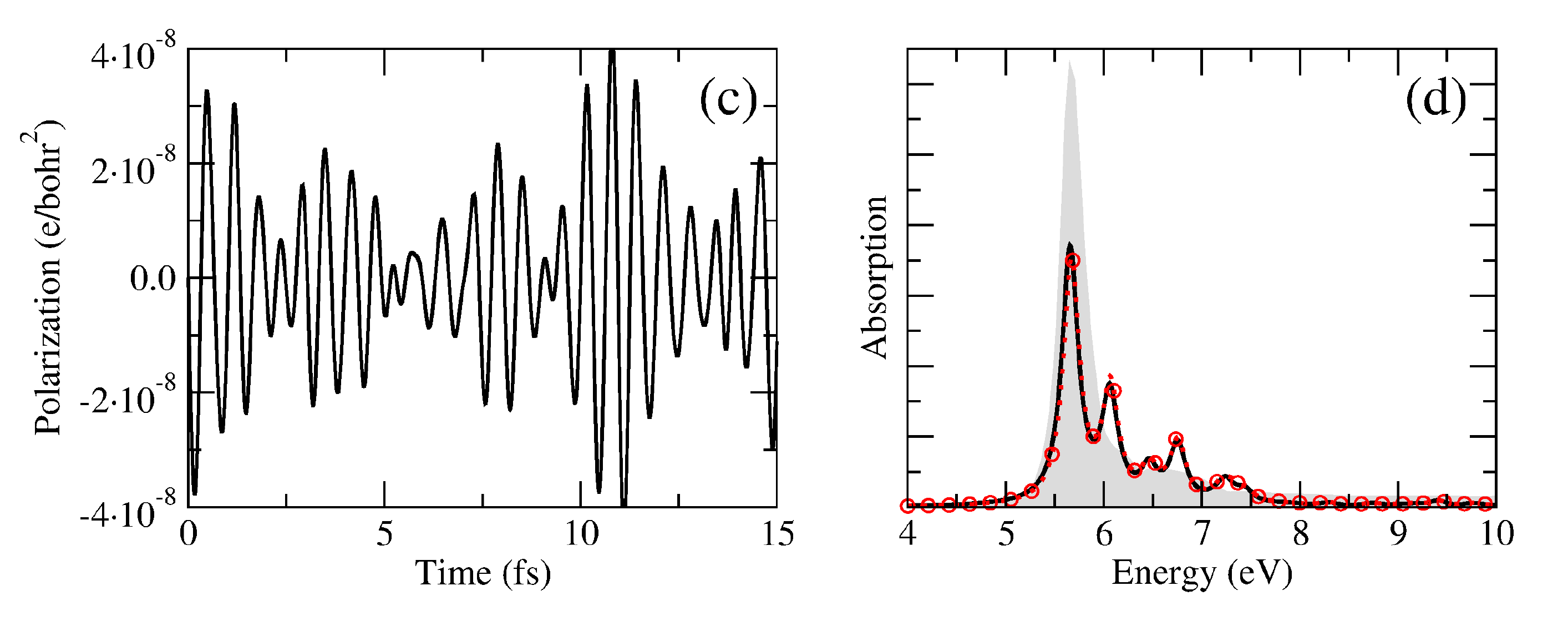}
\caption{\footnotesize{[Color online]  Time dependent polarization in hBN, panel (a), obtained obtained solving Eq.~\eqref{eq:td_rho} within the HSEX approximation. In panel (b) its Fourier transform, red circles, matches the absorption computed within BSE, black line.   (reprinted  with permission from Ref.~\onlinecite{attaccalite2011real}. Copyright (2011) by the American Physical Society.)}
\label{fig:hBN_TD-SEX}}
\end{figure}

The implementation of the external field in the velocity gauge (equivalent to the velocity gauge in linear response) is currently under testing before its final release.
Equation~\eqref{eq:td_rho} represents a set of equations, one for each $\kk$-point in the BZ, coupled via the functional dependency of $\Delta\Sm^{Hxc}$ on the whole $\rhom$. Different options of the self--energy are available, by setting the \HXCkind\, variable to: {\small\ttfamily{IP}}, {\small\ttfamily{Hartree}}, {\small\ttfamily{DFT}}, {\small\ttfamily{Fock}}, {\small\ttfamily{Hartree+Fock}}, {\small\ttfamily{SEX}}, or {\small\ttfamily{Hartree+SEX}}.
For \HXCkind={\small\ttfamily{IP}} one has $\Delta\Sm^{Hxc}=0$. For local \HXCkind\, like {\small\ttfamily{Hartree}} and {\small\ttfamily{DFT}}, $\Delta\Sm^{Hxc}$ can be computed on the fly from the real--space density $n(\rr,t)$ and the approach is in practice equivalent to TD-DFT to linear order in the field. For non--local \HXCkind\, like {\small\ttfamily{Hartree+Fock}} (HF) and {\small\ttfamily{Hartree+SEX}} (HSEX), the self--energy is written in the form $\Delta\Sm^{Hxc}[\rhom]=\Kmm^{Hxc}\cdot\rhom$ with $\Kmm^{Hxc}$ computed before starting the real--time propagation. The calculation of $\Kmm^{Hxc}$ can be either done in a preliminary run, with the matrix--elements stored on disk and then reloaded, or on-the-fly before starting the real-time propagation. In case the HSEX approximation is used the resulting spectrum is equivalent to a BSE calculation in the limit of small perturbations, as shown both analytically and numerically in Ref.~\onlinecite{attaccalite2011real}.
Thus, to linear order, TD-SEX is is able to properly capture excitons, which can be hardly described within TD-DFT. The comparison between the two approaches is reported in Fig.~\ref{fig:hBN_TD-SEX}.

When local self--energies are computed directly from the real space density,
the numerical cost is mainly due to the projection of the potential on the KS--basis set at each iteration. This step is avoided in real--space TD-DFT where, however, the wave--functions on the real--space grid need to be propagated. The relative computation cost of the two strategies depends critically on the size of the real-space grid vs the number of KS function used.
Instead when non local self-energy are used, the computational cost is mostly due to the preliminary calculation of the kernel $\Kmm^{Hxc}$. This step has, roughly, the same computational cost of a standard BSE run and requires to store $\Kmm^{Hxc}$ in memory (disk or RAM). The subsequent real--time propagation is instead very fast. In some cases it is convenient to use this scheme also for local self--energies. However, regardless of the self--energy used, only variations of the self--energy which are linear in the density matrix are described when using $\Kmm^{Hxc}$.

To run simulations and compute the spectra as described in the present section the \yambort\, and \ypprt\, executables need to be used.

\subsubsection{Double-grid in real time \label{sec:RT_RHO_dgrid} }

As for the BSE case (see Section~\ref{sec:BSE_numerical_dgrid}), also the real--time propagation can be done taking advantage of a double--grid in $\kk$-space. Similarly to the BSE, the matrix elements, i.e. the dipoles and $\Kmm^{Hxc}$, are computed using the wave--functions on the coarse grid, while energies and occupations are defined on the fine grid. At variance with the BSE implementation however the matrix elements on the coarse grid are then extrapolated onto the fine grid with a nearest-neighbour technique since $\rhom$ is then defined and propagated on the double grid. This is different in spirit from the inversion solver. It would be equivalent to define the excitonic matrix (or $L$ propagator) on the double grid. Instead, in the double-grid approach within BSE the excitonic matrix remains defined on the coarse grid, while the fine grid enters only in the definition of $L^0$, as described in sec.~\ref{sec:BSE_numerical_dgrid}. 

\subsection{Nonlinear optics}\label{sec:RT_NL}
%================================================================

\begin{figure}[t]
\centering
\includegraphics[width=0.47\textwidth]{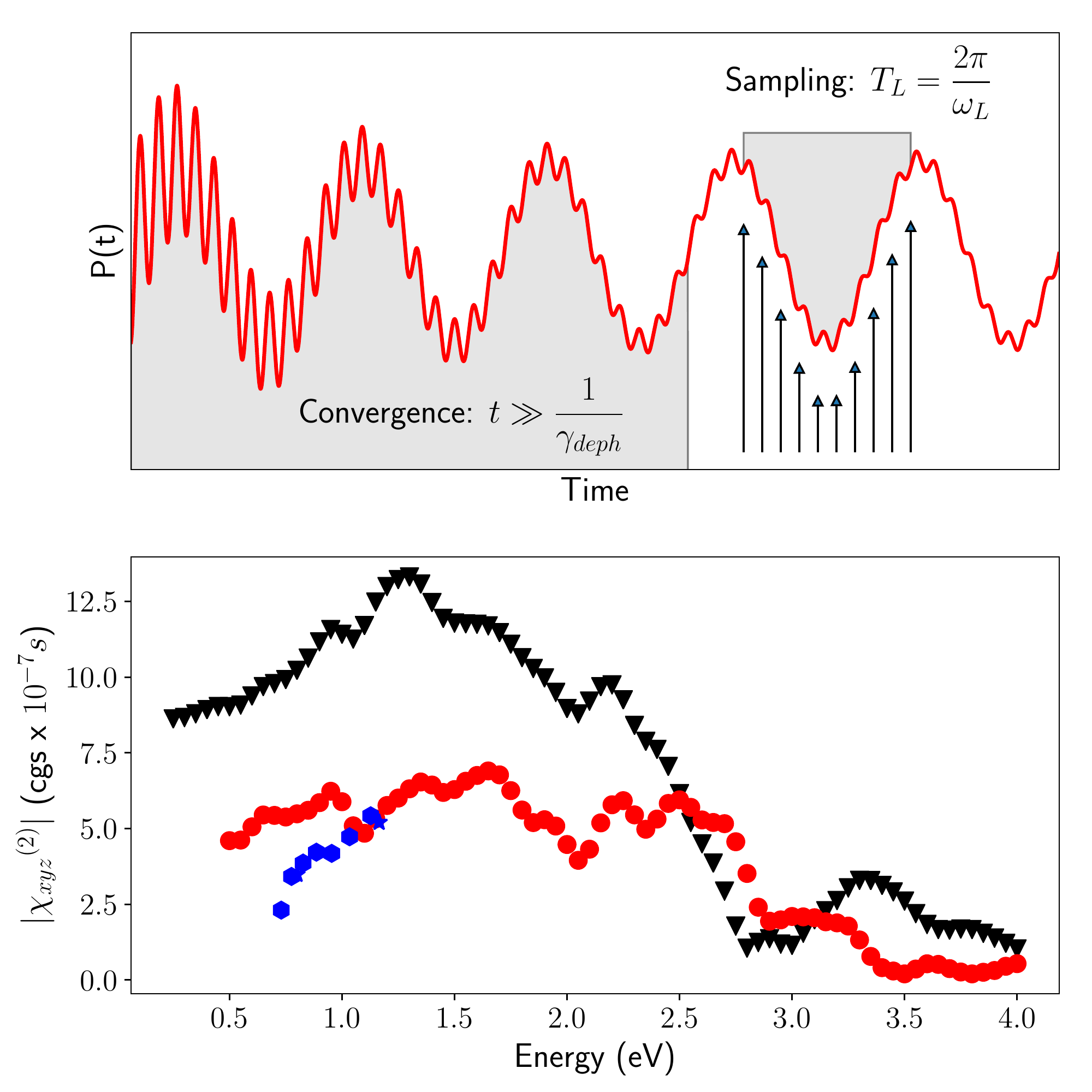}
\caption{\footnotesize{[Color online]  Top panel: schematic representation of real-time simulation for the non-linear response. Bottom panel: magnitude of $\chi^{(2)}(-2\omega,\omega,\omega)$ for bulk CdTe calculated within the QPA (black triangles) and TDH (red circles). Each point corresponds to a real-time simulation at the given laser frequency. Comparison is made with experimental results from Refs.~\onlinecite{Shoji:97,Jang:13} (blue stars and hexagons).  \label{fg:CdTeSHG}
(reprinted with permission from Ref.~\onlinecite{Attaccalite2013}. Copyright (2013) by the American Physical
Society.)}} \end{figure}

Alternative to the time-evolution of the density matrix, it is possible to perform the time-evolution of the Schr\"odinger equation for the periodic part of the Bloch states projected in the eigenstates of the equilibrium Hamiltonian: $|v_{m\textbf{k}} \rangle$.
Here we briefly  present the actual implementation in \yambo{} and how it can be used to obtain non-linear optics response, for more details see Ref.~\cite{Attaccalite2013}.
The EOM for the valence band states reads:
\begin{equation}
i\hbar  \partial_t\, | v_{m\kk} \rangle = \left( h^{\text{rt}}_{\kk}[\rho] +i \efield \cdot \tilde \partial_\kk\right) |v_{m\kk} \rangle \label{eq:td_wfs} \;.
\end{equation}
where the effective Hamiltonian $h^{\text{rt}}_\kk$ has been  introduced in sec.~\ref{sec:RT_RHO} and $\rho(t)$ is constructed starting from $| v_{m\kk} \rangle$.
%The specific form of $ H^{\text{MB}}_{\mathbf k}$ is presented below.
The second term in Eq.~\eqref{eq:td_wfs}, $\efield \cdot \tilde \partial_\textbf{k}$, describes the coupling with the external field $\efield$ in the dipole approximation. As we imposed Born-von K\'arm\'an periodic boundary conditions, the coupling takes the form of a $\textbf{k}$-derivative operator $\tilde \partial_\textbf{k}$. The tilde indicates that the operator is ``gauge covariant'' and guarantees that the solutions of Eq.~\eqref{eq:td_wfs} are invariant under unitary rotations among occupied states at $\textbf{k}$ (see Ref.~\cite{souza_prb} for more details).

Propagating the single particle wave-functions presents advantages and disadvantages with respect to the density matrix. The major advantage is that the coupling of electrons with the external field, within the length gauge, is now written in terms of Berry's phase, which is exact to all orders also in extended systems~\cite{resta1994macroscopic}.
%
%The evaluation of the Berry's phase from density matrix or Green's function remains a difficult task, therefore for application where coherence light excitations is requires as non-linear optical response the  time-dependent Schr\"odinger equation is preferable.  On the other hand, density matrix and Green's functions allow to include correlation effects beyond the static screened exchange in the real-time dynamics, that it is more difficult within a wave-function formalism.
%
Moreover, from the evolution of $| v_{m\textbf{k}} \rangle$ in Eq.~\eqref{eq:td_wfs} also the time-dependent polarization~\cite{king1993theory} $\PP_\parallel$  along the lattice vector $\mathbf a$ can be computed in terms of the Berry phase:
\begin{equation}
\PP_\parallel = -\frac{ef |\mathbf a| }{2 \pi \Omega_c} \text{Im log} \prod_{\textbf{k}}^{N_{\textbf{k}}-1}\ \text{det} S\left(\textbf{k} , \textbf{k} + \mathbf q\right), \label{berryP} 
\end{equation}
where $S(\textbf{k} , \textbf{k} + \mathbf q) = \langle v_{n\textbf{k}}|v_{m\textbf{k} + \textbf{q}}\rangle$ is the overlap matrix between the valence states, $\Omega_c$ is the unit cell volume,  $f$ is the spin degeneracy, $N_{\textbf{k}}$ is the number of $\textbf{k}$ points along the polarization direction, and $\mathbf q = 2\pi/(N_{\textbf{k}} {\mathbf a})$.
The resulting polarization can be expanded in a power series of the field $\efield_j$ as:
\begin{equation}
\textbf{P}_i = \chi^{(1)}_{ij} \efield_j + \chi^{(2)}_{ijk}  \efield_j \efield_k +  \chi^{(3)}_{ijkl} \efield_j \efield_k \efield_l + O(\efield^4)  \; ,
\end{equation}
where the coefficients $\chi^{(i)}$ are functions of the frequency of the perturbing fields and of the outgoing polarization. From the Fourier analysis of the $\textbf{P}_i$ it is possible to extract all the non-linear coefficients (see Ref.~\cite{Attaccalite2013} for more details).
As in Sec.~\ref{sec:RT_RHO}, the level of approximation of the so-calculated susceptibilities depends on the effective Hamiltonian that appears in the right hand side of Eq.~\eqref{eq:td_wfs}.
Different choices are possible, namely, the independent particle approximation (IPA), the time-dependent Hartree, the real-time Bethe-Salpeter equation (RT-BSE) framework, or TD-DFT.
%
%****************************************************
\begin{comment}
Here we describe the latter, then removing the different terms from the Hamiltonian it is possible to recover or the other approximations.\cite{Attaccalite2013} In this framework the Hamiltonian $h^{MB}_{\mathbf k}$ reads:
%
\begin{equation}
  \label{eq:hmb}
  h^{MB}_{\mathbf k} \equiv h^{\text{KS}}_\textbf{k} [\rho^{eq}] + \Delta H_\textbf{k} + V_h(\mathbf r)[\Delta\rho] + \Sigma_{\text{SEX}}[\Delta \gamma] \;,
\end{equation}
%
where $H^{\text{KS}}_\textbf{k}$ is the Kohn-Sham Hamiltonian of the unperturbed (zero-field) \ks\ system, $\Delta H_\textbf{k}$ is the scissor operator that has been applied to the \ks\ eigenvalues, the term $V_h(\mathbf r)[\Delta\rho]$ is the time-dependent Hartree potential\cite{Attaccalite2013} and is responsible for the local-field effects originating from system inhomogeneities. The term $\Sigma_{\text{SEX}}$ is the screened-exchange self-energy that accounts for the electron-hole interaction,\cite{strinati1988application} and is given by the convolution between the screened interaction $W$ and $\Delta \gamma$. In the same equation:
$$\Delta \rho \equiv \rho(\mathbf r;t)-\rho(\mathbf r;t=0)$$ 
is the variation of the electronic density  and:
$$\Delta \gamma \equiv \gamma(\mathbf r,\mathbf r';t) - \gamma(\mathbf r,\mathbf r';t=0)$$
is the variation of the density matrix induced by the external field $\efield$.\\
\end{comment}
%****************************************************
%
This approach has been successfully applied to study second-, third- harmonic generation and two-photon absorption in bulk materials and nanostructures \cite{PhysRevB.95.125403,Attaccalite2013,PhysRevB.98.165126}.
As before, in the limit of small perturbation Eq.~\eqref{eq:td_wfs} reproduces the optical absorption calculated with the standard GW + BSE approach \cite{strinati1988application}.

Since the exact polarization is available, the approach based on Eq.~\eqref{eq:td_wfs} not only reduces to TD-DFT when local functionals are considered, it also includes TD density polarization functional theory (DPFT) as a special case. Thus specific approximations for both the microscopic and the macroscopic part of $\Delta\Sigma^{Hxc}$ are available, which, within TD--DPFT, are expressed as functionals of $\rho$ for the microscopic part ($v^{Hxc}[\rho]$) and of $\PP$ for the macroscopic part ($\efield^{Hxc}$) as discussed in Refs.~\cite{gruning2016dielectrics,gruning2016performance}.\\
A comparison between TD-DPFT in the real-time framework and the solution of the Bethe-Saleper equation for different zinc-blende compounds has been recently published by A. Riefer et al.~.\cite{WGschmith} While for linear response  the different functionals give a satisfactory result,\cite{gruning2016performance} for the second harmonic generation the situation is less clear. This is probably due to the fact all exchange-correlation kernels implemented in \yambo{} and tested in the previous papers were derived in the linear response regime.

In non-linear optics simulations the system is excited with a laser at given frequency $\omega$ and dephasing term $\lambda_{deph}$ is added to the Hamiltonian. After a time $T \gg \lambda_{deph}$, sufficient to damp out the eigenmodes of the system, the signal is analyzed to extract the non-linear response functions, see Fig.~\ref{fg:CdTeSHG} and Ref.~\cite{Attaccalite2013}.
To run simulations and compute the spectra as described in the present section the \yambonl\, and \yppnl\, executables need to be used.

%%%%%%%%%%%%%%%%%%%%%%%%%%%%
\section{Parallelism and performance \label{sec:parallel} }
%%%%%%%%%%%%%%%%%%%%%%%%%%%%

\begin{figure}
    \centering
    \includegraphics[clip,width=0.40\textwidth]{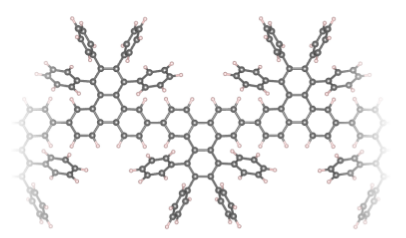}
    \includegraphics[clip,width=0.45\textwidth]{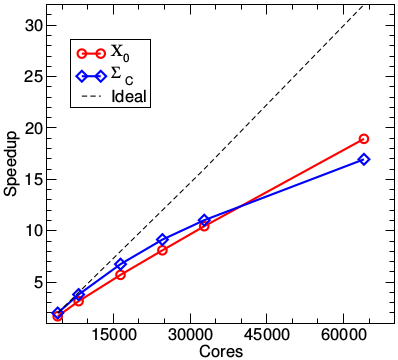}
    \caption{\yambo\ parallel performance. Upper panel: chemical structure of the precursor polymer of a chevron-like Graphene nanoribbon. Lower panel: \protect\yambo{} speedup of the linear response ($\chi_0$) and self-energy ($\Sigma_c$) kernels during a GW run. The scaling (obtained using 16 MPI tasks per node and 4 OpenMP threads/task) is shown up to 1000 Intel KNL nodes on Marconi at Cineca, A2 partition, corresponding to a computational partition of about 3 PetaFlops. The dashed line indicates the ideal scaling slope. (Adapted with permission from ~\cite{denk2017nanoscale}. Copyright (2017) by the Royal Society of Chemistry).
    \label{fig:scaling}}
\end{figure}

During the last years, the evolution of supercomputing technologies pushed towards the adoption of architectural solutions based on many-core platforms. This was due mainly to energetic constraints that did not permit to increase the single core performance, imposing
the need for alternative solutions. 
Two main paradigms arose: on one side, the emergence of hybrid architectures exploiting GPU accelerators. 
On the other side, homogeneous architectures increased the performance per node, by increasing the number of cores, starting the
many-core era. In the latter approach, the main advantage is the possibility to rely on well-known and largely adopted software paradigms,
in contrast to the GPU programming model, where the porting required to adopt ad-hoc languages such CUDA or OpenCL, having a deep impact on the sustainability of the software development. However, even if the many-core paradigm can appear easier to adopt, getting a satisfactory
performance on such architectures may be very challenging. In fact, in order to exploit as much as possible the features of a
many-core node, it is mandatory to use both a shared memory and a distributed memory approach. The first is able to leverage the single node
power with an efficient usage of the available memory, while the second one can be used to scale-up on the nodes available on a cluster facility or a multi-purpose processor. 

From \yambo{} version 4.0, a deep refactoring of the parallel structure has been put in place in order to take full advantage of nodes with many-cores and a limited amount of memory per core. In particular, a MPI multi-level (up to 3--5 according to the runlevel) approach has been adopted, together with an OpenMP coarse grain implementation. An example of the measured parallel performance reaching up to the use of 1000 Intel KNL nodes in a single run is shown in Fig.~\ref{fig:scaling}. We refer to the performance page~\cite{yambo_web_performance} on the \yambo{} website for a more complete description and up-to-date data. 

\begin{figure}
    \centering
    \includegraphics[clip,width=0.50\textwidth]{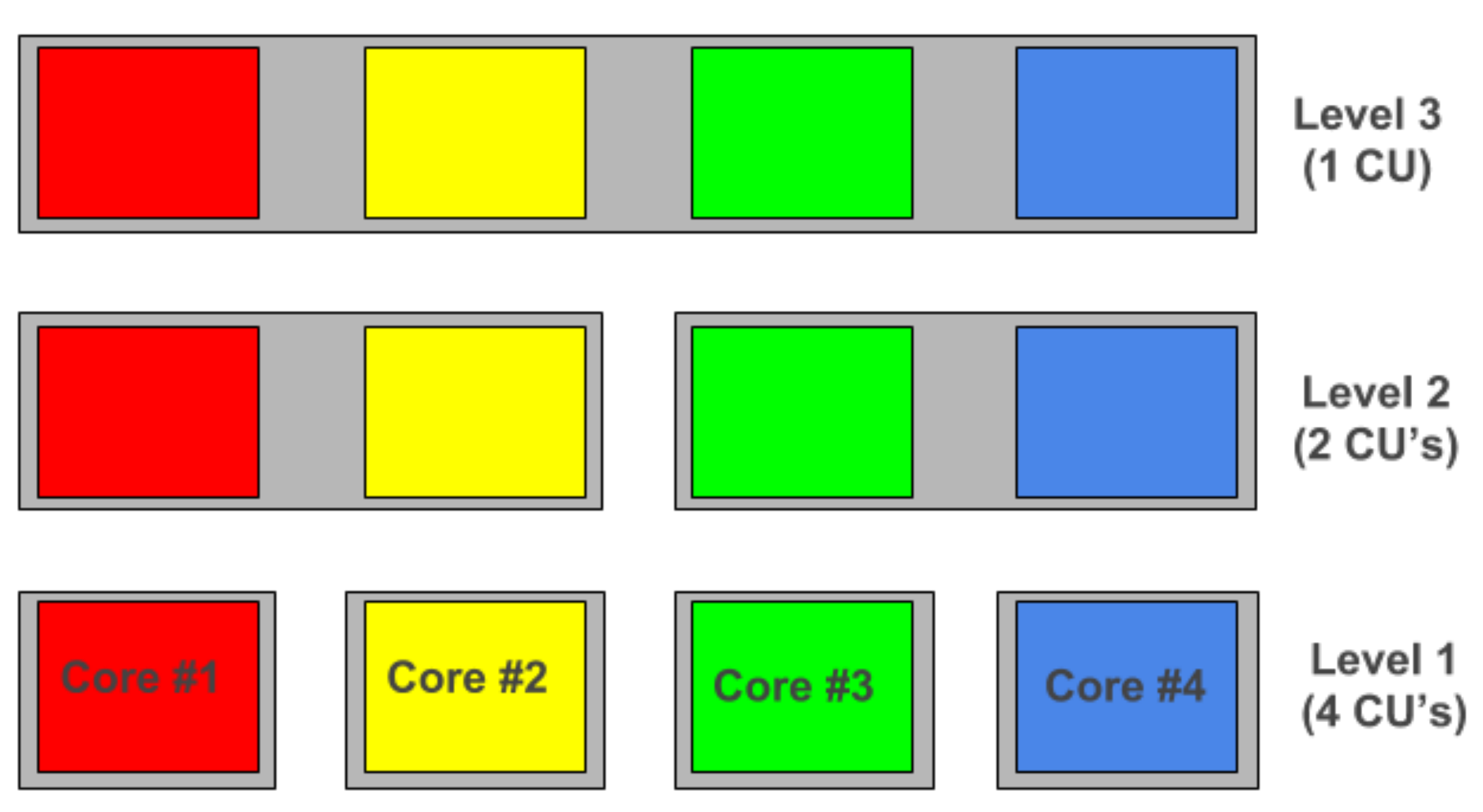}
    \caption{\yambo\ parallel structure in the specific case of 4 cores. Each core is member, at the same time, of three different groups composed of different number of cores. These groups, represented with gray boxes, are the actual computing units of \yambo{}. Therefore each core workload is dictated by the computing units directives and changes depending on the group the core belongs to.
    \label{fig:parallel_structure}}
\end{figure}

This novel multi level distribution of the cores is schematically shown in Fig.~\ref{fig:parallel_structure} in the case of 4 cores.  Instead of using the core as elemental parallel unit, \yambo{} adopts the concept of computing units\,(CU). A CU is composed of a varying number of cores. The work--load distribution is done among CU's rather than cores. Each core workload is decided by the workload of the CU that encloses it. To be more clear let's take the simple case of 4 cores shown in 
Fig.~\ref{fig:parallel_structure}. In this case we have three possible levels of grouping with respectively 4, 2 and 1 CU's. The core workload is assigned to the CU's rather that to the single core. This reduced enormously the inter--core communications and allows the distribution of a very large number of cores. Technical details of the implemented parallelism will be discussed in the next sections.

Finally we also mention that work is in progress to port yambo on GPUs, using CUDA Fortran as a first step. We are currently porting few low-level routines on GPUs taking advantage of CUDA libraries: notably the ones computing the matrix elements of Eq.\eqref{eq:rhotw}. This allows to have a preliminary porting on GPUs of dipoles, Hartree-Fock, linear response, GW, and BSE kernels. Based on the results obtained we will decide subsequent strategies.

\subsection{General structure}

The multilevel MPI structure of \yambo{} is reflected in the input file where, for each computational kernel ({{\small\ttfamily runlevel}} here) there are two related input variables: the first one, {{\small\ttfamily runlevel\_ROLEs}}, sets on which parameters the user wants to distribute the MPI workload, while the second {{\small\ttfamily runlevel\_CPU}} defines how many MPI tasks will be associated to such parameters. As an example 

\vspace{3pt}
{{\small\ttfamily X\_finite\_q\_ROLEs="q.k.c.v.g"}} 

{{\small\ttfamily X\_finite\_q\_CPU="2.3.5.2.1"}}
\vspace{3pt}

\noindent
is a possible input %file 
for running on $2\times 3\times 5\times 2 \times 1= 60$ MPI tasks, with the $\qq$--points distributed on 2 tasks, the $\kk$--points on 3, the conduction and valence bands on 5 and 2 respectively. One more level of parallelism ($g$) is present, acting and distributing the response matrix over space degrees of freedom (plane waves). The order of the parameters in the {{\small\ttfamily runlevel\_ROLEs}} variables is irrelevant.
On top of that, more input variables are available to handle parallel linear algebra (e.g via scalapack and blacs libs) and to select the number of {\tt OpenMP} threads (on a runlevel basis if needed).
Such hierarchical organization makes it possible to have MPI communication only within the subgroups,
thus avoiding, whenever possible, to deal with {\tt all2all} communications.

If the user does not wish to deal with the complexity of such multi-level parallelization a default layout is provided. However, the fine-tuning of the MPI/OpenMP related variables can (further) reduce the load imbalance, improve the memory distribution or decrease the total time-to-solution. For this reason, in the Sec.~\ref{subsec:PAR_LinearResp}--\ref{subsec:PAR_LinearAlgebra} specific suggestions for best parallel exploitation of each runlevel are provided.

\subsection{I/O: parallel and serial}  \label{subsec:PAR_IO}
\yambo{} stores binary data using the {\tt netcdf} library. Depending on the configuration flags, data can be stored in classic {\tt netcdf} format (file size limit of 2 GB, activated with {{\small\ttfamily --enable-netcdf-classic}}), 64-bit {\tt netcdf} format (no file size limit, default) or HDF5 format (requires at least {\tt netcdf} v4 linked to HDF5, activated with {{\small\ttfamily --enable-netcdf-hdf5}}). Since version $4.4$, in case the HDF5 format is specified, parallel I/O can also be activated ({{\small\ttfamily --enable-hdf5-par-io}}) to store the response function in $\GG$ space or the kernel of the BSE. For the $\GG$--space response function, parallel I/O avoids extra communication among the MPI tasks and also reduces the amount of memory allocated per task. For the BSE case, parallel I/O makes it possible to load the kernel computed from a previous calculation using a different parallel scheme and/or a different number of MPI tasks. Indeed the calculation of the kernel matrix elements is very time consuming but has a very efficient memory and load distribution. In contrast, the solution of the BSE eigenproblem is less time consuming but also less efficiently distributed. It is thus suggested to first compute the kernel matrix on a large number of cores and then to solve the BSE on fewer tasks as a follow-up step. 

\subsection{Linear Response} \label{subsec:PAR_LinearResp}
According to Eq.~\eqref{eq:X_IP}, the computation of the response function in $G$-space can be distributed over 5 different levels: $\qq$--points, $\kk$--points, conduction and valence bands ($c,v$), and $\GG$-vectors. The distribution over the $\qq$--points would be the most natural choice, since the response functions at different $\qq$ are completely independent. However, it may lead to significant memory duplication (multiple sets of wavefunctions are managed at the same time) and load imbalance since the number of possible transitions varies from point to point. It is usually not recommended unless a large number of $\qq$--points has to be considered. Instead, it is usually more effective to parallelize over $\kk$, and bands ($c$,$v$) indexes.  This requires slightly more MPI communication (due to a MPI reduction at the end of the calculation) but is very efficient in terms of speedup (almost linear) and in terms of memory distribution (especially for $c,v$, since usually wavefunctions are the leading memory contribution). While transitions are evenly distributed (balanced workload), \yambo{} sorts and groups transitions which are (almost) degenerate in energy (see App.~\ref{App:Resp-function-sum}) thus, in practice, small imbalances can still be present. 

Further, the distribution over the $\GG$-vectors is the most internal one, and requires more communication among MPI tasks (unless parallel I/O is activated). It can be useful for systems with a very large number of $\GG$-vectors (such as low dimensional systems or surfaces) to distribute the response function and ease memory usage. Finally, the computation of $\chi^0$ can also benefit from {\tt OpenMP} parallelism. 
The distribution over threads has been implemented at the same level of the MPI parallelism (i.e. over transitions), resulting in a very good scaling while reducing the memory usage per core. Note however that some memory duplication (a $M(\GG,\GG')$ workspace matrix per thread) has to be paid to make the implementation more efficient. The {\tt OpenMP} parallelism of $\chi^0$ (including dipoles) is governed by the input variables:

\vspace{3pt}
{{\small\ttfamily X\_Threads=8}} 

{{\small\ttfamily DIP\_Threads=8}}
\vspace{3pt}

\noindent
both defaults being set to 0, i.e. controlled as usual by the {\tt OMP\_NUM\_THREADS} environment variable.
Once the independent-particle response function ${\chi^0_{\GG\GG'}(\qq,\omega)}$ has been computed, a Dyson equation is solved for each frequency to construct the RPA response function. This can be done either by distributing over different frequencies or by using parallel linear algebra (see Sec.~\ref{subsec:PAR_LinearAlgebra}).

\subsection{Self-Energy: HF-Exchange and GW}
%========================
Following Eq.~\eqref{eq:sigma_c}, the HF and GW correlation self-energies can be parallelized with MPI over three different layers: $\qq$-points (\verb+q+); bands in the Green's function (\verb+b+) [see $m$ in Eq.~\eqref{fig:hBN_TD-SEX}]; and quasiparticle corrections $\Sigma_{n n' \kk}$ to be computed (\verb+qp+). 
{\tt OpenMP} parallelism here acts at the lowest level, dealing with sums over $\GG$ and $\GG'$, i.e. spatial degrees of freedom.
The following variables can be modified to fine-tune the self-energy parallelization (here shown for 60 MPI tasks and 8 OpenMP threads): 

\vspace{3pt}
{{\small\ttfamily SE\_ROLEs="q.qp.b"}}

{{\small\ttfamily SE\_CPU="1.4.15"}}

{{\small\ttfamily SE\_Threads=8}} 
\vspace{3pt}

Since the sum over $\qq$-points in Eq.~\eqref{eq:sigma_c} is over the whole BZ, the $\qq$-parallelism for the self-energy may be even more unbalanced than that for the response function (here every $\qq$-point needs to be expanded to account by symmetry for its whole star) and is recommended only when a large number of $\qq$-points is available. Instead, the parallelism over bands $b$ tends to distribute evenly both memory and computation, at the price of a mild MPI communication, thereby resulting a natural choice (when enough bands are included in the calculation). $qp$-parallelism distributes the computation but tends to replicate memory (wavefunctions are not further distributed).
In general, the {\tt OpenMP} parallelism is extremely efficient for the GW self-energy without having to pay for any extra memory workspace.

\subsection{Bethe Salpeter Equation}
%========================

In the solution of the BSE most of the CPU time is spent in building up the excitonic matrix, or more precisely, its kernel. The input flags which control the parallel distribution of the workload needed to build the kernel are {\small\ttfamily eh.k.t}. To distribute the workload, first all possible transitions $cv\kk$, i.e. from valence band $v$ to conduction band $c$ at the k--point $\kk$, are split into transition groups (TGs). Then for each pair of TGs a block of the BSE matrix is created $B_{ij}=\{T_i\rightarrow T_j\}$. Defined $N_t$ the total number of TG, then the BSE matrix will be divided into $N_b=N_t^2$ blocks. In the Hermitian case (as in the Tamm-Dancoff approximation), only $N_b=N_t(N_t+1)/2$ blocks will be computed.
The parallelization flags for the BSE define both $N_t$ and $N_b$, and how the resulting blocks are distributed among the MPI tasks.
Indeed $N_t=n_{eh}n_k^{ibz}$ where $n_{eh}$ is the number of MPI tasks assigned to the {\small\ttfamily eh} field and $n_k^{ibz}$ is the number of $\kk$--points in the IBZ. 
This means that even setting {\small\ttfamily eh}=1 a minimum number of $\kk$--based TGs (k-TGs) is always created, which is eventually split into subgroups when $n_{eh}>1$.
 It is important to note that 
 k-TGs are defined using the k-sampling in the the IBZ, while the BSE matrix is defined in the whole BZ, resulting in groups of non-uniform size.  However, the symmetry operations relating matrix elements within a given k-TGs are taken into account by \yambo.
As a consequence, in systems where $n_k^{ibz}\neq n_k^{bz}$ the use of $n_{eh}>1$ is discouraged, as the splitting of k-TGs over different MPI-tasks implies that symmetry-related matrix-elements can be assigned to different MPI-tasks and need to be recomputed. 
Once $N_t$ and hence $N_b$ are defined, transitions and blocks are distributed among the MPI tasks as explained in the following example. 

Suppose we have a system with $18$ $\kk$--points in the IBZ, and we adopt the parallelization strategy $2.3.3$ for {\small\ttfamily eh.k.t} in the case the BSE is Hermitian. Then $N_t=2\times 18=36$ and $N_b=666$.
Thus, in our example we are using in total $2\times 3\times 3$ MPI-tasks. The {\small\ttfamily eh.k} fields define $2\times 3=6$ MPI-groups which split the 36 transition-groups. Thus, each MPI-group has to deal with 6 transition-groups.
For each transition group $T_n$, there are $N_t$ blocks $B^n_{ij}=T_i\rightarrow T_j$ for the Hermitian case, where the $T_n$ appears as initial ($T_i=T_n$) or final ($T_j=T_n$) state. Most of the blocks belongs to two transition-groups (only the blocks $B_{ii}$ belong to one transition-group).
This means that each MPI-group builds half of the $B_{ij}$ ($6*35/2$) plus all $B_{ii}$ (6) blocks. These $111$ blocks are divided according to the {\small\ttfamily t} field and thus each MPI-task will be assigned to $37$ blocks.

\subsection{Linear Algebra} \label{subsec:PAR_LinearAlgebra}
%========================
Dense linear algebra is extensively used in \yambo{}. Among the most time-consuming tasks we have identified the inclusion of local field effects~\cite{onida2002} in the RPA response function 
\begin{multline}
    \chi_{{\bf{G}}{\bf{G}'}}^{RPA}(\qq,\omega)= \chi^{0}_{ {\bf{G}}{\bf{G}'}} (\qq,\omega) + \\\chi^{0}_{ {\bf{G}}{\overline{\bf{G}}}}(\qq,\omega) \frac{4\pi}{|\qq+\overline{\bf{G}}|^2}
     \chi_{\overline{\bf{G}}{\bf{G}'}}^{RPA}(\qq,\omega).
    \label{eq:X_rpa}
\end{multline}
The solution of Eq.~\eqref{eq:X_rpa} can be cast in the form of a matrix inversion. Indeed:
\begin{multline}
    \chi_{{\bf{G}}{\bf{G}'}}^{RPA}(\qq,\omega)=
        \left[ \delta_{\GG,\overline{\GG}} - \chi^{0}_{ {\bf{G}}{\overline{\bf{G}}}}(\qq,\omega) \frac{4\pi}{|\qq+\overline{\bf{G}}|^2}\right]^{-1}\\
        \chi_{\overline{\bf{G}}{\bf{G}'}}^{0}(\qq,\omega).
        \label{eq:X_rpa_inv}
\end{multline}
Eq.~\eqref{eq:X_rpa_inv}, and the solution of the BSE (diagonalization), which can be considered prototype kernels.

Once a finite basis set is adopted, the operators involved are represented as (dense) $N\times N$ matrices, with $N$ easily reaching few-to-tens of thousands or more, making standard linear algebra tasks (such as matrix multiplication, inversion, diagonalization) quite intense.
We have therefore implemented dense parallel linear algebra by exploiting the {\tt ScaLAPACK} library~\cite{scalapack} within the MPI parallel structure of \yambo.
Concerning the RPA response, this means that on top of the MPI parallelism over $\mathbf{q}$-vectors, multiple instances of parallel linear algebra are run at the same time (one per $\mathbf{q}$ vector) to compute $\chi^{\text{RPA}}$.

The behavior of the \yambo{} parallel linear algebra is governed by the variables:
\begin{verbatim}
  runlevel_nCPU_LinAlg_INV=64
  runlevel_nCPU_LinAlg_DIAGO=64
\end{verbatim}

where ({\tt runlevel} could be, for example, the RPA response function or the BSE).
Given the relevance, the calculation of the IP response function ${\chi^0_{\GG\GG'}(\qq,\omega)}$ has also been block-distributed over $\GG,\GG'$ vectors ($g$-parallelism in Sec.~\ref{subsec:PAR_LinearResp}), both in terms of computation and memory-usage.

When using the SLEPC diagonalization method to obtain the BSE spectra, the memory distribution of the eigensolver (not to be confused with the memory distribution of the BSE matrix discussed in Sec.~\ref{sec:BSE_numerical_slepc}) is handled by the SLEPC library itself.
For more details, the reader is referred to the SLEPC specific literature~\cite{hernandez_slepc}.

%%%%%%%%%%%%%%%%%%%%%%%%%%%%
\section{Scripting and Automation \label{sec:scripting} }
%%%%%%%%%%%%%%%%%%%%%%%%%%%%

\begin{figure}
    \centering
    \includegraphics[width=0.50\textwidth]{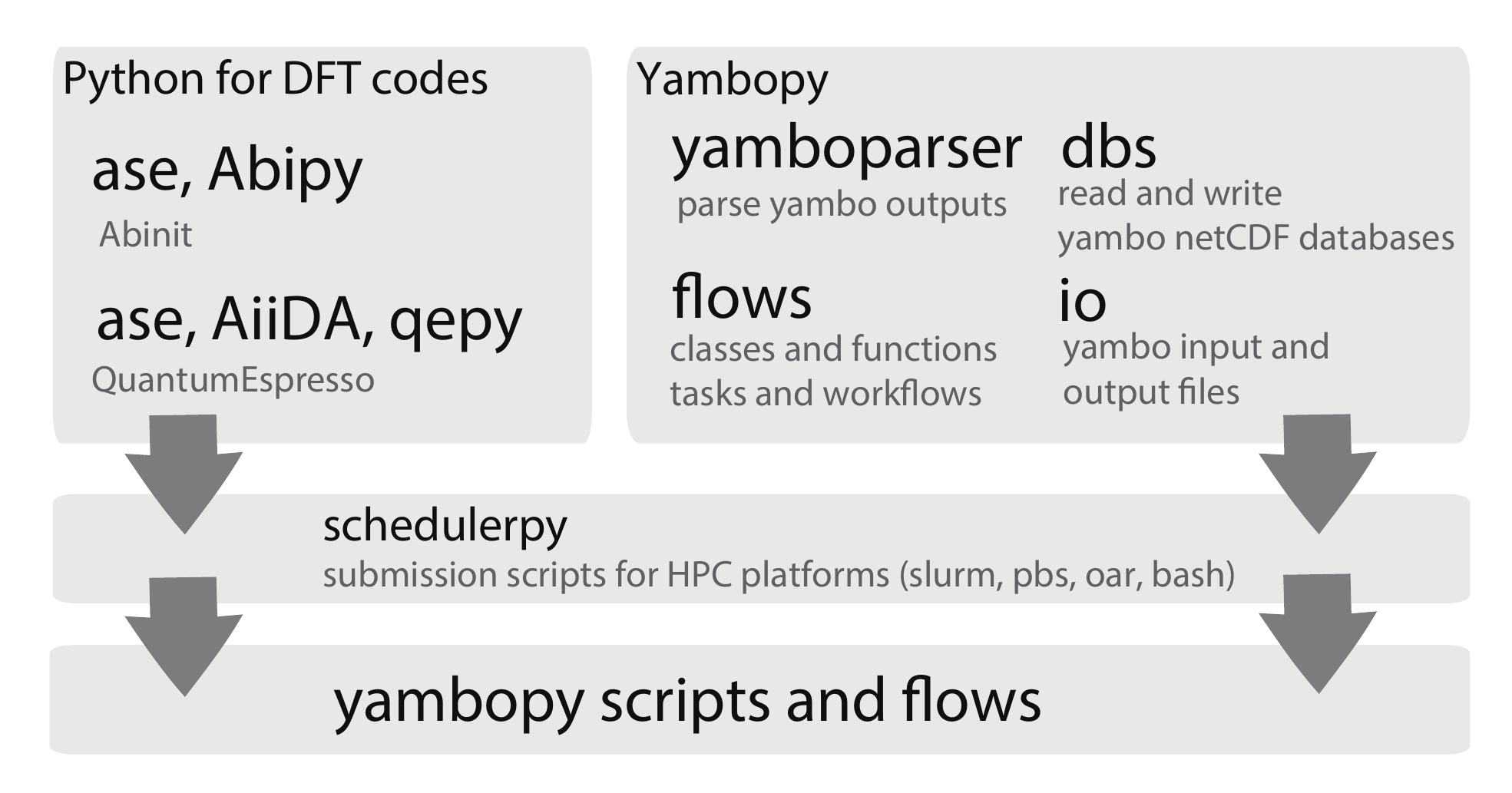}
    \caption{Structure of the \yambopy\, package. The {\tt qepy} and {\tt schedulerpy} packages are distributed as part of \yambopy.
    \label{fig:yambopy}}
\end{figure}

As a pure many-body code, \yambo{} works as a sort of ``quantum engine'' that takes as input DFT calculations and material-specific parameters, producing very large amounts of temporary data (e.g. the response function) and outputs numerical results. Even a single calculation can produce enormous amounts of data. It is therefore necessary to carefully select or extract the relevant information to be stored for future analysis or sharing, possibly ensuring reproducibility. 

In addition, the final quantities of interest (e.g. GW band gaps or BSE spectra) are often the results of a complex and tedious sequence of operations, involving transferring data from different codes (e.g. from \qe\ to \yambo) or repeating calculations with different parameters (e.g. for convergence tests). The benefit of having platforms to organize, simplify and accelerate many-body perturbation theory calculations is obvious.
Two parallel efforts are being developed to facilitate the use of \yambo, namely \yambopy\, and the \yambo~interface with the \aiida{} platform \cite{Pizzi2016}. 
In Fig.~\ref{fig:yambopy} a schematic representation of how such platforms interact with \yambo is shown.

%========================
\subsection{\Yambopy\, \label{subsec:yambopy} }
%========================

\Yambopy\, is a community project to develop Python classes and scripts to express, automate, and analyze calculations with the \yambo{} code.
A typical \yambo{} workflow involves a few steps: generating the KS states with a DFT code, preparing the \yambo{} databases and then running the \yambo{} code.
These workflows become more complicated when performing convergence tests or when repeating the same calculations for multiple materials.
With \yambopy\, the user can express these \yambo{} workflows on a python script that can then be shared and reproduced among different users.
We currently provide python classes to read and write the input files as well as the output of the \yambo{} code.
A lightweight python interface for the \qe\ Suite is also provided in the {\tt qepy} package distributed along with \yambopy.
For more comprehensive python interfaces for the DFT codes see ASE~\cite{Larsen2017}, Abipy~\cite{abipy_web} for the Abinit code, or \aiida{}~\cite{Pizzi2016}.

The {\tt YamboIn} class provided by \yambopy\, is used to read and write the base \yambo{} input file generated by \yambo{} and modify it in a programmatic way.
With this tool it is possible, for example, to create a set of input files by changing a single variable inside a {\tt for} loop.
We also provide classes to read the \yambo{} NetCDF databases in Python (for a complete list see the on-line documentation~\cite{yambopy}).
These classes provide methods to manipulate and represent the data using the matplotlib~\cite{Hunter2007} library
giving great flexibility for the interpretation and analysis of the results.
Running these workflows on a HPC context requires to write job submission scripts for different job schedulers, this can be done using the
{\tt schedulerpy} package also accompanying \yambopy.

For quick access to some features from the command line, we provide the \yambopy\, shell command.
The script is automatically installed with \yambopy\, in such a way that some functionalities of \yambopy\, are directly callable from the command line.
This script has features to plot the convergence tests for GW and BSE calculations, excitonic wave functions, and dielectric functions among others.

To ensure software quality and usability we provide \yambopy{} as an open-source code along with documentation and automatic testing.
A detailed documentation of the classes, features, and a tutorial are available on the \yambopy\, website\cite{yambopy}.
We keep a public git repository hosted on {\tt Github} where the users can download the latest version of the code as well as contribute with patches, features, and workflows.
Sharing the workflows among users allow us to avoid repeated technical work and greatly simplifies the use of the \yambo{} code.
Continuous integration tests are done using the Travis-CI platform~\cite{travisci} leading the code to be tested at each commit and thus enforcing its reliability.
\yambopy\, is a project under development, and will be described on its own in a future work.

%========================
\subsection{\yambo{} within the \aiida{} platform\label{subsec:aiida}}
%========================

The \aiida{} platform is a materials' informatics infrastructure which implements the so-called ADES model (Automation, Data, Environment and Sharing) for computational science~\cite{Pizzi2016}. The \aiida{} plugins and workflows for \yambo\, are publicly available on {\tt GitHub}~\cite{github_aiida-yambo}, while online documentation and tutorials are available on Read the Docs~\cite{readthedocs_aiida-yambo}.

\subsubsection{The \yambo{}-\aiida{} plugin}

Input parameters and scheduler settings are stored as code-agnostic \aiida{} data-types in a database, then converted by the \yambo-\aiida{} plugin into \yambo{} input files and transferred to a computational unit (e.g a remote workstation or an HPC cluster). The \aiida{} daemon submits, monitors and eventually retrieves the output files of the \yambo{} calculation, the relevant information is then parsed and stored by the plugin.  While the relevant data is properly stored in a suitable database, the raw input and output files are also stored locally in a repository. Therefore inputs, calculations and outputs are all stored as nodes of a database connected by directional links, preserving the full data provenance and ensuring reproducibility. 

The \yambo-\aiida{} plugin currently supports calculations of quasi-particle corrections (e.g. at the COHSEX or GW level) and optical properties (e.g. IP-RPA). \qe{}, one of the main DFT codes interfaced with \yambo, is also strongly supported with specific plugins and workflows for \aiida{}~\cite{github_aiida-espresso}. Some of the parsing functionalities of the plugin are powered by the \yambopy\, package \cite{yambopy}. Different types of calculations can be performed, either starting from \qe{} or from \py{} or from a previous (possibly unfinished) \yambo{} run.

\subsubsection{\aiida{} workflows: automated GW}

The \yambo-\aiida{} package provides automated workflows that capture the knowledge of an experienced user in performing e.g. GW calculations within the plasmon-pole approximation, accepting minimal inputs such as a DFT calculation or a crystal structure, and returning as outputs a set of quasi-particle corrections. The \yambo-\aiida{} plugin repository hosts four \aiida{} workflows \cite{Pizzi2016} of increasing complexity and abstraction: \texttt{YamboRestart}, \texttt{YamboWf}, \texttt{YamboConvergence} and
\texttt{YamboFullConvergence}, that perform different but mutually interdependent tasks, with the latter depending on the former in the listed order.

\texttt{YamboRestart} is a low level \aiida{} workflow that takes a DFT calculation (or a prior \yambo\, calculation) as input, performs GW or BSE calculations, and returns the results. \texttt{YamboRestart} interacts directly with the \yambo\, plugin, coping with common failures that may occur during a \yambo{} GW run such as insufficient maximum wall-time and out-of-memory issues: the workflow adjusts the scheduler options as well as the parallelization choices accordingly and resubmits the calculations.

\texttt{YamboWf} is a higher-level \aiida{} workflow that uses \texttt{YamboRestart} and the \qe-\aiida{} plugin to manage end-to-end a GW calculation from the DFT step to the completion of the \yambo{} run. In contrast to \texttt{YamboRestart}, which starts from an already existing calculation (either DFT or \yambo{}), \texttt{YamboWf} does not need to start from any calculation and performs all steps, including all necessary DFT, data interfaces, and \yambo\, calculations.  

\texttt{YamboConvergence} is built on top of \texttt{YamboWf} and automates the convergence of QP corrections (by focusing on the quasiparticle gap) with respect to a single parameter. A one-dimensional line search in the parameter space is used. The convergence is determined by comparing a series of the most recent calculations (four of them are used by default), and ensuring the change between all four successive calculations is less than the convergence tolerance. The deviation from convergence is estimated by fitting the gap to a function of the form $f(x)=c + \frac{a}{x+b}$.  

\texttt{YamboFullConvergence} iterates the above procedure over the main variables governing the convergence of GW calculations, namely the $\mathbf{k}$-point grid, the number of $\GG$-vectors used to represent the response function ($\chi_0$ cutoff), and the number of bands included in the sum-over-states for both the polarizability and the correlation self-energy. Additionally, the possibility to further reduce the FFT grids with respect to the one used at the DFT level is also considered.
A beta version of this workflow has been made available on {\tt GitHub} for testing and fine-tuning of the algorithm.

%========================
\subsection{Test-suite and benchmark-suite \label{subsec:test-suite} }
%========================

A new important tool introduced to improve and stabilize the development of the \yambo{} code is the test-suite. The \yambo{} test-suite is stored in a dedicated repository (yambo-tests) on {\tt GitHub} and contains a series of tests which can be run in an automated manner. The repository is freely accessible after registering as a ``\yambo\, user" on the {\tt GitHub}. While the test-suite is mainly aimed at developers, users can also benefit from accessing its input and reference files and automatically checking if their compiled version of \yambo{} works properly. 

The test-suite is governed by using a Perl script, {\tt driver.pl}. This script uses internal Perl modules to perform several tasks: it automatically compiles the \yambo{} code (a precompiled version can also be used), it runs the code and checks the output against reference files stored in the test-suite repository. 

The code can be run in serial, parallel with {\tt OpenMP} threads and checking parallel I/O and/or parallel linear algebra. At least two different groups of tests are available: smaller (and faster) tests which are run on a daily basis and longer tests which are used for a deeper testing of the code before a release.

The same driver can also be used to run \yambo{} benchmarks. Benchmarks tests are a particular group of materials that, describing complex nano--structures (a 1D polymer or carbon--based ribbon) or a water cell, require a large number of reciprocal space vectors and/or $\kk$--points. As a consequence these systems are suitable to be executed using a large number of cores on parallel machines. In this case the test-suite can collect the results and loop on different parallel configurations testing their performances. More importantly the test-suite organizes the results in machine dependent folders that can be, eventually, post-analyzed.  

The results of the night runs of the test-suite are publicly available on the web-page~\cite{yambo_test-suite} and can be inspected without having access to the machines that run the tests. This is very useful in order for any development to reproduce a specific error to be fixes.
%
%%%%%%%%%%%%%%%%%%%%%%%%%%%%%
\section{Conclusions and Perspectives\label{sec:conclusions} }
%%%%%%%%%%%%%%%%%%%%%%%%%%%%%
%%

This paper describes the main development lines of the \yambo{} project since the 2009 reference paper~\cite{Marini2009}. 
\Yambo{} is a scientific code supported and continuously developed by a collaborative team of researchers. The long list of authors of this work 
attests to the involvement of
numerous experienced and young developers in addition to the four founders~\cite{Marini2009}.

The \yambo{} team currently comprises a balance of renowned scientists, with long-standing experience in ab-initio approaches, and young researchers. 
We welcome students and post-docs with new ideas. This combination makes possible the growth of a software suite which is formally rigorous and able to address topics at the frontiers of materials science.
By exploiting the power of many body perturbation theory at equilibrium and out-of-equilibrium within a state-of-the-art ab initio framework, the code is able to make predictions of the electronic and optical properties of novel materials, and moreover to provide interpretation of cutting-edge experiments ranging from ultrafast electron dynamics to nonlinear optics.

The involvement of parallel computing experts (two members of the Italian National Supercomputing Center CINECA co-authored this paper, for example) ensures that the code is also efficient and portable to the latest supercomputing architectures. As a result, new features added to the code immediately benefit from the native parallelized environment.

The modular structure of the code and the interface to external supporting software (\aiida\, and \Yambopy) complete the picture providing the end-user with a wealth of tools that cover the actual preparation, calculation and post-processing of data.  The \yambo{} suite thus provides all the ingredients for an advanced and computationally powerful approach to theoretical and computational material science. 

Indeed, despite being born as a code for MBPT, thus tailored for sophisticated calculations on simple materials, \yambo{} can nowadays be used to study complex materials and interfaces as well. This means in practice that, while the first versions of the code were designed to run on unit cells containing very few atoms (like  bulk silicon), nowadays \yambo{} can be easily used to study unit cells with 10 to 20 atoms and can be pushed on HPC centers up to hundreds of atoms~\cite{Attaccalite2011hBN,Amato2016,Giorgi2018}. The number of atoms which can be dealt is, thus, approximately one order of magnitude less than advanced DFT codes. The exact limit is mainly imposed by the power of HPC facilities. We are also working on a dedicated section on the \yambo{} web-site with detailed information on time scaling of different runlevels across the releases of the code.

What lies in \yambo{}'s future? We expect that the future development of \yambo{} will be driven by the need to interpret new experiments. This will be achieved through the implementation of advanced computational algorithms and physical methodologies and will increasingly exploit interoperability with other software.
Projects under current development include extension of GW to start from hybrid functionals, the possibility to use ultrasoft pseudopotentials, alternative schemes to avoid empty states, BSE at finite $q$, and incorporation of exciton-phonon coupling, to name just a few. These new developments will become available to general users in the near future.
The code's efficiency will be continuously improved in order to tackle problems that remain computationally cumbersome. We expect that \yambo{} will be further restructured in order to adapt to heterogeneous architectures (GPUs and accelerators) and to fully exploit the computational power of future pre- and ``exascale'' machines.
Further developments are (and hopefully will be) also driven by the participation in European initiatives and projects. At present \yambo{} is part of a user-based European infrastructure~\cite{nffa} and a member of the suite of codes selected for the exascale transition~\cite{max}. 

In conclusion, \yambo{} is a lively community project characterized by a continuous technical and methodological development. The substantial development between the 2009 reference paper~\cite{Marini2009} and today demonstrates its enormous potential. The aim is to provide the scientific community with a tool to perform cutting edge simulations in a computationally efficient environment.

%%%%%%%%%%%%%%%%%%%%%%%%%%%%
\section{Acknowledgments}
%%%%%%%%%%%%%%%%%%%%%%%%%%%%
%

The \yambo\, team acknowledges financial support from a number of sources, notably including H2020 funding, as follows. 
This work was in part supported by the project MaX -- MAterials at the eXascale -- by the European Union H2020-EINFRA-2015-1 and H2020-INFRAEDI-2018-1 programs (Grant No. 676598 and No. 824143, respectively).
This work was also supported by the European Union H2020-INFRAIA-2014-2015 initiative (Grant No. 654360, project NFFA -- Nanoscience Foundries and Fine Analysis). We also would like to acknowledge support from cost action CA17126. We thank CECAM and Psi-K network for financial and practical support related to the organization of \yambo{} training and tutorial events.

\yambo{} developers have benefit from direct interactions with personnel from INTEL (Hans Pabst) and NVIDIA (Massimiliano Fatica, Everett Phillips, Josh Romero), especially concerning parallel performance and porting.
We also acknowledge {\tt GitHub} for providing free hosting to the \yambo{} project, including several aspects ranging from inner-core developments, GPL releases, and testing.

We acknowledge PRACE for awarding us access to computing resources on Fermi and Marconi machines at CINECA, Italy and on Piz Daint at CSCS, Switzerland. We also acknowledge the access to computational resources obtained by national programs, such as ISCRA by Italian MIUR.

The \yambo\ developers thanks the \abinit\ team, and in particular Matteo Giantomassi and Xavier Gonze, for: (i) first providing, together with the \abinit\ source of versions 6 and 7, a patch for supporting multi-channel projectors and (ii) later for coding the new structure of the WFK files in NETCDF format in version 8, which allowed the development of the latest a2y interface.

%%%%%%%%%%%%%%%%%%%%%%%%%%%%
\appendix

\section{Glossary}
\begin{tabular}{p{2cm}|p{6.6cm}}
BSE & Bethe--Salpeter equation \\
CBM & Conduction bands minimum \\
DFT & Density Functional Theory \\
DFPT & Density functional perturbation theory \\
EOM & equation of motion\\
EP & election-phonon \\
GGA    & Generalized Gradient Approximation \\
GW & Green's function (G) / \\
   & Screened Coulomb interaction (W)\\
HAC & Heine--Allen--Cardona\\
HDF & Hierarchical Data Format\\
HPC    & High Performance Computing \\
HF & Hartree-Fock \\
IPA    & Independent Particles Approximation \\
KB & Kleinman--Bylander \\
KS & Kohn--Sham \\
LDA    & Localized Density Approximation \\
MBPT & Many-body perturbation theory \\
MPI  & Message Passing interface  \\
netCDF  & Network Common Data Form\\
OMS & on-mass-shell \\
OpenMP & Open Multi-Processing\\
PPA & Plasmon--Pole approximation \\
PETSc  & Portable, Extensible Toolkit \\
       & for Scientific Computation\\
QP & Quasiparticle \\
SF & spectral function \\
COH    & COulomb Hole \\
SEX   & Screened EXchange \\
SLEPc    & Scalable Library for \\
         & Eigenvalue Problem Computations \\
UPF & Unified Pseudopotential Format\\
VBM & Valence bands maximum \\
XC & exchange--correlation
     
\end{tabular}

%============================
\section{Evaluation of the response function} \label{App:Resp-function-sum}
%============================

To compute the response function in $\GG$ space in an efficient way, Eq.~\eqref{eq:X_IP} is evaluated by splitting the sum in an internal frequency independent term running over all transitions and an external frequency dependent term running over groups of transitions as follows: 
\begin{align}
 \chi_{ {\bf{G}}{\bf{G}'}}^0 (\qq,\omega)=
\sum_{\nt\mt\kkt} F_{\nt\mt\kkt}(\omega,\qq) \sum_{n'm'\kk' \in D_{\nt\mt\kkt}(\qq)} R^{n'm'\kk'}_{\GG \GG'}(\qq),
\label{eq:X_IP_split}
\end{align}
where
\begin{eqnarray}
F_{nm\kk}(\omega,\qq)&=&
    \Big[ \frac{1}{\omega -(\epsilon_{m\kk} - \epsilon_{n\kk-\qq}) -i\eta} \nonumber \\
\label{eq:X_IP_split1}
&&\phantom{C_{nm}}
           -\frac{1}{\omega -(\epsilon_{n\kk-\qq} -\epsilon_{m\kk}) +i\eta}  \Big] \\
R^{nm\kk}_{\GG \GG'}(\qq)&=&\frac{f_s}{N_{\bf{k}}\Omega} \,
         f_{m\kk}(1-f_{n\kk-\qq}) \times \nonumber \\
\label{eq:X_IP_split2}
&&\phantom{\frac{f_s}{\Omega N_{nm\kk}} }
         \rho_{nm\kk}(\qq, \GG)  \rho^{\star}_{mn\kk} (\qq,\GG').
\end{eqnarray}
The internal sum runs over degenerate poles $\{n'm'\kk' \in D_{\nt\mt\kkt}(\qq)\}$ while the external sum runs over only one member of the degenerate group.
Poles are set to be degenerate if
\begin{eqnarray}
(\epsilon_{n\kk-\qq} -\epsilon_{m\kk})-(\epsilon_{n'\kk'-\qq} -\epsilon_{m'\kk'}) < \epsilon_{thresh}.
\end{eqnarray}
The degeneracy threshold is controlled via the input variable \\
{{\small\ttfamily CGrdSpXd= 100.           \# [Xd] [o/o] Coarse grid controller}}  \\
The default $100.$ means the degeneracy threshold is ${\epsilon_{thresh}=10^{-5}}$ Hartree. Reducing the value of the input variable the threshold is increased. Only in case the input value is set to zero the size of the groups is set to 1 and the external sum runs over all transitions.

%============================
\section{Sum-over-states terminators}
\label{App:terminators}
%============================
For the sake of completeness, here we report the sum-over-states terminator expressions introduced in Ref.~\cite{bruneval2008} and implemented in \yambo.
Introducing
\begin{eqnarray}
\label{eq:rhotw-term}
\tilde{\rho}_{m\kk}(\GG,\GG')&=&\langle m\kk | e^{i(\GG'+\GG)\cdot\hat{\mathbf{r}}} | m \kk \rangle, \\
\tilde{F}_{m\kk}(\omega,\bar{\epsilon}_{\chi_0})&=&
    \Big[ \frac{1}{\omega -(\epsilon_{m\kk} - \bar{\epsilon}_{\chi_0}) -i\eta} \nonumber \\
\label{eq:F-term}
&&\phantom{C_{nm}}
           -\frac{1}{\omega -(\bar{\epsilon}_{\chi_0} -\epsilon_{m\kk}) +i\eta}  \Big]   
\end{eqnarray}
the correction to the independent particle response function $\chi$, see Eq.~\eqref{eq:X_term_def}, reads:

\begin{multline}
\label{X-term}
 \Delta\chi_{ {\bf{G}}{\bf{G'}}}({\bf{q}},\omega) = 
 \sum_{m\kk}
 \tilde{F}_{m\kk}(\omega,\bar{\epsilon}_{\chi_0})
  \\ 
  \left[ \frac{f_s\,f_{m\kk}\,\tilde{\rho}_{m\kk}(\GG,\GG')}{N_{\bf{k}} \Omega}
  -\sum_{n \le N_b'}R^{nm\kk}_{\GG \GG'}(\qq) \right].
\end{multline}

In Eq. \eqref{X-term}, the parameter $\bar{\epsilon}_{\chi_0}$ denotes the extrapolar energy for the polarizability, while $N_b^{\prime}$ is the number of conduction band states included in the calculation. Finally, as in sec.~\ref{sec:linear_resp}, $f_s$ is the spin occupation factor, while $n$ and $m$ are band indexes.

%============================
\section{Covariant dipoles} \label{App:cov-dips}
%============================
%
In extended system the coupling of electrons with external fields is described in terms of Berry phase~\cite{souza_prb}. In this formulation the dipole operator is replaced by the derivative in $\bf k$-space, $\mathbf{r} = i \frac{\partial}{ \partial \kk} $. In case of a finite $\bf k$-points sampling the $\bf k$-derivative is replaced by a finite-difference representation, described in Refs.\cite{souza_prb,Attaccalite2013}. In the limit of linear response, it is possible to derive from this  representation a new formula for the dipole matrix elements as:
\begin{equation}
\langle m\kk |\mathbf{r} | n \kk \rangle = {\rm w}_{mn  \kk} + {\rm w}^+_{mn  \kk} + O(\Delta \kk^4), 
\end{equation}
with
\begin{equation}
        {\rm w}_{mn  \kk} = \frac{ie}{2} \sum_{i=\alpha}^3\, (\mathbf{r} \cdot {\bf a}_\alpha) \frac{ 4 D_{mn}(\Delta \kk_\alpha) - D_{mn}( 2 \Delta \kk_\alpha)}{3} , 
\label{eq:wkhat2}
\end{equation}
where ${\bf a}_\alpha$ is the crystal lattice versor. The $D_{mn}$ factors are 
\begin{align}
D_{mn}(\Delta {\kk}_\alpha) = \frac{{P}_{mn}(\kk+\Delta {\kk}_\alpha) 
- {P}_{mn}(\kk- \Delta \kk_\alpha)}{2 \Delta {\kk}_\alpha}, \label{eq:wkhat} 
\end{align}
with
\begin{align}
\Delta \kk_\alpha = \frac{2 \pi}{ |{\bf a}_\alpha| N_{\kk_\alpha^\parallel}},
\label{eq:Delta}
\end{align}
and
\begin{multline}
   {P}_{mn}(\kk + \Delta {\kk}_\alpha) = \sum_l^{\text{occ}} \, \left[S (\kk , \kk + \Delta {\kk}_\alpha)\right]_{ml} \times \\ 
  \left[ S^{-1}(\kk , \kk + \Delta {\kk}_\alpha)\right]_{ln}. \label{eq:proj}
\end{multline}
In Eq.~\eqref{eq:Delta}  $N_{\kk_\alpha^\parallel}$ is the number of $\kk$-points along the reciprocal lattice vector ${\bf b}_\alpha$, $S (\kk , \kk + \Delta \kk_\alpha)_{ml}$ is the overlap matrix between the orbitals $m$ and $l$ at $\kk$ and $\kk+\Delta \kk_\alpha$ points and $[S^{-1}(\kk , \kk + \Delta \kk_\alpha)]_{ln}$ is the inverse of the overlap matrix between the valence bands. \\ 
$P_{mn}(\kk_i + \Delta \kk_\alpha) $ are the matrix elements of the operators projecting the orbitals of the  $\kk_i + \Delta \kk_\alpha$ and $\kk_i - \Delta \kk_\alpha$ bands on  $\kk_i$ in such a way to cancel the phase factor and then the derivative is performed.

%%%%%%%%%%%%%%%%%%%%%%%%%%%%

%
% bibliography
%
%\section*{Bibliography}
%\bibliographystyle{iopart-num}
\bibliographystyle{aip}
\bibliography{paper}

\begin{thebibliography}{100}
\bibitem{yambo_web_note}
Content will eventually be updated and ported to a more user-friendly wiki
  style site.
\end{thebibliography}

\end{document}